\documentstyle[12pt,epsf,epsfig,amsmath]{article}
\newfont{\thiplo}{msbm10 scaled\magstep 2}
\newfont{\gothic}{eufb10 scaled\magstep 2}
\newfont{\unc}{eurb10} 
\newskip\humongous \humongous=0pt plus 1000pt minus 1000pt
\def\caja{\mathsurround=0pt}
\def\eqalign#1{\,\vcenter{\openup1\jot \caja
        \ialign{\strut \hfil$\displaystyle{##}$&$
        \displaystyle{{}##}$\hfil\crcr#1\crcr}}\,}
\newif\ifdtup

\def\eqright #1\cr{\noalign{\hfill$\displaystyle{{}#1}$}}
\def\eqleft #1\cr{\noalign{\noindent$\displaystyle{{}#1}$\hfill}}

\def\oldreffmt#1{\rlap{[#1]} \hbox to 2\parindent{}}

\def\figfmt#1{\rlap{Figure {#1}} \hbox to 1in{}}

\def\sectioneq{\def\theequation{\thesection.\arabic{equation}}{\let
\holdsection=\section\def\section{\setcounter{equation}{0}\holdsection}}}%

\newcounter{holdequation}

\def\auto{\eqno(\refstepcounter{equation}\theequation)}
\def\begineq #1\endeq{$$ \refstepcounter{equation}\eqalign{#1}\eqno
	(\theequation) $$}
\def\contlimit{\,{\hbox{$\longrightarrow$}\kern-1.8em\lower1ex
\hbox{${\scriptstyle (a\rightarrow0)}$}}\,}
\def\centeron#1#2{{\setbox0=\hbox{#1}\setbox1=\hbox{#2}\ifdim
\wd1>\wd0\kern.5\wd1\kern-.5\wd0\fi
\copy0\kern-.5\wd0\kern-.5\wd1\copy1\ifdim\wd0>\wd1
\kern.5\wd0\kern-.5\wd1\fi}}
\def\centerover#1#2{\centeron{#1}{\setbox0=\hbox{#1}\setbox
1=\hbox{#2}\raise\ht0\hbox{\raise\dp1\hbox{\copy1}}}}
\def\centerunder#1#2{\centeron{#1}{\setbox0=\hbox{#1}\setbox
1=\hbox{#2}\lower\dp0\hbox{\lower\ht1\hbox{\copy1}}}}
\def\lsim{\;\centeron{\raise.35ex\hbox{$<$}}{\lower.65ex\hbox
{$\sim$}}\;}
\def\gsim{\;\centeron{\raise.35ex\hbox{$>$}}{\lower.65ex\hbox
{$\sim$}}\;}
\def\st#1{\centeron{$#1$}{$/$}}

\def\super#1{\ifmmode \hbox{\textsuper{#1}}\else\textsuper{#1}\fi}
\def\textsuper#1{\newcount\holdspacefactor\holdspacefactor=\spacefactor
$^{#1}$\spacefactor=\holdspacefactor}

\def\getcite#1,{\advance\citenumber by1
\def\getcitearg{#1}\def\lastarg{@}
\ifnum\citenumber=1
\ref{#1}\let\next=\getcite\else\ifx\getcitearg\lastarg\let\next=\relax
\else ,\ref{#1}\let\next=\getcite\fi\fi\next}

\def\pom{{\rm P\kern -0.53em\llap I\,}}
\def\spom{{\rm P\kern -0.36em\llap \small I\,}}
\def\sspom{{\rm P\kern -0.33em\llap \footnotesize I\,}}

\relax
\def\contlimit{\,{\hbox{$\longrightarrow$}\kern-1.8em\lower1ex
\hbox{${\scriptstyle (a\rightarrow0)}$}}\,}
\def\upon #1/#2 {{\textstyle{#1\over #2}}}
\relax
\renewcommand{\thefootnote}{\fnsymbol{footnote}} 

\def\mainhead#1{\setcounter{equation}{0}\addtocounter{section}{1}
  \vbox{\begin{center}\large\bf #1\end{center}}\nobreak\par}
\sectioneq
\def\subhead#1{\bigskip\vbox{\noindent\bf #1}\nobreak\par}

\def\til#1{\centeron{\hbox{$#1$}}{\lower 2ex\hbox{$\char'176$}}}
\def\tild#1{\centeron{\hbox{$\,#1$}}{\lower 2.5ex\hbox{$\char'176$}}}
\def\sumtil{\centeron{\hbox{$\displaystyle\sum$}}{\lower
-1.5ex\hbox{$\widetilde{\phantom{xx}}$}}}

\def\p{\unc p}

\newcommand{\bit}{\begin{itemize}}
\newcommand{\eit}{\end{itemize}}

\newcommand{\beq}{\begin{equation}}
\newcommand{\eeq}{\end{equation}}
\newcommand{\beqa}{\begin{eqnarray}}
\newcommand{\eeqa}{\end{eqnarray}}

\begin{document} 

\begin{titlepage} 

\rightline{\vbox{\halign{&#\hfil\cr
&ANL-HEP-PR-02-19\cr
&\today\cr}}} 
\vspace{0.25in} 

\begin{center} 
 
{\large\bf THE CHIRAL ANOMALY AND  HIGH-ENERGY}

{\large\bf SCATTERING IN QCD}\footnote{Work 
supported by the U.S.
Department of Energy, Division of High Energy Physics, \newline Contracts
W-31-109-ENG-38 and DEFG05-86-ER-40272} 
\medskip

Alan. R. White\footnote{arw@hep.anl.gov }

\vskip 0.6cm

\centerline{High Energy Physics Division}
\centerline{Argonne National Laboratory}
\centerline{9700 South Cass, Il 60439, USA.}
\vspace{0.5cm}

\end{center}

\begin{abstract} 

Infra-red properties of the triangle anomaly and the ``anomaly pole'' 
are elaborated and applied to the study of high-energy scattering in QCD, 
when the gauge symmetry is partially
broken to SU(2). 
It is shown that the chiral flavor anomaly provides a wee-gluon
component for Goldstone bosons that combines
with interactions due to the U(1) anomaly to produce an infra-red
transverse momentum scaling divergence in scattering amplitudes.  
After the divergence is factorized out, as a wee gluon condensate
in the infinite momentum pion, 
the remaining physical amplitudes have confinement and
chiral symmetry breaking. A lowest-order contribution 
to the pion scattering amplitude is calculated in detail.
Although originating from very complicated diagrams, the amplitude 
has a remarkable (semi-)perturbative simplicity. The momentum structure is
that of single gluon exchange but zero transverse momentum quarks 
inject additional spin and color structure via anomaly interactions.
 
\end{abstract}

\renewcommand{\thefootnote}{\arabic{footnote}} \end{titlepage}

\mainhead{1. INTRODUCTION}

Any solution of the full regge limit of QCD must, 
almost certainly, involve a resolution of the unsolved problem of 
matching perturbation theory with confinement.
Since the limit involves large energies 
it's description should not be too far from perturbation theory. Conversely,
since small momentum transfers are involved, both
confinement and chiral symmetry breaking must be manifest in 
the contribution of physical $t$-channel states. In this paper
we will show that a transition from perturbation theory to confinement
can indeed occur in the regge region.

For some time we have pursued what might be called a ``semi-perturbative''
description of the QCD pomeron. In doing so we have 
made extensive use of the formalism of (multi-)regge theory, 
which many authors currently studying 
the pomeron make little or no reference to. In this paper we endeavor to keep,
at least the most unfamiliar parts of, this formalism to a minimum. 
Nevertheless, we can summarize the reasons why we believe that 
regge poles and regge theory must play a fundamental role in solving the 
regge limit of QCD as follows. 

In general, 
multiparticle $t$-channel unitarity has been shown to be satisfied 
when the only $J$-plane singularities are regge poles and the regge cuts
generated by them - provided the regge cut discontinuities satisfy 
``reggeon unitarity'' \cite{gpt}-\cite{arw00}. No other solution is known.
It is well established\cite{fkl}-\cite{arw93} 
that when the gauge symmetry of QCD is spontaneously broken,
multi-regge limits of quark and gluon
amplitudes are described perturbatively by reggeon diagrams 
containing massive gluon and quark regge poles. Both $t$-channel (reggeon) 
unitarity and $s$-channel unitarity are satisfied. General arguments
imply that the small
transverse momentum part of the massless theory
can be obtained smoothly from the massive theory. In which case, the 
unitarity properties of the massive theory, including reggeon unitarity, 
should persist in the 
massless theory. (Note that the BFKL pomeron, which is not
a regge pole and also does not satisfy $s$-channel unitarity, is a large
transverse momentum phenomenon that  
appears only when a subclass of diagrams is isolated and
summed to all orders - without a transverse momentum cut-off.)

The critical pomeron\cite{cri} is an abstract
solution of reggeon unitarity, obtained
via reggeon field theory (RFT), that produces asymptotically 
rising cross-sections. A single regge pole and the
corresponding regge cuts are the only $J$-plane singularities.
Since the critical pomeron retains the factorization
properties of a single regge pole, if it
occurs in QCD it will be associated\cite{arw84} 
with a ``universal wee-parton distribution'' in hadrons. This 
universality property
allows wee partons to carry vacuum properties which, in combination 
with rising cross-sections, should lead to the 
maximal asymptotic applicability of the parton model.

We expect the occurrence of the
critical pomeron in QCD to be of crucial importance, therefore, both for 
the satisfaction of $t$-channel unitarity and for the maximal validity of the
parton model. To see that it can indeed occur we
have proposed\cite{arw93,arw84,arw80} starting with the gluon and quark
reggeon diagrams of spontaneously-broken QCD.
With a transverse momentum cut-off imposed, 
the gauge symmetry can be restored in stages
and RFT can be used to analyze the infra-red divergences that occur. 
We have anticipated that the only 
additional ingredient beyond the perturbative regge behavior of gluons 
and quarks will be chirality transitions produced by the fermion anomaly.
Quarks will, therefore, play an essential role.

We have now shown that chirality transitions occur\cite{arw011}-\cite{arw98} 
in effective triangle diagram reggeon interactions obtained by 
placing quark lines on-shell in large quark loops. These interactions 
appear in the reggeon vertices that couple different reggeon channels
(in a general multi-regge limit). In particular, they occur in 
the triple-regge vertex\cite{gw} that couples three distinct
reggeon channels - each
carrying a separate transverse momentum. 
Such vertices include the couplings of bound-state
reggeons (e.g. pions and nucleons) together with their couplings
to the physical pomeron. Effectively, therefore, vertices of this kind 
determine the bound-states of the theory and their high-energy scattering
amplitudes.

Our expectation has long been that
when the gauge symmetry is restored first
to SU(2), giving ``color superconducting QCD'', 
SU(2) color confinement will be due to the appearance of a 
condensate in reggeon states  produced by 
infra-red divergent ``wee-gluon'' configurations coupling through  
anomaly interactions. The resulting pomeron 
could then be in a supercritical phase\cite{arw91} 
of RFT,  implying that the 
critical pomeron would occur as the full SU(3) gauge symmetry is restored
(provided the transverse cut-off can be removed first - a strong requirement).
 A-priori, however,
to understand in detail how the anomaly interactions  
produce the condensate, and determine both hadron states 
and the pomeron, it is necessary to self-consistently
construct the full multi-regge S-matrix. 
This is a very complicated project to carry out. We  
outlined, essentially, how it could be done in \cite{arw98}, 
although we did not then have the full knowledge of anomaly vertices 
that we now have.
 
In this paper, as an intermediate step 
before attempting to construct the full multi-regge
S-Matrix, we approach the problem from a different stand-point.
We use a procedure that is less rigorously 
formulated (as will become apparent)
than the multi-regge approach. However, it leads directly to 
explicit results and provides a straightforward 
understanding of the physics that is involved. Also 
the terminology used is, we hope, more widely familiar.
The new approach is not only sufficient to show how,
in infinite momentum scattering, anomalies determine 
both the physical states and the exchanged pomeron,
but it also allows us to obtain explicit high-energy
scattering amplitudes. In fact,
we directly calculate
the on-shell (massless) pion amplitude rather than the amplitude for
spacelike reggeized pions to scatter, that multi-regge theory would
lead us to try to calculate.

We start directly from infra-red properties of the triangle diagram.
It is well-known\cite{gth1}-\cite{ach} 
that, when the quarks involved are massless, 
the chiral flavor anomaly requires that an
``anomaly pole'' appear in the vertices for an  
axial current to couple to
pairs of vector currents carrying light-like momenta. If there is
confinement and the chiral symmetry is broken spontaneously this pole
becomes a physical Goldstone pole. In Section 2 we 
study in detail how the pole 
is generated in the triangle diagram and show that, in the 
momentum configuration involved, one propagator carries 
zero momentum (and undergoes a chirality transition)
while the other two carry the external light-cone momenta. We also 
show that while the tensor coupling of the anomaly pole necessarily vanishes 
on-shell at finite momentum, an on-shell 
coupling potentially exists at infinite momentum.

In Section 3 we show that in color superconducting
QCD the role of light-cone momenta in producing  
the anomaly pole implies the existence of
crucial wee-gluon effective couplings for Goldstone bosons
at infinite momentum.
(The massive gluons produced by the color symmetry breaking are
essential for the existence of these couplings.)
As a result, 
the quark/antiquark ``pion'' and quark/quark (or antiquark/antiquark) 
``nucleon'' Goldstone states that appear\cite{kog}  
have just the (massless) wee gluon content that we envisaged emerging from
our general reggeon diagram analysis. The presence of the wee gluons 
then leads directly to the contribution of U(1) anomaly reggeon
interactions in the high-energy scattering of pions (and nucleons). 
An overall logarithmic divergence is produced that selects anomaly
mediated scattering as the dominant physical process. The divergence
can be factorized off as the expected ``condensate'' within the scattering
pions - with the residue being the physical scattering amplitude.
The ``anomaly pole'' is manifest as a transverse
momentum $\delta$-function that factorizes the momentum
dependence of the divergent wee gluon
interactions and 
the ``parton interaction'' of the massive sector of the theory. 

The lowest-order contribution to the pion scattering
amplitude has a remarkable simplicity. The momentum structure is
just that of lowest-order gluon exchange. However, zero momentum quarks 
inject spin and color structure (via anomaly interactions) that
modifies the signature and color symmetry properties of the amplitude.
Because of the complexity of the initial
diagrams and the resulting reduction process we limit the 
presentation, in this paper, to an ``existence proof'' that demonstrates
how the kinematical and 
dynamical properties of the chiral flavor and U(1) 
anomalies actually combine with 
transverse momentum divergences to produce physical amplitudes.
To do this we follow the
reduction process through in detail for just one of the diagrams involved. 

In Section 4 we discuss both the conclusions that can be drawn from our 
results and the further work that needs to be done to establish the
relationship of the critical pomeron to QCD. We also discuss some more 
general issues of principle.

\newpage

\mainhead{2. PROPERTIES OF THE TRIANGLE GRAPH}

In our previous paper\cite{arw01} we based our infra-red anomaly analysis 
on the rather abstract discussions of \cite{cg} and \cite{bfsy}.
In this paper we will use 
explicit evaluations of the triangle graph (in particular kinematic 
configurations) that exist in the literature\cite{hor,ach}. In the following 
we summarize and expand 
the results and properties we will use. We will particularly 
emphasize the important role of (both external and internal)
light-cone momenta in the infra-red properties that we exploit. 

\subhead{2.1 Invariant Amplitudes and Ward Identities}

We consider the elementary triangle diagram amplitude
$$
\Gamma_{\mu \alpha \beta}(k_1,k_2)~
=~ {1 \over (2 \pi)^4} \int {  d^4 p~ Tr \{ \gamma_5
\gamma_{\mu} ~ (\st{k}_2
- \st{p})~  \gamma_{\alpha}~ ~( - \st{k}_1 + \st{k}_2 - \st{p}) ~
\gamma_{\beta}~ (-\st{k}_1 - \st{p} ) \} 
\over  (p + k_1 - k_2)^2  (k_2 - p)^2 
 (p + k_1)^2 }
\auto\label{tamp}
$$
where the notation is illustrated in Fig.~2.1.
\begin{center}
\leavevmode
\epsfxsize=2.3in
\epsffile{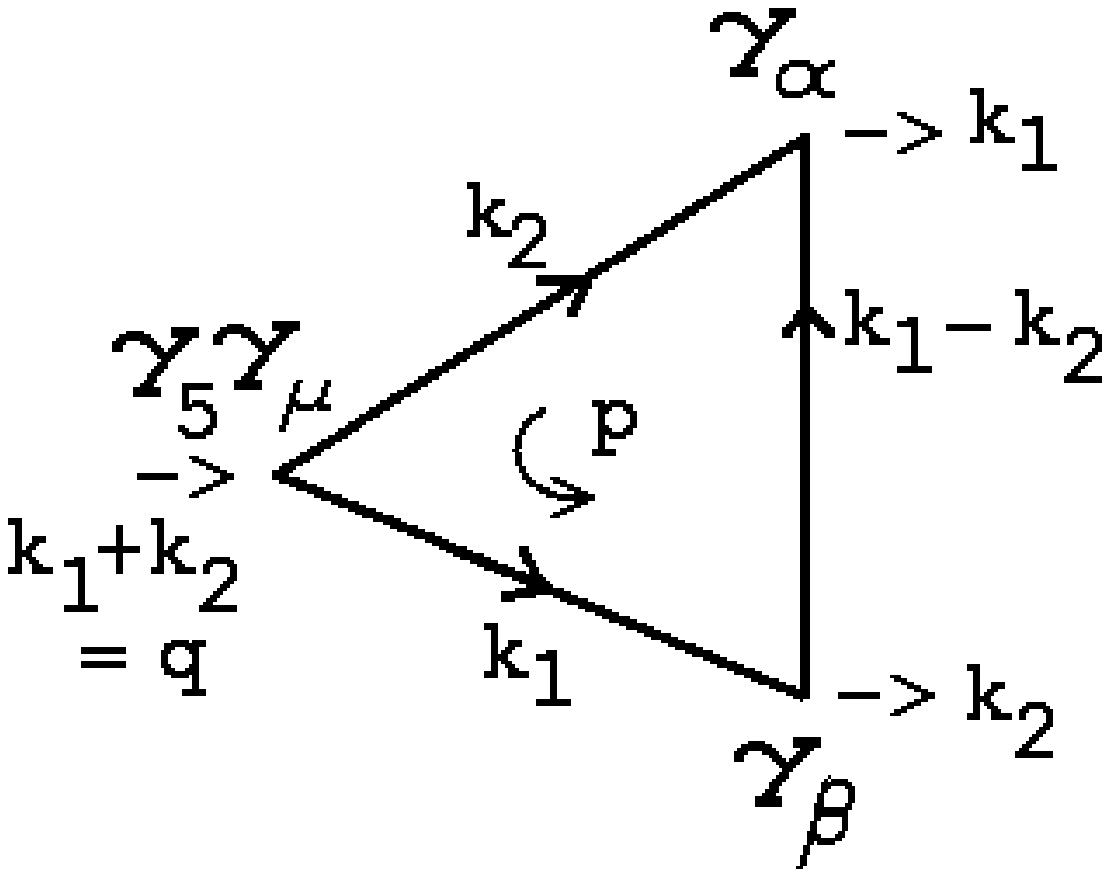}

Fig.~2.1 Triangle Diagram Notation
\end{center}
The significance of routing the external momenta as we have done will be 
discussed shortly. The amplitude 
$$
T_{\mu \alpha \beta}(k_1,k_2)~=~\Gamma_{\mu \alpha \beta}(k_1,k_2)
~+~ \Gamma_{\mu \beta \alpha}(k_2,k_1)
\auto\label{tamp12}
$$
is the lowest order interaction of the axial vector current 
$A_{\mu}(q)$, where 
$A_{\mu} = \bar{\psi}\gamma_5 \gamma_{\mu}\psi$ and
the vector currents, $V_{\alpha}(k_1)$ and $V_{\beta}(k_2)$,
where $V_{\mu} = \bar{\psi}\gamma_{\mu}\psi$ and $\psi$ is a massless
fermion field. 

$T_{\mu \alpha \beta}$ can be decomposed into invariant amplitudes by writing 
$$
\eqalign{T_{\mu \alpha \beta}(k_1,k_2) ~&= ~ A_1~
{\hbox{\large $\epsilon$}}_{\sigma\alpha\beta\mu}~ k_1^{\sigma}  ~+~ A_2~ 
{\hbox{\large $\epsilon$}}_{\sigma\alpha\beta\mu} ~k_2^{\sigma} 
~+~A_3~
{\hbox{\large $\epsilon$}}_{\delta \sigma\alpha\mu}~ 
k_{1\beta}k_1^{\delta} k_2^{\sigma}  \cr
~~~& +~A_4~  {\hbox{\large $\epsilon$}}_{\delta \sigma\alpha\mu}
~ k_{2\beta}k_1^{\delta}
k_2^{\sigma}~+~A_5~  {\hbox{\large $\epsilon$}}_{\delta \sigma\beta\mu}
~k_{1\alpha}k_1^{\delta}
k_2^{\sigma}~+~A_6~ {\hbox{\large $\epsilon$}}_{\delta \sigma\beta\mu} 
~ k_{2\alpha}k_1^{\delta}
k_2^{\sigma} }
\auto\label{inde}
$$
Bose symmetry implies 
$$
T_{\mu \alpha \beta}(k_1,k_2)~= ~ T_{\mu \beta \alpha}(k_2,k_1)
\auto\label{bsym0}
$$
and so requires that
$$
\eqalign{A_1(k_1,k_2)~&=~-A_2(k_2,k_1) \cr 
A_3(k_1,k_2)~&=~-A_6(k_2,k_1) \cr 
A_4(k_1,k_2)~&=~-A_5(k_2,k_1) } 
\auto\label{bsym}
$$
In addition, the vector Ward identities
$$
k_1^{\alpha}~\Gamma_{\mu \alpha \beta}~=0 ~,~~~
k_2^{\beta}~\Gamma_{\mu \alpha \beta}~=0
\auto \label{vwi}
$$
require 
$$
A_2~=~k_1^2~A_5 ~+~k_1\cdot k_2 ~A_6
\auto\label{vwi1}
$$
$$
A_1~=~k_2^2~A_4 ~+~k_1\cdot k_2 ~A_3
\auto\label{vwi2}
$$

A-priori, a term of the form 
$$
 A(k_1,k_2)~
{\hbox{\large $\epsilon$}}_{\delta \sigma\alpha\beta}~ k_1^\delta 
k_2^{\sigma}~(k_1 + k_2 )_{\mu}  
\auto\label{exa}
$$
with $A(k_2,k_1)= - A(k_1,k_2)$ ,
could be added to (\ref{inde}). Such a term separately satisies the 
Ward identities (\ref{vwi}). However, because of the identity
$$ 
\eqalign{~~~{\hbox{\Large $\epsilon$}}_{\delta\sigma\alpha\beta} 
 k_1^{\delta} k_2^{\sigma} [k_1 + k_2]_{\mu} ~=~&
- ({\hbox{\Large $\epsilon$}}_{\delta \sigma\alpha\mu}k_{1\beta}
 -{\hbox{\Large $\epsilon$}}_{\delta \sigma\beta\mu}k_{2\alpha} 
 -{\hbox{\Large $\epsilon$}}_{\delta \sigma\beta\mu}k_{1\alpha}
 +{\hbox{\Large $\epsilon$}}_{\delta \sigma\beta\mu} 
k_{2\beta}) k_1^{\delta}k_2^{\sigma}\cr
 &
~+~{\hbox{\Large $\epsilon$}}_{\sigma\alpha\beta\mu}k_2^{\sigma}~
 (k_1\cdot k_2 - k_1^2)
~-~{\hbox{\Large $\epsilon$}}_{\sigma\alpha\beta\mu}k_1^{\sigma}~ 
(k_1\cdot k_2 -k_2^2)}
\auto\label{epid}
$$
(\ref{exa}) can be re-expressed in the form (\ref{inde}).
As we will see, the identity (\ref{epid}) can be used\cite{hor}
to simplify (\ref{inde})
in many special kinematic situations. Note that, with $A$ a constant, 
(\ref{exa}) does not satisfy
Bose symmetry.  Nevertheless, in asymmetric momentum configurations
it can, effectively, appear with a constant coefficient. This
will be important for the discussion later in this Section. 

We define the integral (\ref{tamp}) as the limit $m \to 0$ of that in which 
a fermion mass $m$ is added. (\ref{inde}) and (\ref{bsym}) hold also when 
$m \neq 0$ and ultra-violet regularization 
can be carried out with (\ref{vwi1}) and (\ref{vwi2})
maintained. Indeed, it is well-known that the Ward identities (\ref{vwi})
can be regarded as a consequence of routing the external
momenta as we have done in Fig.~2.1. An ``anomaly ''
then appears in the Ward identity for the axial current.
Since only the $A_1$ and $A_2$ terms in (\ref{inde}) contribute to  
the axial current divergence, the anomaly has to appear in these 
terms. In fact, ultra-violet regularization of
(\ref{tamp}) directly produces the contribution 
$$
T_{\mu \alpha \beta}(k_1,k_2) ~= ~ {1\over 4 \pi^2}~
{\hbox{\large $\epsilon$}}_{\sigma\alpha\beta\mu}~ k_1^{\sigma}  ~+~ 
{1\over 4 \pi^2}~ 
{\hbox{\large $\epsilon$}}_{\sigma\alpha\beta\mu} ~k_2^{\sigma} ~~+~~\cdots
\auto\label{uvco}
$$ 
leading to the divergence equation (when $m=0$)
$$
(k_1 + k_2)^{\mu}~T_{\mu \alpha \beta}~=~
{1\over 2 {\pi}^2 }~{\hbox{\Large $\epsilon$}}_{\delta\sigma\alpha\beta} 
~k_1^{\delta} k_2^{\sigma}
\auto\label{awi}
$$

The (coefficient on the) right-hand side of (\ref{awi})
is commonly referred to as ``the anomaly''. 
Even though the anomaly occurs only in $A_1$ and $A_2$ the vector Ward 
identities (\ref{vwi1}) and (\ref{vwi2}) require related terms to
appear in the other $A_i$.
We will be particularly concerned with the infra-red
behavior of the massless $A_i$ that is required\cite{cg,bfsy}.

\subhead{2.2 Explicit Formulae for the $A_i$}

No (non-integral) analytic expression for the full amplitude
(\ref{tamp}) exists in the literature. However, it is possible to 
give explicit expressions in limited kinematic configurations.
For example, when $k_1^2=k_2^2~$ the imaginary parts of each of the
invariant amplitudes $A_i$ is given in \cite{hor}. For our purposes we 
will use the following set of formulae, given in \cite{ach}, for the full 
amplitudes.

When $k_1^2 = 0$  ($k_2^2, q^2 < 0,~ m^2 >0$)~,
$$
\eqalign{A_6~&=~-A_3~= ~- ~{1 \over 2\pi^2}{1 \over k^2_2 -q^2} \biggl(
{k_2^2 \over k_2^2 -q^2 } L_1 ~-~ {m^2 \over k_2^2 -q^2 } L_2 ~- ~1\biggr) \cr
A_4~&= ~ {1 \over 2\pi^2}{1 \over k^2_2 -q^2 } ~ L_1 \cr
A_2~&= ~ {1 \over 4\pi^2}\biggl(
{k_2^2 \over k_2^2 -q^2} L_1 ~-~ {m^2 \over k_2^2 -q^2 } L_2 ~- ~1\biggr)\cr
A_1~&= ~ {1 \over 4\pi^2}\biggl(
{k_2^2 \over k_2^2 -q^2 } L_1 ~+~ {m^2 \over k_2^2 -q^2 } L_2 ~+ ~1\biggr)\cr
A_5 &= -A_4 - {3 \over \pi^2}k_2^2 {d \over d k_2^2}\biggl( {1 \over   
k_2^2 - q^2 } L_1 \biggr) + 
{3 \over 2\pi^2}k_2^4 \biggl({d \over d k_2^2}\biggr)^2\biggl( {1 \over   
k_2^2 - q^2 } L_1 \biggr) \cr
& ~~~~ + {3 \over 4\pi^2}k_2^2 {d \over d k_2^2}\biggl( {1 \over   
k_2^2 -q^2  } L_2 \biggr) +
{1 \over 2\pi^2}m^2 k_2^2 \biggl({d \over d k_2^2}\biggr)^2
\biggl( {1 \over k_2^2 -q^2 } L_2 \biggr)}
\auto\label{k2q}
$$
where
$$
\eqalign{ ~~~~L_1 ~&=~ -~ \rho~ \ln{ \rho +1 \over \rho -1} ~+~
 \beta ~\ln{ \beta +1 \over \beta -1} ~~~~~~~~~~~~~~~~~~~~~~~~~~~~~~~~~~\cr
 L_2 ~&=~ - ~\rho ~\ln^2{ \rho +1 \over \rho -1}~ + ~
 \beta ~\ln^2{ \beta +1 \over \beta -1} \cr
\rho^2~&=~ 1 ~- ~ 4m^2/q^2 ~,~~ ~~ \beta^2~=~ 1 ~- ~ 4m^2/k_2^2 }
\auto\label{L1L2}
$$
Note that the simple relationship between $A_6$ and $A_2$ 
in (\ref{k2q}) is required by the
Ward identity (\ref{vwi1}) which, when $k_1^2=0$, becomes
$$
A_2~=~k_1 \cdot k_2~A_6 ~=~{q^2 - k_2^2 \over 2} ~A_6
\auto\label{vwi11}
$$

If the limit $ m^2 \to 0$ is taken in (\ref{k2q}) the result is
$$
\eqalign{~~~~A_1~&=~{1\over 4{\pi}^2} \biggl({k_2^2 \over k_2^2 -q^2 }~ln{k_2^2
\over q^2} ~+~1 \biggr)~~~~~~~~~~~~~~~~~~~~~~~~~~~~~~~~~~~~~~\cr 
A_2~&=~{1\over 4{\pi}^2} \biggl({k_2^2 \over k_2^2 -q^2 }~ln{k_2^2
\over q^2} ~-~1 \biggr) \cr
A_3~&=-A_6~=~{1 \over 2{\pi}^2 }{1 \over k_2^2 -q^2}
 \biggl({k_2^2 \over k_2^2 - q^2}~ln{k_2^2
\over q^2} ~-~1 \biggr) }
\auto\label{k1m0}
$$
While the Ward identity (\ref{vwi1}) does not determine $A_5$ in this limit, 
$A_4$ can be obtained from (\ref{vwi2}). 

If instead the limit $ k_2^2 \to 0$ is taken, with $m^2 >0$, the result is
$$
\eqalign{~~~~A_6 ~&= ~-A_3 ~= ~ {1 \over 2 \pi^2} ~{1 \over q^2} \biggl( 1 +
  ~{m^2 \over q^2} ~\ln^2{ \rho +1 \over \rho -1}~\biggr) ~~~~~~~~~~~~~~~~
~~~~~~~~~~~~~~~\cr
A_4 ~&=~-A_5 ~=~-~ {1 \over 2 \pi^2} ~{1 \over q^2} \biggl( 2 -
 ~ \rho ~\ln{ \rho +1 \over \rho -1}~\biggr) }
\auto\label{k1k20}
$$
$A_1~=-A_2$ can be obtained from the vector
Ward identities and using (\ref{epid}) gives\cite{ach} 
$$
\eqalign{T_{\mu \alpha \beta} ~&= ~ A_6~q_{\mu}
{\hbox{\Large $\epsilon$}}_{\alpha\beta\sigma\delta}~ k_1^{\sigma}
k_2^{\delta} \cr  & ~+~ (A_4 + A_6)~ 
({\hbox{\Large $\epsilon$}}_{\mu\alpha\sigma\delta} ~k_1^{\sigma} 
k_2^{\delta}k_{2\beta}~-~
{\hbox{\Large $\epsilon$}}_{\mu\beta\sigma\delta} ~k_1^{\sigma} 
k_2^{\delta}k_{2\alpha}) }
\auto\label{Tk0}
$$
where $A_4$ and $A_6$ are given by (\ref{k1k20}). Note that the first 
term has the form of (\ref{exa}). This is consistent just because 
$k_1^2= k_2^2 =0, ~q^2\neq 0$ is not possible in a symmetric momentum 
configuration. Also the anomaly is produced by the first term alone
while, within the momentum configuration that we are discussing,
each term separately satisfies the vector Ward identities.  

When $k_2^2 \to 0$, with $q^2$ fixed, 
(\ref{k1m0}) gives
$$
A_{1,2} ~\to~\pm~ {1 \over 4 \pi^2 }~, 
~~~~ A_3~\to ~ {1 \over 2 {\pi}^2}~  {1 \over q^2}
\auto\label{k20}
$$
That is, a pole appears in $A_3$ ($= - A_6$).
If, instead, we integrate over spacelike values of $q^2$, we obtain 
$$
\int~ dq^2~A_3(q^2,k_2^2)~f(q^2,k_2^2)~
\centerunder{$\longrightarrow$}{\raisebox{-6mm}{$k_2^2 \to 0$}} ~ ~
{1\over \pi}~f(0,0)~=~ \int~ dq^2~{1\over \pi}\delta(q^2)~f(q^2,0)
\auto\label{dfn}
$$
(provided $f(q^2,k^2_2)$ is regular at $q^2,k_2^2 = 0$). 

The pole that appears in $A_3$ (and $A_6$) is the ``anomaly pole'' 
discussed by a number of authors\cite{cg}-\cite{ach}. The coefficient 
coincides with that of the anomaly and it is possible to give general
arguments\cite{cg,bfsy} that this pole is directly required by (\ref{awi}). 
The simplest way to see that this might be the case 
is to note that if $k_1^2=k_2^2=0$ 
and $A_4$ and $A_5$ are not (sufficiently) singular
the identities (\ref{vwi1}) and (\ref{vwi2}) reduce to the very simple
form  
$$
A_3~=~ {2 \over q^2}~A_1 ~, ~~~~ A_6~=~ {2 \over q^2}~ A_2  
\auto\label{vwi3}
$$
(\ref{uvco}) then leads directly to (\ref{k20}). In fact, we will
see below how the momentum routing of Fig.~2.1 that produces the 
ultra-violet anomaly  (\ref{uvco}) is also responsible 
for the numerator that accompanies the anomaly pole. 
 
The amplitudes $A_4$ and $A_5$ will play very little role in our discussion.
It is well-known that these amplitudes do not contribute at $k_1^2=k_2^2=0$
when $T_{\mu\alpha\beta}$ is contracted with physical polarization tensors.
Our analysis will also be concerned with momentum configurations and
components of $T_{\mu\alpha\beta}$ such that these amplitudes do not 
contribute. Note that (\ref{k1k20}) implies, 
and it is straightforward to check directly
from (\ref{k2q}), that the limits $m \to 0$ and $k_1^2, k_2^2 \to 0$
do not commute for $A_4$ and $A_5$. A property that we will avoid in our 
analysis.

Finally we emphasize that if we keep only the pole
terms in $T_{\mu\alpha\beta}$, as we will eventually do, then the vector
Ward identities will necessarily be violated, for at least some momenta. 
As elaborated in \cite{arw98} the reggeon Ward identities that are necessary
to avoid infra-red divergences in regge limit amplitudes depend on Ward
identities being satisfied for all momenta. We will see below
that when the Ward identities are
satisfied only by a limited range of momenta, infra-red divergences occur 
that, nevertheless, produce gauge-invariant amplitudes.

\subhead{2.3 Interpretation}

The results of the previous sub-section extend straightforwardly to the
case when there are gauge and flavor symmetries and 
$T_{\mu \alpha \beta}$ is a three-point amplitude for currents  
defined in terms of appropriate combinations of fermion fields.
The anomaly in (\ref{awi}) is then a number determined by adding  
all contributing triangle diagrams. Most importantly, as is very well known,
the ultra-violet anomaly in all (flavored)
axial current Ward identities remains unchanged as gauge field interactions 
are included\cite{awb}. As a result, the general arguments alluded to 
above\cite{cg,bfsy} (and more directly the identities
(\ref{vwi3}) ) determine that a pole
with coefficient given by the anomaly is always present in the special
kinematic configuration $k_1^2=k_2^2=m^2=0$. 
As first argued by 't Hooft\cite{gth1}, if there is confinement and there are 
no physical massless fermions, this pole 
has to be reproduced by a Goldstone boson pole. As we will discuss in the 
next Section,
this will provide the basis for our use of the chiral flavor anomaly 
to extract ``infinite momentum'' 
pion couplings to physical current
components that produce scattering amplitudes. 
For the U(1) anomaly there will be no Goldstone boson pole
but instead the $\delta$-function (\ref{dfn}), produced by 
the integration of the anomaly pole over $q^2$, 
will contribute in an essential manner to infinite momentum amplitudes.
 
To interpret the pole in
(\ref{tamp}) in terms of Landau singularities we note the following. 
The expressions for the $A_i$ given above demonstrate 
that, when a fermion mass $m$ is present, 
only two-particle normal thresholds 
are present in each invariant channel. These thresholds are 
responsible for the $\ln{q^2}$ and $\ln{k_2^2}$ 
factors that are present in (\ref{k1k20}). The pole at 
$q^2 = k_2^2$, which is superficially present in each of the $A_i$,
cancels when the logarithms have their physical
sheet values. On unphysical sheets of the logarithms a pole is present
and corresponds to the triangle Landau singularity. When $k_2^2 \to 0$
followed by $q^2 \to 0$ the physical sheet thresholds coincide at the 
point of interest and the unphysical sheet singularity is able to 
enhance the thresholds. 

The simplest example of this last discussion is provided by (\ref{k1k20})
which gives  
$$
\eqalign{~~A_6(q^2,m^2)& ~~
\centerunder{$\longrightarrow$}{\raisebox{-6mm}{$q^2 \to 0$}}
 ~~ ~ {1 \over 2 \pi^2} ~{1 \over q^2} \biggl( 1 +
  ~{m^2 \over q^2} ~\ln^2{ [1 + ({-q^2 \over m^2})^{1\over 2} + \cdots}] 
~\biggr) ~~~~~~\st{\longrightarrow}~~\infty ~~~~~~~~~~~
~~~~~~~~~~~~~~~\cr
A_4(q^2,m^2) &~~ \centerunder{$\longrightarrow$}{\raisebox{-6mm}
{$q^2 \to 0$}} 
~~~ {1 \over 2 \pi^2} ~{1 \over q^2} \biggl( 2 -
 ~ \rho ~\ln{[ 1 + {2 \over \rho} + \cdots] }\biggr)~
~~~~~~~~~~~\st{\longrightarrow}~~\infty}  
\auto\label{k1k201}
$$
and so, for $m^2 \neq 0$, the pole is absent. The only finite $q^2$ 
singularity in either amplitude is the threshold at $q^2 = 4m^2$. If 
we continue around this threshold then $\rho \to - \rho$ and so
$$
\ln{ \rho +1 \over \rho -1} ~\to ~ \pi i - \ln{ \rho +1 \over \rho -1}
~~~~~~~~~~~~~~~~~
\auto\label{k1k202}
$$
and the pole at $q^2 = 0$ is present. It is present on the physical sheet
only at $m^2 = k_1^2=k_2^2 =0$.  

We conclude that the Goldstone boson pole appears, in very special kinematics, 
because an unphysical singularity enters the edge of the physical region
in the massless limit. 
It occurs in $A_3$ (and $A_6$), and not in the other 
$A_i$, because the unphysical singularity is a double pole rather than
a single pole. It would be interesting to determine more explicitly
how this feature
relates to the momentum routing ambiguity associated with the anomaly.

\subhead{2.4 Internal Momentum Analysis}

In the next Section we will want to derive anomaly pole couplings from
the reduction of more complicated diagrams to triangle diagrams and also 
to separate the anomaly pole from
the ultra-violet anomaly contribution. For these purposes  
it is important to determine the internal momenta $p$  in (\ref{tamp}) 
that generate the pole. We will see that 
light-cone momenta play a crucial role. Note that an external 
light-cone momentum is necessarily involved since 
if $k_1^2 = k_2^2 = q^2 = 0$ then, necessarily, 
$k_1 \parallel k_2 \parallel k_+$ where $k_+$ is light-like. 
We first consider reaching the $q^2=0$ limit via the momentum 
configuration
$$
\eqalign{k_1 &= (k_+/\sqrt{2},k_+/\sqrt{2},0,0) ~~~~ ~~
\equiv ~~~~k_1^+= k_{1-}= k_+~, ~~k_1^- = 0~ ,~~k_{\perp} =0 \cr
\hbox{$\vspace{0.2in}$}
k_2 &= (-k_-/\sqrt{2},k_-/\sqrt{2},0,0) ~~~~ 
\equiv ~~~~k_2^+= 0~,~~ k_2^- =k_{2+} = -k_-~ ,~~k_{\perp} =0 }
\auto\label{k+k-}
$$
in which $k_1^2 = k_2^2 = 0$ and $q^2 = - 2 k_+ k_-$. 

We will shortly understand the anomaly pole contribution to (\ref{tamp})
as produced by external momentum numerator factors together with a pole
produced (by the denominators) in a part of the integration region
that includes zero internal momentum. At first sight,
(\ref{k+k-}) is not a very sensible configuration to
discuss. If we consider the pole contribution of $A_3$ 
to $ T_{32-}$, for example, this has the form
$$ 
 T_{23-}~=~-~ {\hbox{\Large $\epsilon$}}_{\sigma\delta 2 3}~ 
{k_1^{\sigma} k_2^{\delta}~k_{1-} \over 2 \pi^2 q^2} 
~~=~-~{k_+^2 k_- \over 2 \pi^2 q^2}~~=~ {k_+ \over 4 \pi^2 }
\auto\label{A3an}
$$
and so there is no divergence as $q^2 \to 0$. At best we can obtain a finite
contribution by taking $q^2 \sim k_- \to 0$, with $k_+$ kept finite.
As a consequence, in the momentum configuration (\ref{k+k-}), the anomaly pole 
contribution can not be distinguished from other non-singular contributions. 
However, for our initial goal of obtaining a simple
understanding of the origin of the denominator 
pole the momentum configuration (\ref{k+k-}) will be very useful. (Indeed, it
will play a key role throughout the paper.)

If we drop the numerator terms
in (\ref{tamp})and keep only the $k_+$ and $k_-$ - dependence  
we obtain 
$$
\eqalign{I(k_+,&k_-,m^2)~ =~ I(q^2, m^2)~= ~ \int~ dp_+ dp_- d^2 p_{\perp}  
~ [2p_+(p_- - k_-) - p_{\perp}^2 -m^2 + i\epsilon ]^{-1} \cr
&[2(p_+ - k_+)(p_- - k_-) - p_{\perp}^2  -m^2+ i\epsilon]^{-1} 
[2(p_+ -  k_+)p_- - p_{\perp}^2 - m^2 +i\epsilon]^{-1} }
\auto\label{a1}
$$
We will find that $I(q^2, m^2)$ is finite as $\epsilon \to 0$ only when 
$m^2 \neq 0$. This is not surprising since 
$I(q^2,m^2)$ is closely related to $A_4$ and $A_6$, as given by
(\ref{k1k20}). As we already noted above, the $k_1^2, k_2^2
\to 0$ limit commutes with the massless limit for $A_6$, but not for
$A_4$. As a result, we expect that for part of  $I(q^2,m^2)$
the limit $m^2 \to 0$ will not exist. ( Of course, the numerator terms 
in (\ref{tamp}) will play a central role in determining the nature of the
divergence that occurs.) However, the pole term we
are looking for appears, with the same (anomaly) coefficient, in both
kinematic terms in (\ref{Tk0}) as $m^2 \to 0$. We therefore 
anticipate that the momentum
region generating it will be unambiguous in this limit.

We will first evaluate $I(q^2,m^2)$ exactly. After we determine  
the origin of the pole we will give a more direct argument to locate the
contributing momentum region. 
We begin by making the (scaling) change of variables 
$$
p_+ ~=~x_+ k_+~, ~~~ p_-~=~x_- k_-~,~~~ p_{\perp}~=~ 
(k_+ k_-)^{1\over 2} x_{\perp} 
\auto\label{cgv}
$$
and also write $ m^2 ~ - i\epsilon 
~=~2 k_+ k_- ~\mu ~=~-q^2~\mu$. If we carry out the angular 
$x_{\perp}$ integration (which gives a factor of $2 \pi~$)
and write $y~= ~x_{\perp}^2 /2$ we then have  
$$
\eqalign{  I(q^2,m^2)~&=~{\pi \over 4~ q^2}~I(\mu)~=~ 
~{\pi \over 4~ q^2} 
~\int_{-\infty}^{+\infty} dx_+~\int_{-\infty}^{+\infty} d x_- 
\int_{0}^{+\infty}dy ~~ \times \cr
& {1 \over [(x_- -1)x_+ -y  - \mu][ (x_- - 1)(x_+ -1) -y - \mu]
[ x_-(x_+ -1)- y - \mu]}}
\auto\label{imu}
$$
The propagators can be separated via partial fractions and the $y$ - 
integration can then be carried out to give
$$
\eqalign{ I(\mu)~=~ \int_{-\infty}^{+\infty}& dx_+ 
~\int_{-\infty}^{+\infty} d x_-  ~~
{1 \over (x_- - 1)(x_- -x_+)}
 \ln{[(x_- -1)x_+ - \mu]}\cr
& ~-~{1 \over (x_+ - 1) (x_- - x_+)} \ln{[ x_-(x_+ -1) - \mu]}\cr
&~+~ {1\over (x_- - 1)(x_+ - 1)}\ln{[ (x_- - 1)(x_+ -1)- \mu]}
}
\auto\label{imu1}
$$

We can evaluate (\ref{imu1}) by contour integration in the $x_-$ - plane 
as follows.
The three logarithmic branch points are on the same side of the 
$x_-$ - integration and the contour can be closed to zero 
unless $1 > x_+ > 0$. (Note that
if the numerators of (\ref{tamp}) were present then we could not 
close the contour without obtaining a contribution from the 
large $x_-$ - region.) When  $1 > x_+ > 0$ the logarithmic 
branch cuts lie as illustrated in Fig.~2.2. 
\begin{center}
\epsfxsize=5.5in
\epsffile{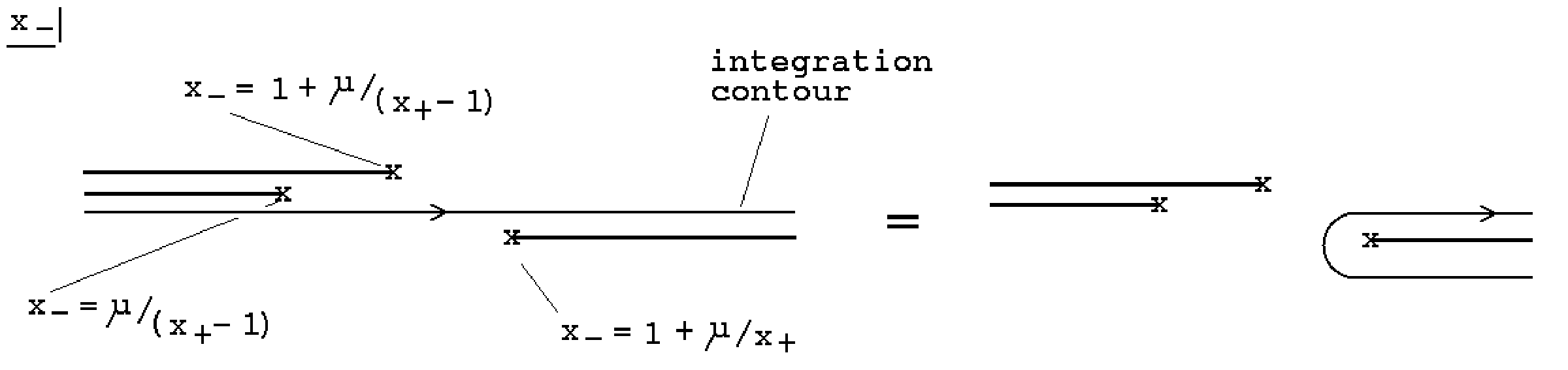}

Fig.~2.2 The $x_-$ - plane ( $\mu~= ~ [m^2 -i\epsilon]/q^2$ ). 
\end{center} 
In this case, the contour can be closed 
around the one branch cut, as illustrated, and
$I(\mu)$ is then given as an integral over just this discontinuity, i.e.
$$
\eqalign{ I(\mu)~&=~ 2\pi~i ~\int_{0}^{1}dx_+ \int_{1 + \mu/x_+}^{\infty} 
dx_-   ~~ {1 \over (x_- - 1) (x_- - x_+)}\cr
&=~ 2\pi ~i ~\int_{0}^{1}dx_+~{1 \over 1-x_+}~
\ln{\big[1 + x_+(1-x_+)/ \mu \bigr]}}
\auto\label{imu20}
$$
which an integration by parts allows us to rewrite as 
$$
\eqalign{ &I(\mu)~=~ 2\pi ~i ~\int_{0}^{1}dx_+~
\ln{\big[ 1-x_+ \bigr]}~~
{1 -2x_+ \over \mu + x_+(1-x_+)}\cr
&~\centerunder{$\longrightarrow$}{\raisebox{-5mm}{$\mu \to 0$}}
~~ 2\pi ~i ~\int_{0}^{1}dx_+~
\ln{\big[ 1-x_+ \bigr]}~~\biggl[
{1\over x_+ }~-~ {1 \over (1-x_+)} \biggr]
}
\auto\label{ipts}
$$

The first term in (\ref{ipts}) is finite while the second one has a 
logarithmic divergence of the kind we expected to find. As we discussed, we 
expect this divergence to be modified by,  and to be dependent on,
the numerator terms that we are presently ignoring. The first term we 
expect to be closely related to the anomaly pole term. If we consider the 
behaviour of the integrands of both terms near $x_+ =0$ then we note that
the first term has a constant term in it's Taylor expansion while the second 
does not. If we extract this term as a piece that is independent of how we
handle the divergence of the second term we obtain
$$
I(\mu)~ =  2\pi~i~~\int_{0}^{1}~dx_+ ~\big[1~+~ O(x_+) \bigr]
~~ = ~ 2\pi~i ~+ ~ \cdots 
\auto\label{imu2}
$$
giving
$$
I(q^2,m^2)~~\centerunder{$\longrightarrow$}{\raisebox{-5mm}{$\mu \to 0$}}
~~ {\pi^2 ~i \over2 q^2} ~~ + ~~ \cdots
\auto\label{imu4}
$$ 
If an additional function $R(p_+,p_-,p_{\perp})$ (produced by propagator
numerators, for example)
were present in the
integrand of $I(q^2,m^2)$ then, if we 
again use the limit $k_- \to 0 $ to obtain $q^2 \to 0$, the 
pole residue would simply contain an additional factor of  $R(0,0,0,0)$.

Note that if we cut off the $x_-$ - integration at $x_- = \lambda_-$ we 
obtain an extra contribution to $I(0)$ of the form
$$
\eqalign{ I(0)~&=~ 2\pi ~i ~\int_{0}^{1}dx_+~{1 \over 1-x_+}~
\ln{\biggl[{\lambda^- - x_+ \over \lambda_- -1} \biggr]} \cr
&=~ 2\pi ~i ~\int_{0}^{1}dx_+~\ln{\bigl[ 1-x_+\bigr]}~
\biggl[{1 \over \lambda^- - x_+} - {1 \over \lambda_- -1} \biggr]}
\auto\label{x-co}
$$
in which the integrand has no constant term in it's expansion around 
$x_+=0$ and so, in this sense, does not modify the anomaly term
extracted in (\ref{imu2}). Therefore the anomaly term originates
close to the lower end-point for the $x_-$ - integration (i.e. 
$x_- =p_- /k_- \sim 1$) 
and is, indeed, independent of how we treat the large $x_-$ region.

That the integration by parts, to obtain (\ref{ipts}), 
is necessary to clearly expose
the anomaly term is a consequence of the contour integration we used. We
can extract the same term more directly from $I(q^2,m^2)$ as follows. First 
we write
$$
\int_{0}^{+\infty}~ {dy \over [ (x_- - 1)(x_+ -1) -y - \mu]}
~~~\centerunder{$\to$}{\raisebox{-4mm}{$\mu \to 0$}}~~~
\ln{(x_+ - 1)} ~+\cdots
\auto\label{lny}
$$
giving, if we undo the scaling of $x_-$,
$$
\eqalign{  I(q^2,m^2)~&\to~~ 
~{\pi \over 8~ k_+} 
~\int_{0}^{1} dx_+~ln{(x_+ -1)} \int_{-\infty}^{+\infty} d p_- 
 \cr
& {1 \over [(p_- -k_-)x_+  - m^2/2k_+][ p_- (x_+ -1) - m^2/2k_+]}
~~+~\cdots }
\auto\label{lny1}
$$
We can then close the $p_-$ contour around the second pole to obtain,
in the limit $m^2 \to 0$, 
$$
\eqalign{I(q^2)~&= ~{\pi^2i \over 4~ k_+} 
\biggl( ~~\int_{0}^{1} dx_+~{ln{(1 - x_+)}  
\over k_-~ x_+ }~\biggr) ~~+~\cdots \cr
&= ~ {\pi^2i \over 2~ q^2 }~~+~\cdots }
\auto\label{lny2}
$$
which reproduces (\ref{ipts}), and hence (\ref{imu4}), directly. 

Note that the denominator $k_-~ x_+$ in (\ref{lny2})
is provided by the propagator that carries only the $k_-$ 
external momentum. The factors $k_-^{-1}$ and $x_+^{-1}$ represent the 
separate ``particle'' and ``antiparticle'' poles of this propagator
and both contribute in an essential manner. 
$k_-^{-1}$ produces the $(q^2)^{-1}$
pole in the final result. The residue of the pole at $x_+ = 0$, multiplied
by $\ln{(1-x_+)}$ (which is the integrated propagator 
contribution obtained from (\ref{lny}) ), 
is integrated to produce the final anomaly coefficient.
That both particle and antiparticle poles 
contribute to the anomaly pole is
a very important point that we will elaborate on shortly. 

To determine that (\ref{imu4}) is indeed the anomaly coefficient that we want
we must reintroduce the propagator numerators,
that we have so far neglected, and evaluate them
at zero internal momentum. In the configuration (\ref{k+k-}) the 
external momentum numerators contribute the combination of  
light-like momenta and $\gamma$ - matrices
shown in Fig.~2.3.
\begin{center}
\leavevmode
\epsfxsize=2.8in
\epsffile{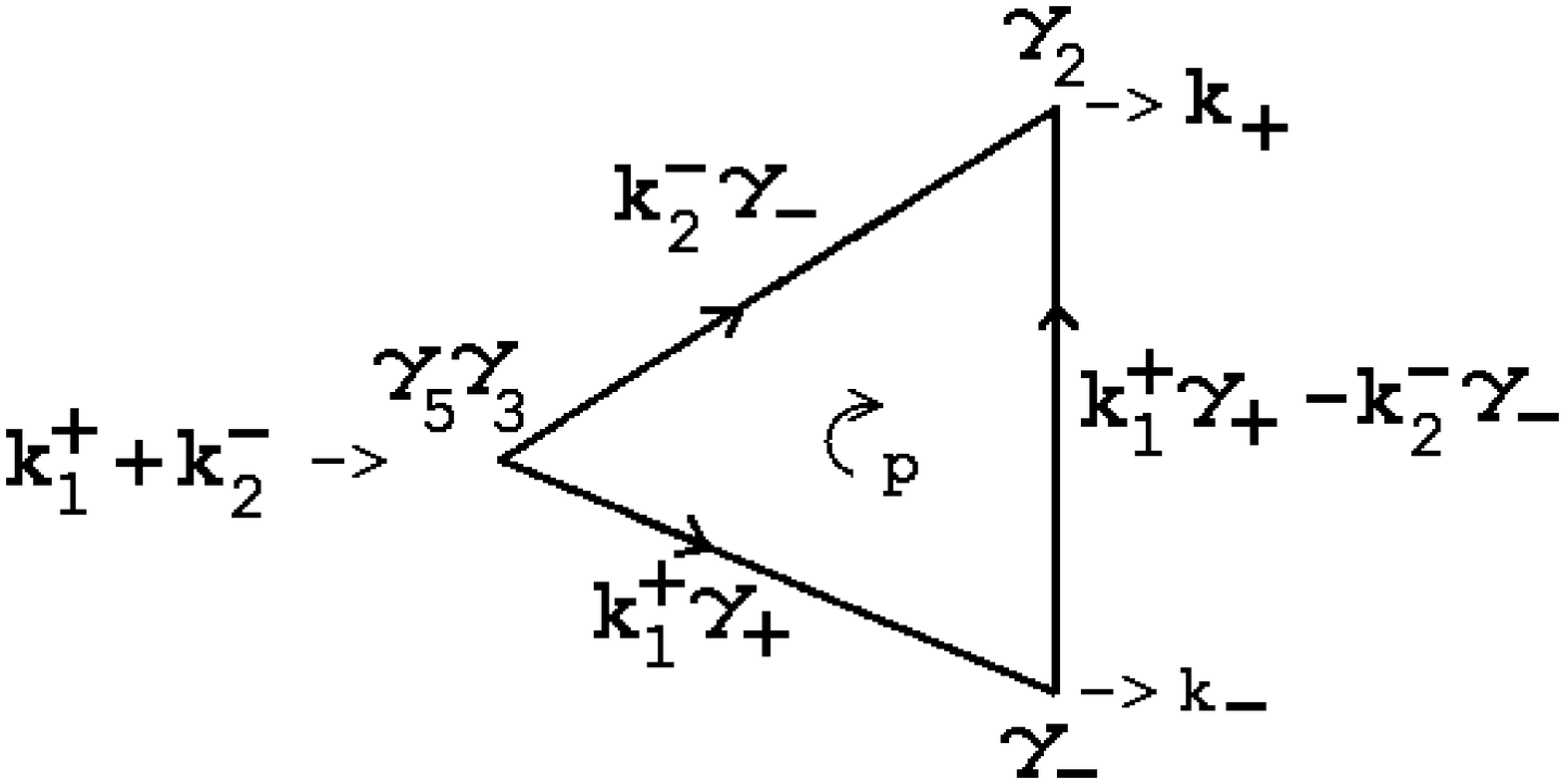}

Fig.~2.3 Vertices and Propagator Numerators for  $\Gamma_{32-}(k_+,k_-)$.  
\end{center} 
The corresponding contribution to $\Gamma_{32-}(k_+,k_-)$ is 
$$
\eqalign{Tr\{\gamma_5 
\gamma_3 ~& [k_2 \cdot \gamma]~ \gamma_2~ [k_1\cdot \gamma]
~\gamma_- ~[(k_1-k_2) \cdot \gamma]\} \cr
~&= ~Tr \{\gamma_5
\gamma_3 ~ [k_2^- \gamma_-]~  \gamma_2~ [k_1^+ \gamma_+ 
- k_2^- \gamma_-] \gamma_-
[k_1^+ \gamma_+] \} \cr
& = ~Tr \{\gamma_5 \gamma_3~\gamma_- \gamma_2~  \gamma_+ \gamma_-
\gamma_+ \} ~k_+^2 k_-  \cr
& = ~- 2~Tr \{\gamma_5 \gamma_3~\gamma_2 \gamma_-~ \gamma_+\} ~k_+^2 k_- }
\auto\label{gma1}
$$
The well-known identity for a product of three orthogonal
$\gamma$ - matrices
$$
\gamma_{\alpha}\gamma_{\beta}\gamma_{\lambda}~=~
g_{\alpha\beta} \gamma_{\lambda} ~+~ g_{\beta\lambda} \gamma_{\alpha} 
~-~ g_{\alpha\lambda} \gamma_{\beta} + i \epsilon_{\mu\alpha\beta\gamma}
\gamma^{\mu} \gamma_5
\auto\label{3ga}
$$
then gives  
$$
\eqalign{- 4~Tr \{i \gamma_5^2 ~+ 
~\gamma_5 \gamma_2\gamma_3 \} ~k_+^2 k_-  
 & = ~ 4i ~Tr \{ \gamma_5^2 \} ~k_+^2 k_- ~~~~~\cr
& = 16 i ~k_+^2 k_- }
\auto\label{gma11}
$$
Combined with (\ref{imu4}), 
this gives the desired contribution of the anomaly pole
(after taking into account the factor of
$1/(2\pi)^4$ in the original integral (\ref{tamp}) ).

If we return to the original momenta
we see from (\ref{lny})-(\ref{lny2}) that 
the relevant integration region for the anomaly pole is
$$
 i)~~ p_{\perp}^2 ~ 
\centerunder{$<$}{\raisebox{-1mm}{$\sim $}}
 ~~ q^2 ~~~~
ii)~~0 ~~\leq ~~p_+ ~ \leq k_+ ~~~~
iii) ~~ p_- ~~ \sim  ~ ~k_-  ~\to ~0 
\auto\label{intr01}  
$$
and that any additional factors in the integrand (besides the propagator
denominators) are to
be evaluated at zero internal momentum.
The surviving external light-cone momentum then flows directly
around two of the three internal propagators.
This will be very important in the next Section. 

In the following, we will use
manipulations analagous to (\ref{gma1}) and (\ref{gma11}), in which the 
numerators carrying the limiting momentum configuration are combined,
to determine whether the anomaly is present in diagrams.
However, as we noted above, the anomaly pole terms are not actually 
singular in the limiting momentum configuration we have discussed.
To consistently isolate anomaly pole contributions to 
$\Gamma_{\mu\alpha\beta}$ it is necessary
to work in a kinematical configuration
where singular contributions are obtained. This is the case
if an additional external transverse momentum $q_{\perp}$ is part of 
the limiting momentum configuration, such that $q^2
\sim q^2_{\perp}$, while the corresponding propagator numerator 
provides a factor that is $O(q_{\perp})$ and vanishes more slowly than
$q^2$. We will, nevertheless, be able to 
apply the above analysis by exploiting the
Lorentz invariance properties of the internal momentum integration.

\subhead{2.5 Frame Dependence of the Anomaly Numerator}

A second momentum configuration that can be used to approach $q^2 = 0$ is
$$
\eqalign{~~~~&k_1 ~= ~(k/\sqrt{2},k /\sqrt{2},0,0) ~~~~ ~~
\equiv ~~~~k_1^+~= k~, k_{1-}~ = 0~ ,k_{\perp} =0 \cr
& k_2 ~= ~(- k/\sqrt{2},- k\cos{\theta}/\sqrt{2}, 0,
- k \sin{\theta}/\sqrt{2})\cr
&\centerunder{$\sim$}{\raisebox{-4mm}{$\theta \to 0$}} ~-~k_1 ~-~ 
(0,0, k \theta/\sqrt{2}, 0)~= ~-~k_1 ~-~(0,0,q,0)   }
\auto\label{k+k-2}
$$
where
$$
q^2 ~=~ (k_1 +k_2)^2 ~
~\centerunder{$\sim$}{\raisebox{-4mm}{$\theta \to 0$}}
\auto\label{qth}
$$ 
In the configuration (\ref{k+k-2}), 
we obtain the largest numerator if we consider
the anomaly contribution of $A_3$ to $ T_{--3}$. This has the form
$$ 
 T_{--3}~=~ {\hbox{\Large $\epsilon$}}_{\sigma\delta - 3}~ 
{k_1^{\sigma} k_2^{\delta}~k_{1-} \over q^2} 
~~=~{~k^2 [k \theta /\sqrt{2} ] \over q^2}~ ~~
\centerunder{$\sim$}{\raisebox{-4mm}{$\theta \to 0$}}~~~
{\sqrt{2}k \over \theta}
\auto\label{A3an11}
$$
and so a divergence is present.
 
In the limit $q \to 0$ , the external momentum flow 
and $\gamma$ - matrix couplings are now as shown in Fig.~2.5
\begin{center}
\leavevmode
\epsfxsize=2in
\epsffile{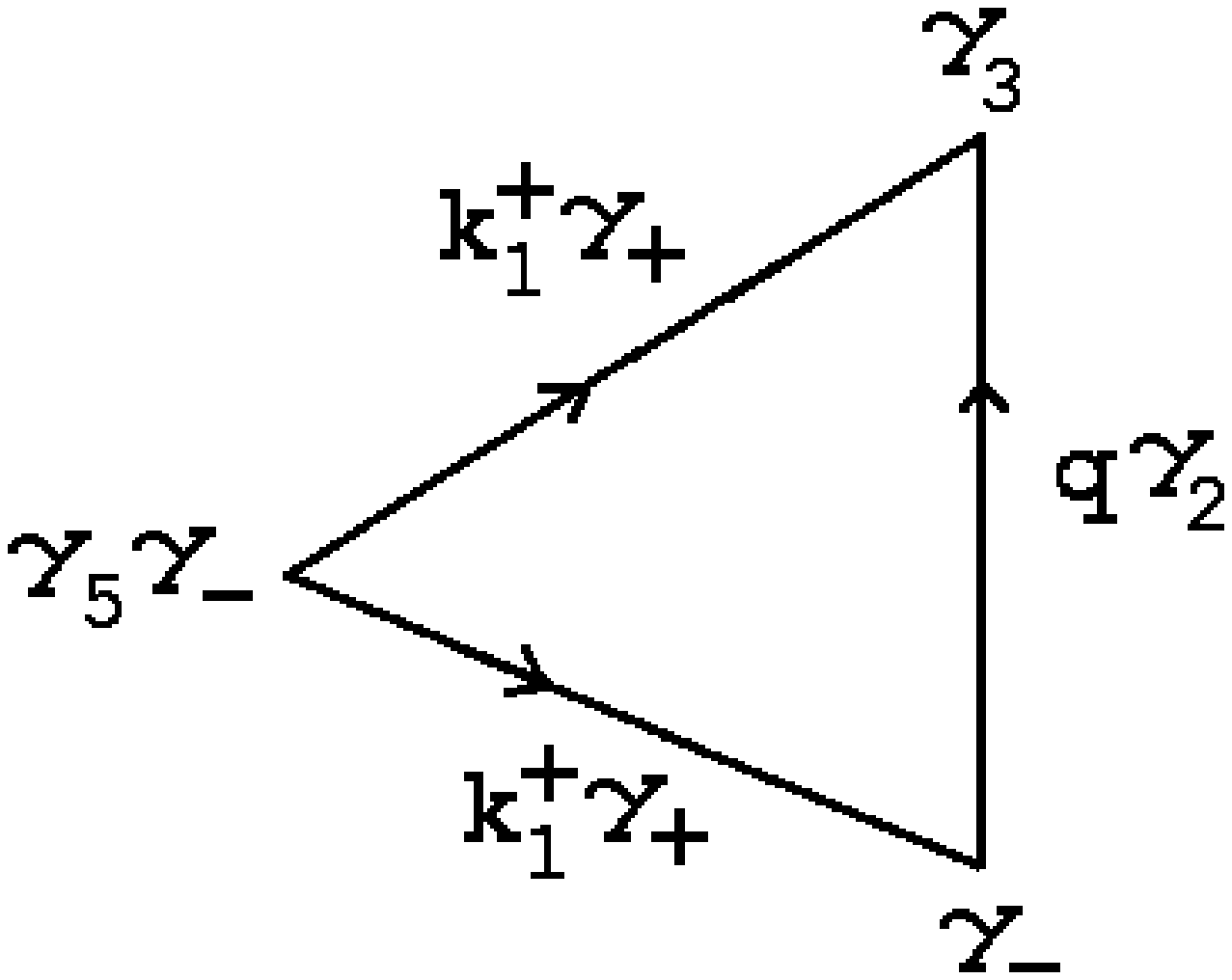}

Fig.~2.4  Vertices and Propagator Numerators for $ T_{--3}$ .
\end{center}  
Essentially the same calculation as (\ref{gma1}) and  (\ref{gma11})
gives the numerator in (\ref{A3an11}) directly from the external
momentum propagator numerators. (Note that the propagator that carries 
zero momentum in the limiting configuration is now that corresponding 
to the vertical line in Fig.~2.4.) It remains, therefore, 
to understand the anomaly pole
as arising from an internal zero momentum configuration.

Since all invariants remain unchanged it must, of course, 
be possible to obtain (\ref{k+k-2}) from (\ref{k+k-}) via a Lorentz 
transformation. This can be done as follows. We first set $k_+=k_-= q$ (which 
can be done trivially via a Lorentz transformation to the ``center of mass
frame''). We then apply a boost 
$a_y(\zeta)$ to obtain
$$
\eqalign{k_1 ~&\to~ \biggl({q  \cosh{\zeta}\over \sqrt{2}} ,
{q \over \sqrt{2}}, 
{q \sinh{\zeta}\over \sqrt{2}},0\biggr) \cr
k_2 ~& \to  \biggl({q  \cosh{\zeta}\over \sqrt{2}}, - {q \over \sqrt{2}}, 
{q \sinh{\zeta} \over \sqrt{2}},\biggr) ~~=~ k_1 ~ - ~(0,\sqrt{2}q,0,0) }
\auto\label{bay}
$$
which, if $q\cosh{\zeta}=k $ is kept finite as $q \to 0$, differs
from (\ref{k+k-2}) only by a rotation. 

If we consider (\ref{tamp}) directly in the momentum configuration 
(\ref{k+k-2}), the numerator contribution giving
(\ref{A3an11}) will be multiplied by a denominator integral that is a Lorentz
invariant. If the reverse Lorentz transformation to that giving  
(\ref{k+k-2}) from (\ref{k+k-}) is applied to the momentum integration
variables then $I(q^2,0)$, as given by (\ref{imu}), will appear and the 
above analysis can be used to extract the anomaly pole, with (\ref{k+k-2}) 
now appearing as the limiting momentum configuration. This implies, of course,
that the limit $q^2 \to 0$ is provided by an internal momentum configuration
that is reached by an infinite boost from the original zero momentum region.   

We can further enhance the anomaly numerator if we instead apply $a_y(\zeta)$
directly to (\ref{k+k-}), but now let $q^2 \to 0$ by taking $k_- \to 0$.
This gives
$$
\eqalign{&k_1 ~\to ~\biggl({k_+\cosh{\zeta} \over \sqrt{2}}, 
{k_+ \over \sqrt{2}}, 
{k_+\sinh{\zeta} \over \sqrt{2}}, 0\biggr) \cr 
&k_2 ~\to ~\biggl({k_-\cosh{\zeta} \over \sqrt{2}},-{ k_- \over \sqrt{2}}, 
{k_-\sinh{\zeta} \over \sqrt{2}}, 0\biggr)}
\auto\label{bay1}
$$
and so, for example, $T_{23-}$ (defined with respect to the 
axes of the new frame) is given by
$$
T_{23- } ~ \sim ~ {\hbox{\Large $\epsilon$}}_{\sigma\delta 23}~ 
{k_1^{\sigma} k_2^{\delta}~k_{1-} \over q^2} 
~\sim ~ {\bigl[-~k_+ k_- \cosh{\zeta}~\bigr]
~ ~ k_+\cosh{\zeta} \over \sqrt{2}q^2}
\auto\label{bay3}
$$
Now, if we let $k_- \to 0$ and take $\cosh{\zeta} \to \infty$ such 
that  $k_- \cosh{\zeta}$ remains finite, the numerator in (\ref{bay3})
$ \to \infty$ while $q^2 \to 0$ and most importantly (as we discuss in 
the next subsection) that part of the numerator contained in square brackets
remains finite. This demonstrates that the anomaly pole can have a finite 
coupling to infinite momentum states. (It is, of course, crucial
for the enhancement (\ref{bay3}) that the tensor component discussed is 
defined with respect to the axes of the new frame.) 

It will be important in succeeding Sections
that both the  component of $T_{\mu \alpha \beta}$ that dominates
and the zero momentum line involved, depend on how the anomaly pole limit is
approached (or, equivalently, the Lorentz frame involved). This is 
because our analysis of anomaly contributions in high-energy scattering
is not Lorentz invariant, but rather we combine contributions that are 
(initially calculated) in different finite and infinite momentum frames.

As noted in \cite{cg}, if we consider the helicities
of the internal massless fermions producing the anomaly pole numerator
we find that the fermion 
that carries zero momentum must effectively flip it's helicity.
Equivalently, it must reverse it's particle/antiparticle identification.
The vertex at one end of the propagator must be that for production 
of a particle while, simultaneously, that
at the other end describes the production of the antiparticle.
This is possible just because, as we discussed above, 
both particle and antiparticle 
poles contribute to a divergence that occurs when the propagator carries 
zero momentum. This process is an integral
part of the formation of a pion pole. 

The  pion scattering amplitude that we derive in the next Section will 
also contain
a zero momentum propagator (within a U(1) anomaly interaction)
which describes a physical zero momentum 
transition. If this process, and that producing the pion pole, 
are to be interpreted as a physical processes the Dirac sea must be 
shifted at the second vertex relative to the first. The
production of the antiparticle has to be reinterpreted as production of a 
state that fills a hole in the sea, i.e. the absorption of an antiparticle.
That is to say, there must be
spectral flow of the Dirac sea during the interaction. 
In a field-theoretic path integral language, this phenomenon 
is what produces a ``chirality transition''
due to a topological background gauge field. However, in our discussion
there is no implication that a topological background field is involved.

We also note that the part of our calculation of the 
``anomaly pole'' in the above that 
involved only the denominators could equally well be applied to 
the calculation\cite{fl0} of a gluon triangle (involving
an effective vertex) that appears in the coupling of a reggeized
gluon to on-shell gluons. This coupling need not satisfy
a gauge invariance Ward identity. 
Of course, the $\epsilon$-tensor structure of the anomaly
that is due to the fermionic numerators will not occur. However, a
particle/antiparticle transition, via a zero momentum propagator,
can be responsible for the helicity transition that occurs.
 
The ultra-violet anomaly is well-known to 
originate from the region 
$$
p_+ ~~\sim ~~ \p_- ~~ \sim ~~ p_{\perp}~~ \to ~~\infty
\auto\label{uvan}
$$
Therefore, in principle,
we can keep the anomaly pole in $T_{\mu\alpha\beta}$ while
dropping the ultra-violet anomaly if we integrate only over the 
momentum region (\ref{intr01}). Isolating
the anomaly pole from the ultra-violet anomaly will be an important part
of our analysis in the following. While we can
suppose that, as a matter of principle, we are restricting the 
integration region, in practise we will simply use
an anomaly pole coupling as discussed in the following
subsection. This violates full 
gauge invariance
but, as we discuss, if we keep only the anomaly
pole term and restrict our analysis to $k_1^2=k_2^2=0,~ q^2 \sim 0$ ,
we will keep the partial gauge invariance 
that is sufficient to produce gauge-invariant amplitudes. 
Nevertheless, the loss of full gauge invariance plays a crucial
role in generating the transverse momentum infra-red divergences that  
are the cornerstone of our confinement dynamics. By manipulating the 
relative contributions of the anomaly pole and the ultraviolet anomaly
we will effectively be regulating the relative
ultra-violet and infra-red spectral flow. 

\subhead{2.6 The Pole Residue as a Goldstone Boson Coupling}

A major question is whether we can use the identification of the 
anomaly pole as a Goldstone boson pole to obtain information about 
the interactions of physical Goldstone bosons.
If we keep just the anomaly pole contributions 
of $A_3$ and $A_6$ to $T_{\mu \alpha \beta}$ we can write
$$
T_{\mu \alpha \beta}(k_1,k_2) ~=~-~{1 \over 2 {\pi}^2 } ~
{({\hbox{\Large $\epsilon$}}_{\delta \sigma\alpha\mu}k_{1\beta}
 ~-~{\hbox{\Large $\epsilon$}}_{\delta \sigma\beta\mu} 
~ k_{2\alpha})~ k_1^{\delta}
k_2^{\sigma} \over (k_1+k_2)^2} ~~~ +~\cdots 
\auto\label{pipo}
$$
This expression does not satisfy the vector Ward identities and does
not have the axial current anomaly. According to the above discussion, it 
is nevertheless obtained if we keep only the integration region 
(\ref{intr01}) in (\ref{tamp}), together with the momentum dependence
of propagator numerators given by the external momenta.

When $k_1^2=k_2^2=0$, we can use the identity (\ref{epid}) to obtain
(what is essentially (\ref{Tk0}) with $m^2 \to 0$)
$$
T_{\mu \alpha \beta}(k_1,k_2) ~=~-~{1 \over 2 {\pi}^2 }~{
[~-{\hbox{\Large $\epsilon$}}_{\delta\sigma\alpha\beta} [k_1 + k_2]_{\mu}
~+~({\hbox{\Large $\epsilon$}}_{\delta \sigma\beta\mu}k_{1\alpha}
 ~-~{\hbox{\Large $\epsilon$}}_{\delta \sigma\alpha\mu} 
k_{2\beta})~]~ k_1^{\delta}
k_2^{\sigma} \over (k_1+k_2)^2~} ~~~+~\cdots 
\auto\label{pipo1}
$$
where the additional omitted terms are those that are less singular as 
$q^2=(k_1 +k_2)^2 \to 0$. (Note that to justify omitting these terms it is
crucial that we consider a component in which there is a singularity at $q^2=0$
and the numerator does not cancel the denominator singularity, as in 
(\ref{A3an}) ). Each term in (\ref{pipo1}) 
separately satisfies the vector Ward identities (for momenta
which satisfy $k_1^2=k_2^2= 0$)
but only the first term has the appropriate factorised form to
provide a pion pole coupled to the axial current $A_{\mu}$. The second term
corresponds to the $A_4$ and $A_5$ contributions in (\ref{inde}) which we
anticipated would not contribute to the tensor components that would
appear in our discussion. Therefore, we might expect that we can use 
$$ 
T_{\mu \alpha \beta}(k_1,k_2) ~=~{1 \over 2 {\pi}^2 }~{
 [k_1 + k_2]_{\mu}~{\hbox{\Large $\epsilon$}}_{\delta\sigma\alpha\beta}
~ k_1^{\delta}
k_2^{\sigma} \over (k_1+k_2)^2} ~~~+~\cdots 
\auto\label{pipo2}
$$
to obtain physical pion pole couplings, anticipating that $[k_1 +k_2]_{\mu}$
provides the coupling to the axial current $A_{\mu}$ while the factor 
${\hbox{\Large $\epsilon$}}_{\delta\sigma\alpha\beta}k_1^{\delta} k_2^{\sigma}$
provides the coupling to currents $V_{\alpha}$ and $V_{\beta}$. (In a 
general current vertex the $1/2{\pi}^2$ in (\ref{pipo2}) will be 
replaced by the 
appropriate anomaly coefficient.) (\ref{pipo2})
not only satisfies the vector
Ward identities but also produces the anomaly in the axial current. 
Remarkably, perhaps, 
we have obtained these properties from (\ref{pipo}) simply by 
restricting to the momentum region 
$$
k_1^2~=~k_2^2~=~0~,~ ~ q^2~ \to~ 0
\auto\label{zmr}
$$
and asking for a factorizable pole residue. Therefore, if we restrict our
discussion to the region (\ref{zmr}) (and to components of 
$T_{\mu\alpha\beta}$ to which the second term in (\ref{pipo1}) does
not give a leading contribution) all desired, factorization, 
gauge invariance and
anomalous divergence properties are contained in (\ref{pipo2}).

While it is well-known that (\ref{pipo2}) describes well the decay of a
physical (massive) pion into physical photons there is, not surprisingly,
an obvious problem with attempting to use it to discuss the
coupling of a pion to dynamical gluon currents. 
It is crucial for our infra-red anomaly analysis
that the ``pion'' is massless. In this case
the ``pion pole'' appears only in the $q^2 \to 0$ limit in which 
$k_1 \parallel k_2 \parallel k_+$ where $k_+$ is light-like. 
Because of the {\Large $\epsilon$}-tensor,
the numerator in (\ref{pipo2}) then vanishes in any 
finite momentum configuration - as we have seen explicitly
above. In general, if the limiting
configuration is approached via a vanishing spacelike momentum $q$ and
$k_+$ is the non-vanishing component of $k$ then, at best,
$$
{\hbox{\Large $\epsilon$}}_{\delta\sigma\alpha\beta}~ k_1^{\delta}
k_2^{\sigma}~~\sim~~k_+ q
\auto\label{qdep}
$$
which, of course, still vanishes as $q \to 0$.  
The fundamental reason for this is that 
(\ref{pipo2}) is 
antisymmetric in $k_1$ and $k_2$ and, because of Bose symmetry, 
can only describe the contribution of
antisymmetric momentum configurations of the kind we have discussed.
For consistency, it 
must vanish at the symmetric point where $k_1^2 = k_2^2 = q^2 =0$.
The conclusion is, clearly, that  we can not obtain a finite coupling
as $q^2 \to 0$ and the limit onto the (massless)
pion mass-shell is taken. Therefore,
the anomaly provides no information about physical, finite momentum, 
massless pion-gluon interactions.

However, we see from (\ref{bay3}) that 
if we go to an ``infinite momentum frame'' we can 
keep components of $q$ finite, even though $q^2 \to 0$ and the ratio
$q/k_+$ goes to zero. If we use (\ref{pipo2}), instead of 
(\ref{bay3}),  to evaluate $T_{23-}$ in this frame, we obtain 
$$
T_{23- } ~ \sim ~ {\hbox{\Large $\epsilon$}}_{\sigma\delta 3-}~ 
{k_1^{\sigma} k_2^{\delta}~k_{12} \over q^2} 
~\sim ~ {\bigl[-~k_+ k_- (\sinh{\zeta} )~\bigr]
~ ~ k_+\sinh{\zeta} \over \sqrt{2} ~q^2}
\auto\label{bay4}
$$
which, not surprisingly, gives the same leading result as (\ref{bay3}).
(Note that the second term in (\ref{pipo1}) gives a non-leading
contribution.)
The ``infinite momentum'' pion coupling is now given as  
$$
\hbox{\Large $\epsilon$}_{\sigma\delta 3 - }~ 
k_1^{\sigma} k_2^{\delta}~
~\sim  \bigl[~k_+ k_- \sinh{\zeta}~\bigr] 
\auto\label{bay5}
$$  
which, as we noted above, is finite if $k_- \to 0$ with
$ k_- \cosh{\zeta}~$ kept finite. We conclude that,
although the anomaly provides no information about finite momentum 
gluon couplings, it can potentially provide 
information about the ``wee-gluon'', or ``wee-parton'' couplings 
of the infinite momentum pion. We will discuss such couplings in the
next Section. We will find that the current component involved can not
be that of a simple local current but must itself originate from a non-local
interaction that produces an effective local interaction at infinite 
momentum.

\newpage

\mainhead{3. BUILDING COLOR SUPERCONDUCTING PION AMPLITUDES}

\subhead{3.1 The Gluon and Quark Spectrum}

When the gauge symmetry of QCD is sontaneously broken 
from SU(3) to SU(2) the resulting theory
is commonly called 
``Color Superconducting QCD''. Our eventual goal is to give a detailed
construction of high-energy
scattering amplitudes (for Goldstone  bosons) in color superconducting
QCD and then to discuss the restoration of the full gauge symmetry 
using Reggeon Field Theory.
In this paper we want to concentrate on how the kinematical and 
dynamical properties of the chiral flavor and U(1) 
anomalies discussed in the previous Section combine with 
transverse momentum infra-red divergences to produce 
such amplitudes. For this purpose we will use
only general properties of the gluon and quark spectrum, which 
we now discuss, and will make only qualitative comments about color and color
factors.

Some number of quark flavors will be present, 
which we will not specify since we will not give them distinct masses.
The symmetry breaking could be due to the expectation 
value of a complex color triplet scalar field, with
Yukawa couplings generating a mass for SU(2) singlet
quarks. Alternatively, and perhaps preferably, 
since the scalar field itself plays no role in our discussion,
the symmetry breaking could equally well be dynamical and due to a
diquark condensate associated with the additional chiral symmetry breaking
dicussed below. Independently of the nature of the symmetry breaking, 
the complete structure of the broken gauge group,
i.e. all the interactions of massless and massive gluons amongst themselves
together with their interactions with massless and massive quarks, will be 
important. 

The gluon spectrum consists of a massless SU(2) triplet, two massive SU(2)
doublets with mass $\sim M_C$, and a massive singlet with mass $M_C$. 
The quark spectrum consists of a massless SU(2) doublet and a massive singlet
for each flavor, with mass $m_C \sim M_C$. 
Because of the equivalence of quark and antiquark color 
representations, there is an extended chiral symmetry\cite{kog}.
In particular, SU(2) color singlet
axial currents can be formed from pairs of quark fields and pairs 
of antiquark
fields, in addition to the usual quark/antiquark currents. We will 
generically refer to the SU(2) singlet quark/antiquark Goldstone bosons 
associated with chiral symmetry breaking as pions and will refer to 
the singlet quark/quark Goldstone bosons as nucleons. 

We will be considering infra-red divergences due to both the massless
quarks and the massless gluons. To discuss these divergences
we should, initially, invoke a
second symmetry-breaking mechanism to give all quarks and gluons masses. 
A second complex triplet scalar could be used for this purpose or the 
symmetry breaking could again be dynamical.
We simply assume that there is an 
an initial mass $M$ for the SU(2) gluons that is taken to zero and
an SU(2) quark mass $m$ that is also taken to zero. When $m \to 0$ the
anomaly pole
discussed in the last Section, will be produced by massless quark loops.
This will be our starting point. When the gluon mass $M \to 0$ also,
there will be an overall
infra-red divergence that  will produce confinement and
select the color zero amplitudes in which the anomaly pole becomes a pion
or nucleon pole. As we will see, our analysis involves only on mass-shell
states and gauge-invariant transverse momentum diagrams. The only breaking
of gauge invariance in our discussion will be that associated with
phase-space cut-offs in anomaly generating diagrams. As we implied in the 
previous Section, gauge-invariance will be preserved 
for those momenta involved in physical amplitudes.

\subhead{3.2 Transverse Momentum Infra-red Divergences}
 
Before discussing anomaly couplings we first summarize, briefly, the
established properties of the gauge-invariant 
massless transverse momentum diagrams that will
be involved. The overall infra-red divergence we discuss in the following
will be produced when these diagrams couple through anomaly generating
effective interactions. 

It is well-known 
from perturbative calculations\cite{fkl}-\cite{arw93} that 
in gauge theories the regge limit is described
by transverse momentum diagrams. When all gluons and quarks
have a mass there are no infra-red divergences and high-order leading and
next to leading log calculations show that these diagrams 
exponentiate (in momentum space) 
to produce regge pole and regge cut behavior.
Both gluons and quarks lie on regge trajectories, i.e. they ``reggeize''. 
Reggeization of the gluon corresponds to the exponentiation
$$
{s \over t - M^2 }~~\equiv~{1 \over t-M^2 }\int{dJ~S^J \over (J-1)}~ 
~~\to ~~~ {s^{1 - \Delta(t)} \over t - M^2 }
~~\equiv~{1 \over t-M^2} \int{dJ~S^J \over (J-1+\Delta(t))}
\auto\label{regg}
$$
where $1 - \Delta(t)$
is the (massive) gluon regge trajectory given (in the leading log
approximation) by
$$
\Delta(-Q^2)~=~ {(Q^2 + M^2) \over 16 \pi^2}
\int {d^2k_1 \over k_1^2 +M^2}
~{ d^2k_2  \over k_2^2 + M^2}~
\delta^2(Q -k_1-k_2)
\auto\label{regt}
$$

As is illustrated by (\ref{regg}), 
momentum space exponentiation corresponds to power series summation in
the $J$ - plane ($J = $ complex angular momentum). We can further illustrate 
this by considering an amplitude for which the leading high-energy
behavior is given by the regge-cut corresponding
to two reggeized gluons. In this case the lowest-order result is (apart
from a normalization factor)
$$
A_0(J,t)~=~ {1 \over J-1}~\int {d^2k_1 \over k_1^2 +M^2}
~{ d^2k_2  \over k_2^2 + M^2}~
\delta^2(Q -k_1-k_2) 
\auto\label{2rdi12}
$$ 
where $t=Q^2$ . 
The momentum space exponentiation corresponding to reggeization of the gluons
is now described by replacing the fixed pole at $J=1$ by the two-reggeon 
propagator 
$$
\Gamma_2~=~{1 \over J-1 +  \Delta(k_1^2) + \Delta(k_2^2)}
\auto\label{2rdi2}
$$
giving
$$
A_0(J,t) ~\to~ A(J,t)~=~ 
\int {d^2k_1 \over k_1^2 +M^2}~{ d^2k_2  \over k_2^2 + M^2}~
{\delta^2(Q -k_1-k_2)
\over J-1 +  \Delta(k_1^2) + \Delta(k_2^2)} 
\auto\label{2rdi}
$$
Further momentum space exponentiation is provided by 
reggeon interactions that, in the $J$ - plane, simply iterate
(\ref{2rdi}) - which we identify as a ``two-reggeon state''. 
The form of the interaction depends on the 
$t$ - channel color of the iterated reggeon state, i.e. 
we can write\cite{bs} (imposing $k_1 + k_2 = k_1'+ k_2'$)
$$
\Gamma_{22}(k_1,k_2,k_1',k_2' )= 
a~(k_1+k_2)^2+b~M^2 - c~R_{22}\left(k_1,
k_2,k_1',k_2'\right)~,
\auto\label{g22}
$$
where $a, b$ and $c$ are color factors (that include an overall normalization
factor) and 
$$                        
\eqalign{
R_{22}(k_1,k_2,k_1',k_2')= ~~
&~{{\left(k^2_1+M^2\right)\left(
{k^2_2}'+M^2\right)+\left(k^2_2+M^2\right)\left(
{k^2_1}'+M^2\right)}\over
{\left(k_1-k_1'\right)^2+M^2}}\cr
+ &~{{\left( k^2_1+M^2\right)\left( {k^2_1}'+M^2\right)+
\left( k^2_2 +M^2\right)\left({k^2_2}'+M^2\right)}
\over {\left( k_1 -
k_2'\right)^2+M^2}}~. }
\auto\label{r22}
$$
The (massive) BFKL equation\cite{fkl} is simply the color zero
reggeon ``Bethe-Salpeter'' 
equation obtained by iterating the reggeon interaction $\Gamma_{22}$ in 
reggeon diagrams.  $\Gamma_{22}$ is not a Fredholm kernel and so
the solution of the BFKL equation need not contain only regge poles. Indeed,
the BFKL pomeron is generated from the large transverse momentum region
and is a fixed cut. For our purposes, we will impose an upper transverse
momentum cut-off and (ultimately) will utilise only the infra-red properties
of the BFKL equation.

In general, it can be shown\cite{asen} that the contributions of all
logarithms (down to an arbitrary non-leading level) can be described
by transverse momentum diagrams. 
Abstract S-Matrix results\cite{gpt}-\cite{arw00} 
on unitarity in the complex angular momentum plane (reggeon unitarity) 
imply that the transverse momentum diagrams can be organized into an elaborate
exponentiation phenomenon in which  
a complete set of reggeon 
diagrams appears, involving all possible $J$-plane multi-reggeon states.
For our present purposes we require only a few infra-red properties that 
existing calculations, combined with general arguments, imply 
are satisfied by the complete set of reggeon diagrams 
(or, equivalently, the complete set of transverse momentum diagrams). 
A more extensive discussion can be found in \cite{arw93}.

When $M \to 0$ infra-red divergences appear in both the
reggeon trajectories and the (integrated) 
reggeon interactions. At first sight the divergence 
$$
\Delta(Q^2) ~~\centerunder{$\longrightarrow$}
{\raisebox{-5mm}{$M^2 \to 0$}} ~~ \ln{M^2}
\auto\label{del0}
$$
exponentiates to zero all reggeon amplitudes 
via the regge pole exponentiation (\ref{regg}). In the $J$ - plane
this exponentiation of divergences is reflected in the vanishing of  
the reggeon propagator (\ref{2rdi2}), and all higher 
multi-reggeon propagators. However, since divergences also appear in the
reggeon interactions, to discuss the $M\to 0$ limit in detail,
it is advantageous 
to undo the reggeon diagram organization and go back to transverse momentum
diagrams. The reggeon interactions and reggeon trajectory 
contributions can be combined into ``kernels'' $
K^I_N(\hdots,k_i,\hdots,{k_j}', \hdots)$, where $I$ denotes
SU(2) color. If the kernels are defined
to include a transverse
momentum conserving $\delta$-function they are dimensionless
(in transverse momentum) and describe the iteration of
dimensionless 
lowest-order ``multigluon transverse momentum states'' $T_N$ where
$$
T_N~=~{1 \over J-1}~ \int~ \prod_{i=1}^N~ {d^2k_i \over k_i^2 }
\auto\label{2rdi1}
$$
For example, 
$$
\eqalign{
K_2^I (k_1,k_2,k_1',k_2')
&~=~ \delta^2(k_1+k_2-k_1'-k_2')~\biggl[\Gamma_{2,2}^I 
(k_1, k_2,k_1',k_2') \cr
&+  ~k_1^2 k_2^2 \left[ \Delta(k_1^2)  
+ \Delta (k_2^2) \right] 
\bigl[ {{1} \over {2}} \delta^2\,  
(k_1 - k_1') + {{1} \over {2}} \delta^2\, (k_1 - k_2')
\bigr] \biggr]\cr
}
\auto\label{ki2}
$$

For simplicity we refer to $T_N$ as a ``multigluon state'' in the
following. In this context a multigluon state will always be the lowest-order
transverse momentum diagram contributing to a multi-reggeon state.  
As such, the multigluon state will carry the color and signature 
properties of the parent multi-reggeon state. Note that gauge 
invariance (in the form of reggeon Ward identities\cite{arw98}) 
implies that the kernels $K^I_N$ have zeroes (when any $k_i$ or ${k_j}'$
vanishes) which, at fixed $Q^2$, prevent the poles in the $T_N$ 
from producing divergences. At fixed $Q^2$, therefore, the 
divergences come only from the trajectory
and interaction terms contained in the kernels. 

When the $t$-channel 
color is non-zero the divergences produced by $\Gamma_{2,2}^I$ do not cancel
those due to the $\Delta(k_i^2)$ terms in (\ref{ki2}) and, in general,
for a multigluon kernel with non-zero color, the interaction
divergences do not cancel the trajectory divergences. As a result
$$ 
T_N~K^I_N~
=~{1 \over J-1}~ \int~ \prod_{i=1}^N~ {d^2k_i \over k_i^2 }
~K^I_N(\hdots,k_i,\hdots,{k_j}', \hdots)~\to ~\infty~,~~~Q^2, I \neq ~0
\auto\label{KIN}
$$
and so the exponentiate of divergences due to reggeization dominates and 
sends the sum of all diagrams in any colored 
channel to zero, as illustrated in Fig.~3.1. 
\begin{center}
\leavevmode
\epsfxsize=4.5in
\epsffile{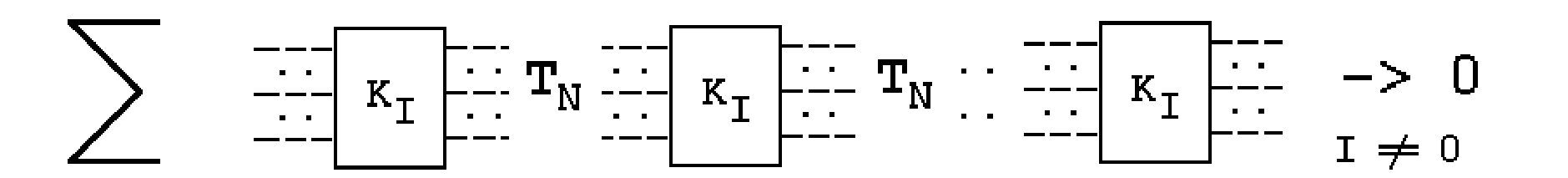}

Fig.~3.1 Iteration of a Massless Gluon Kernel.
\end{center}
When $I=0$ and $Q^2 ~\neq 0$, 
the trajectory and interaction
divergences do cancel. As a result there is no  
exponentiation of divergences. $K^0_2$, as given by (\ref{ki2}), 
is the familiar (massless) BFKL kernel and 
if there is no ultra-violet cut-off on the 
transverse momenta (as we will shortly impose)  
the iteration shown in Fig.~3.1 produces the BFKL pomeron.

The disappearance of all colored multigluon states is not 
confinement since, in the color zero diagrams, 
the gluon poles in the states remain - even though there is
a cancelation of divergences for $Q^2 \neq 0$.
If the iterated diagrams are coupled gauge invariantly
to scattering states then such couplings will also have the 
necessary zeroes to make the complete amplitude finite at fixed $Q^2$. 
This is the infra-red
finiteness property which is extensively exploited in BFKL applications.
Nevertheless, at $Q^2 = 0$ a singularity remains that is   
associated with the multigluon states and whose exact nature depends on 
the behaviour of the kernels as $Q^2 \to 0$. Confinement could be produced
if the $Q^2 = 0$ singularity can be 
absorbed into a ``condensate'', as will be
the case at the end of our analysis. \footnote{In effect, we will 
use the scale 
invariance properties of color SU(2) reggeon diagrams, which generate all of
the conformal symmetry properties of the BFKL pomeron, only to generate a
factorizing infra-red condensate. We then build up the regge pole nature of 
the pomeron through the remaining, massive, part of the gauge group.}
 
In leading-log calculations the infra-red finiteness 
property of the dimensionless 
kernels leads directly to
conformal scale invariance. In general non-leading log contributions the 
introduction of a scale for the gauge coupling 
destroys all scale-invariance properties. If,
however, there is an infra-red fixed point for the gauge coupling 
(as is the case when a large number of massless quarks are present) the
scale invariance properties will still be present in the infra-red region.
In this paper we effectively assume the existence of such a fixed point.
We will also, for the purposes of this paper, impose an upper cut-off
on the transverse momenta. Infra-red finiteness 
then implies that the kernels $K^0_N$ scale canonically
as $Q^2 \to 0$ so that 
$$
\int_{|k_i|^2, |k^\prime_j|^2 ~<~\lambda} ~\prod_i {d^2k_i \over k^2_i}
~\prod_j {d^2k^{\prime}_j \over {k'}^2_j}~
K^0_N ( k_1, \cdots k_N,  k^{\prime}_1, \cdots k^{\prime}_N)
~ \sim ~ ~\int^{\lambda_{\perp}} {dQ^2 \over Q^2}
\auto\label{irscl}
$$
where, as in the above, $Q = \sum k_i = \sum k'_j $ . If (\ref{irscl})
is obtained via the limit $M^2 \to 0$, this divergence would appear
as a factor of $\ln{ [M^2 / \lambda_{\perp}]}$. 

To understand the implications of this last divergence 
we formally rewrite (\ref{irscl}), analogously to (\ref{KIN}),  as 
$$
(J-1)^2  ~T_N ~T'_N ~K^0_N
\auto\label{irscl0} 
$$
and note that infra-red finiteness implies firstly that 
$(J-1)~T'_N~K^0_N$
is finite when the $k_i$ are finite and, also, that $(J-1)T_N~K^0_N$
is finite when the $k'_j$ are finite. Consequently,
there are two contributions to the
divergence in (\ref{irscl}), depending on
whether the $Q^2$ - integration is performed as part of the integration over
the $k_i$ or as part of the integration over the ${k'}_j$. In the first case 
the divergence is obtained from the region $\{ ~k_i~ <<~  k'_j ~~
\forall ~i,j\}$ ,
whereas in the second case it is the region $\{ ~k'_j~ <<~  
k'_i ~~\forall ~i,j\}$. In effect, either the $T_N$ or the $T'_N$ integration
produces the divergence, but not both.

If a color zero multigluon
state is coupled without the Ward identity zero 
(involving the transverse momentum of the complete state) that
is (normally) a consequence of gauge invariance, (\ref{irscl}) is 
a potential source of an infra-red divergence. This will be the case 
for the anomaly couplings that we discuss below. It is important that 
as the kernel 
$K^0_N$ is iterated a divergence always occurs when $Q^2 \to 0$.
The degree of divergence does not increase but rather,
in an integral involving a product of many kernels, there is
a distinct contribution from each $T_N$. The divergent $T_N$ 
can then be isolated and the remaining integrations 
organized, in the complete set of diagrams,
as illustrated in Fig.~3.2. 
\begin{center}
\leavevmode
\epsfxsize=5.5in
\epsffile{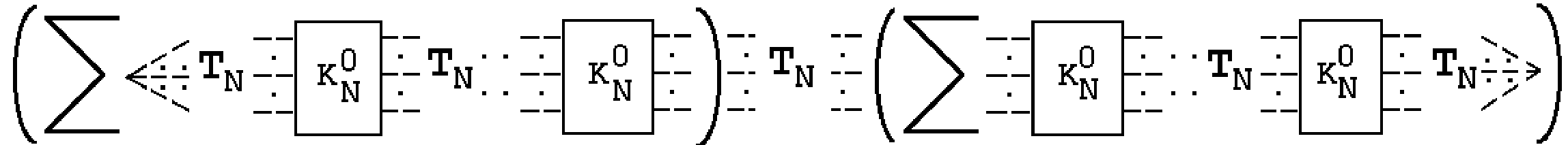}

Fig.~3.2 Isolation of the Divergence Associated with $T_N$.
\end{center}
It follows that the residue of the logarithmic divergence 
can be written in the factorized form
$$
{1 \over J-1 }~ \int~{dQ^2 \over Q^2}  
\int ~\prod_i {d^2k_i \over k^2_i} ~\delta^2(Q - \sum k_i)
~|M^0_N (J, k_1, \cdots k_N, \lambda_{\perp})|^2
\auto\label{cld}
$$
where $M^0_N$ is given by the sum of diagrams shown in Fig.~3.3.
\begin{center}
\leavevmode
\epsfxsize=3.4in
\epsffile{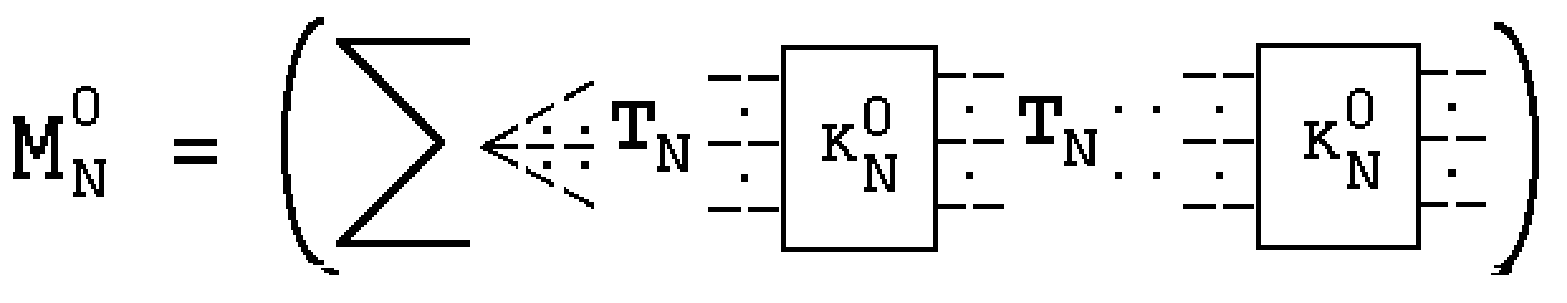}

Fig.~3.3 Diagrams Contributing to $M^0_N$
\end{center}

In the following we will need to know the interaction
between massless multigluon states and the massive (reggeized) gluons that
are also in the theory. For SU(2) color zero we can 
distinguish two classes of multigluon states, as follows. 
First we introduce the color charge 
conjugation operator for both gluons and quarks. For a gluon field, with 
color matrix $A^i_{\alpha, \beta}$, color charge conjugation $C$ gives
$$
A^i_{\alpha\beta}~~\to ~~-~A^i_{\beta\alpha}
\auto\label{gccc}
$$
while a quark with a given helicity is transformed to an antiquark of the
opposite helicity. We can also define the signature $\tau$ of a multigluon
state as $\tau=\pm 1 $ 
for an even/odd numbers of gluons. There are, essentially,
two distinct color zero combinations of gluon fields, i.e.
$$ 
Tr\{ \delta_{ij}~ A^i~A^j\}~, ~~~  Tr\{ \epsilon_{ijk}~ A^i~A^j~A^k\}
\auto\label{cz}
$$
which both have $C= +1$ but can, respectively, create $\tau=+1$ and
$\tau= -1$ states. However, since a multigluon state inherits
the signature of a multireggeon state, $\tau$ must satisfy 
$$
\tau~=~C~P 
\auto\label{tcp}
$$
where $P$ and $C$ are, respectively, 
the behavior of the coupling of the  
multigluon state under the parity and color
charge conjugation operations. 
In perturbation theory such couplings have $P= +1$ for color zero.
$P= -1$ corresponds to ``abnormal'' parity (as would be required
for the coupling of a color zero axial vector - such as the winding-number 
current). From (\ref{cz}) and (\ref{tcp}), it then follows that 
only even signature combinations of gluons can couple. Odd signature
multigluon states can couple only via the abnormal parity 
properties of the anomaly couplings that we discuss next. 
However, a kernel describing the 
interaction of massless and massive gluons will not contain any anomaly
and so, as illustrated in Fig.~3.4, it will vanish 
for odd-signature combinations of massless gluons.
\begin{center}
\leavevmode
\epsfxsize=2.4in
\epsffile{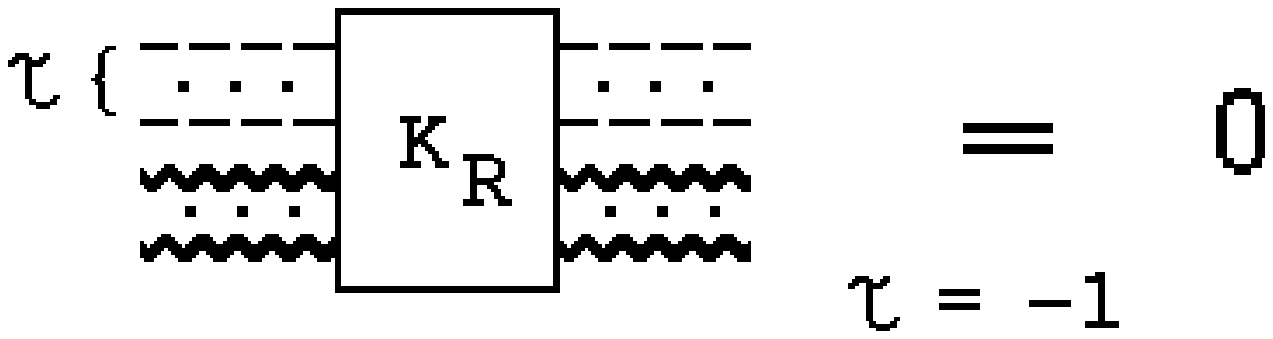}

Fig.~3.4 Interaction of Massive and Massless Gluons.
\end{center}
In the following we will also need to
assume that, at least for $\lambda_{\perp}$ sufficiently
small, when odd signature gluons do couple (via an anomaly coupling) 
and then interact amongst themselves the 
$M^0_N(J, \cdots)$ given by Fig.~3.3 is not singular for $J \geq 1$. This
will justify our extraction of an overall scaling divergence from what, in
lowest-order, is just a simple, odd-signature, multigluon state.

We will need only elementary properties of quark (and antiquark)
transverse momentum
diagrams. Although we will not need to discuss reggeization effects in any
detail, it is important that massless gluons again produce infra-red
divergences in multi-quark transverse momentum 
kernels defined analagously to the
multi-gluon kernels. Again, also, the exponentiation of reggeization implies
that only color zero states survive.
In fact, because our introduction of regge kinematics will be to some
extent artificial, even the use of transverse momentum diagrams for 
quarks will seem, in part, 
to be forced. If the ``full multi-regge'' calculation, to which we refer
at various points in this paper, were to be carried out then quark 
transverse momentum diagrams would appear directly and naturally. Color zero
quark (and antiquark) states would be directly selected by infra-red
divergences.
 
For fermions,
in addition to using light-cone momenta $k_{\pm} = (k_0 \pm k_1)/\sqrt{2}$ ,
it is convenient\cite{kms} 
to use complex momenta $\kappa = k_2 +ik_3$ to describe
transverse momenta and also to use a corresponding notation for transverse 
$\gamma$ - matrices, i.e.
$$
\gamma~=~(~\gamma_2~+~i~\gamma_3~)/\sqrt{2}, 
~~~~\gamma^*~=~(~\gamma_2~-~i~\gamma_3~)/\sqrt{2}
\auto\label{cga}
$$
We then have 
$$
\gamma^2~=~ {\gamma^*}^2 ~=~0~, ~~~~ 
\gamma~\gamma^*~+~ \gamma^*~\gamma ~= ~2
\auto\label{cga1}
$$
 
In the regge limit the transverse part of
an exchanged fermion propagator dominates, i.e. for a massless fermion
$$
{\st{k} \over k^2} ~~\to ~~ {1 \over 2} \biggl(\gamma^*{1 \over \kappa*}
+ \gamma {1\over \kappa} \biggr)
\auto\label{frp}
$$  
where the two terms represent the two different chiralities. For two fermion
exchange the combination of opposite sign chiralities dominates and so the 
transverse momentum state corresponding to (\ref{2rdi1}) is
$$
F_2~=~{1 \over J}~ \int~ d^2\kappa_1 d^2\kappa_2
\biggl(  {\gamma \over \kappa_1 }\otimes {\gamma \over \kappa_2 }
~~+~~ {\gamma^* \over \kappa_1^* }\otimes {\gamma^* \over \kappa_2^* }\biggr)
\auto\label{f2tr}
$$
where the $\otimes$ sign indicates that the two $\gamma$ - matrices are 
separately associated with the two fermion lines.
 
\subhead{3.3 Pion Couplings to Wee Gluons}

We now generalize the light-cone analysis of the 
triangle anomaly pole in the previous 
Section to derive further anomaly pole couplings involving wee gluons.
It will be helpful to describe these couplings before we 
discuss their role in producing high-energy scattering amplitudes.

The massless pion (and nucleon) Goldstone states we create will have two
distinct components, as illustrated in Fig.~3.5(a).
\begin{center}
\leavevmode
\epsfxsize=4in
\epsffile{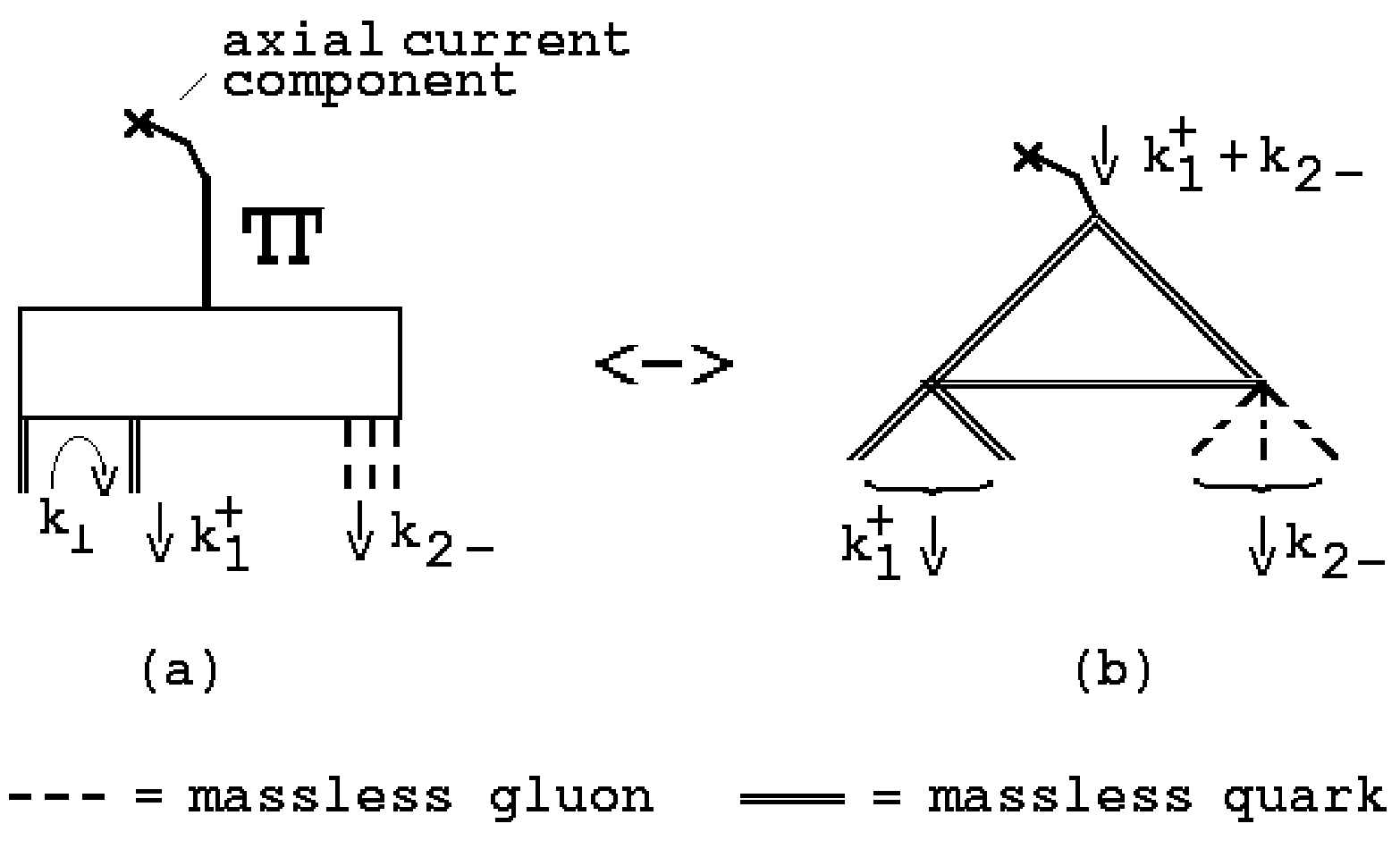}

Fig.~3.5 (a) Pion Components (b) the Anomaly Coupling.
\end{center}
A massless pion, with light-cone momentum $k^+_1$, 
will contain an (odd-signature, color zero) ``wee-gluon'' component with 
light-cone momentum $k_2^-$ (where  $k_2^-/ k^+_1 \to 0$ ) together with 
a massless quark-antiquark pair that carries the flavor
quantum numbers and the light-cone momentum  $k^+_1$. 
The pion coupling to both components will be provided by 
the triangle diagram anomaly as 
illustrated in Fig.~3.5(b). 
We discuss a diagram containing three massless gluons since this
is the simplest color zero, odd signature, multigluon state
of the kind discussed in the previous subsection.
Our discussion will easily generalize
to any number of massless gluons coupling at adjacent points.
The anomaly couplings we  obtain will imply 
that the leading high-energy behavior
in pion scattering arises when
either the quark or the antiquark carries all 
the light-cone momentum $k^+_1$. For our immediate
discussion we will take it to be the quark that carries this momentum.

Fig.~3.5(b) contains two 
``effective vertices'' that are each obtained by placing propagators on-shell
in a larger diagram, as illustrated in Fig.~3.6. As in our discussion
of the elementary triangle diagram we justify keeping only
the anomaly pole part of the diagram by appropriately
restricting the internal momentum
region. As we discuss in the following subsection, the on-shell 
propagators will then arise consistently 
from longitudinal momentum integrations (that are external to the triangle). 
\begin{center}
\leavevmode
\epsfxsize=4in
\epsffile{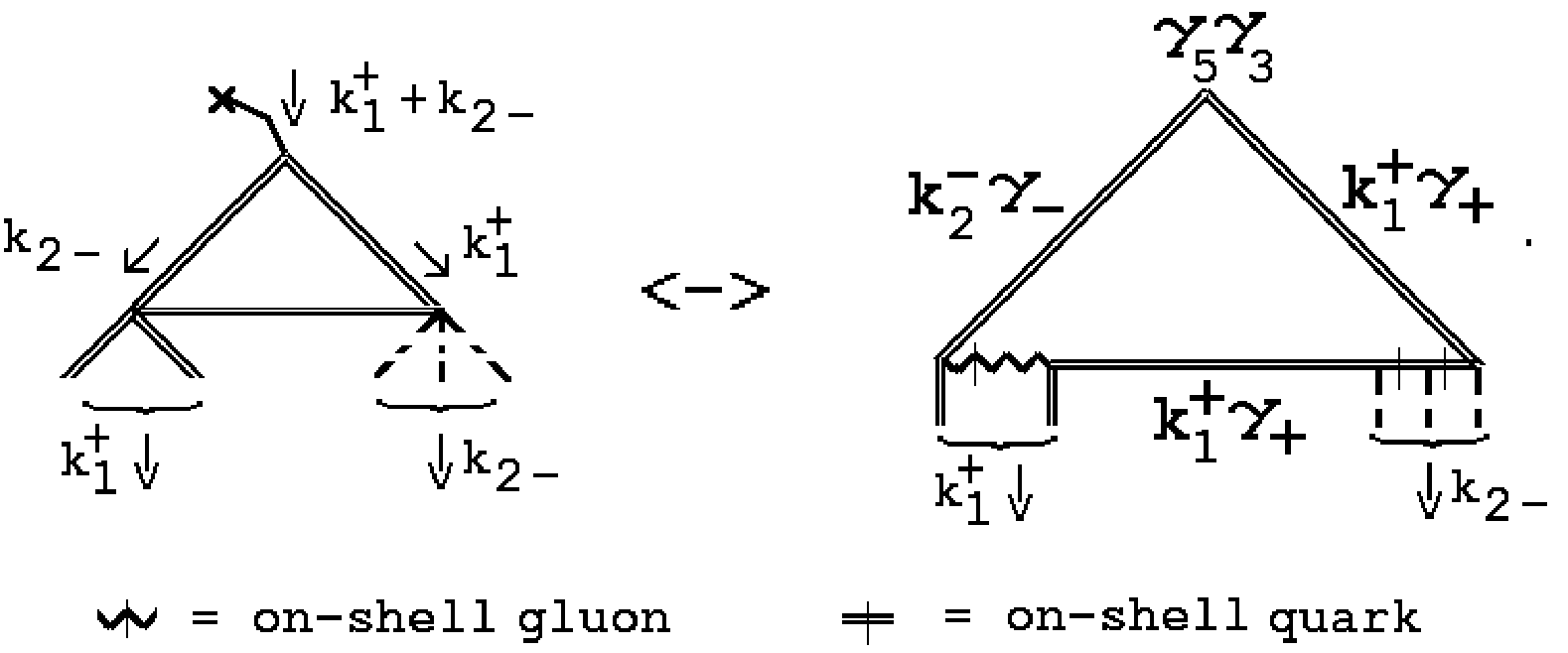}

Fig.~3.6 Reduction to a Triangle.
\end{center}
We again allow light-like 
momenta $k_1^+$ and $k_2^-$ to flow through the diagram and
generate the numerator factors shown.
The pion mass-shell will be approached 
in the limit that we take $k_2^- \to 0$ with $k_1^+$ kept fixed. 
In this limit, therefore, the massless gluons become wee gluons.

The generation
of an effective vertex for the wee gluons is straightforward and is 
illustrated in Fig.~3.7.
\begin{center}
\leavevmode
\epsfxsize=4in
\epsffile{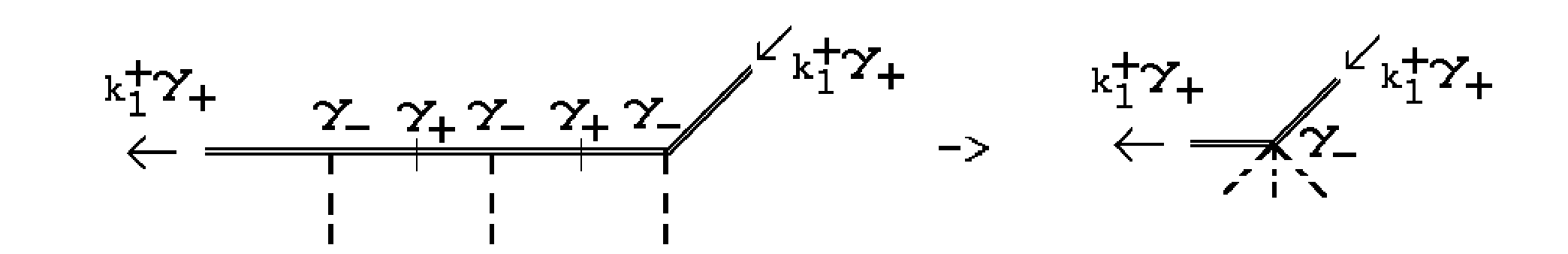}

Fig.~3.7  Generation of an Effective Vertex.
\end{center}
The wee gluons will all carry the same (longitudinal) polarization
and so, as the hatched lines in Fig.~3.7 are placed on-shell,
we will obtain an effective vertex
$$
\gamma_- 
\int ~{dk'_{1^+} ~\gamma_+ k_1^+ \over k'_{1+} k_1^+ + \cdots }
~\gamma_-
\int ~{dk'_{2^+} ~\gamma_+ k_1^+ \over k'_{2+} k_1^+ + \cdots }
~\gamma_- ~
= ~ \gamma_- \gamma_+ \gamma_- \gamma_+ \gamma_-~=~4\gamma_-
\auto\label{rdve}
$$
We will give more details on how these integrations arise later.

The generation of an effective vertex involving the external 
quark/antiquark pair is a little more complicated.
Because the internal quark and antiquark carry distinct quantum numbers
they can interact only by gluon exchange. To obtain a gauge-invariant
transverse momentum diagram the gluon must be on-shell.
In a conventional transverse momentum diagram the
produced quark/antiquark pair would have opposite chiralities (to couple
to the transverse momentum state (\ref{f2tr})). This will 
not be the case in our analysis since the
quark/antiquark pair will carry the 
light-cone momentum $k^+_1$. However, as we 
discuss further in the following,
we expect our analysis to be the continuation to 
light-like pion momentum of spacelike reggeized pion
exchange within which the quark/antiquark pair would appear
as a transverse momentum state.

The interaction needed to 
produce a quark/antiquark pair (with opposite chiralities)
in a transverse momentum state 
has the $\gamma$ - matrix structure shown in Fig.~3.8(a). 
The quark/antiquark interaction that we will need is shown in Fig.~3.8(b).  
In both cases we have included
the (upper) $\gamma$ - matrices that come from the internal
numerators of the triangle diagram as well as the (lower) $\gamma$ - matrices 
associated with the propagating quark/antiquark state. The middle 
$\gamma$ - matrices are the couplings to be 
produced by the exchanged gluon.
\begin{center}
\leavevmode
\epsfxsize=3in
\epsffile{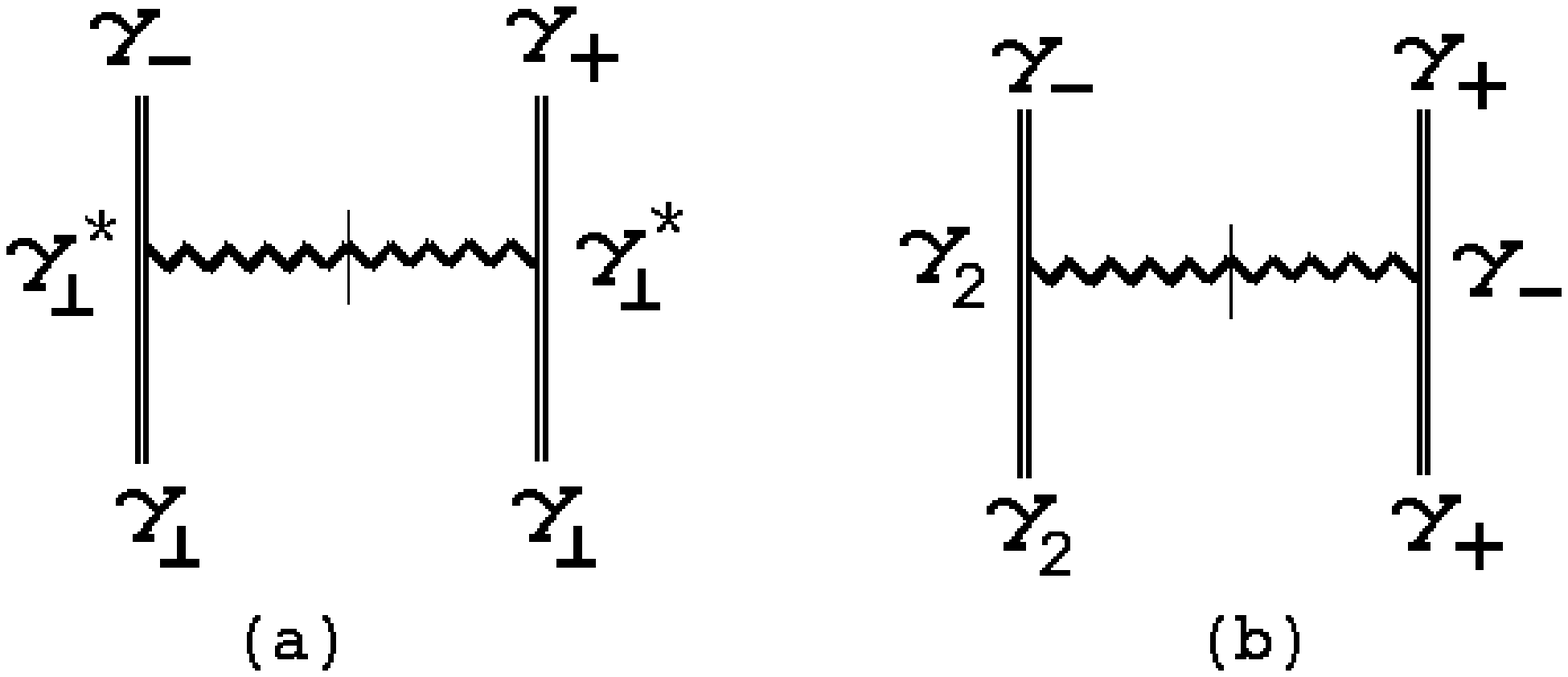}

Fig.~3.8  (a) Production of a Transverse State (b) The Needed Interaction.
\end{center} 
In the ``needed interaction'',
one $\gamma_{\perp}$ interaction will be necessary 
to obtain the anomaly numerator.
In our case we will need 
$\gamma_2$ because we will specifically choose the $\{3\}$ - 
component of the axial current generating the triangle diagrams we utilise
(see Fig.~3.11 below).
The replacement of the second  $\gamma_{\perp}$ interaction by $\gamma_-$
is necessary to allow the quark to carry a light-like momentum implying, of
course, that the spin structure of the
quark/antiquark state can not be symmetric.

The $\gamma$-matrix structure of the 
interaction due to the exchange of an on-shell
massless gluon can be written as
$$
\sum_{i=1}^4~ \gamma^i ~\otimes ~ \gamma_i ~=~ \gamma^+ ~\otimes ~\gamma_+ ~+~
\gamma^- ~\otimes ~ \gamma_-~+~\gamma^*_{\perp} ~\otimes ~\gamma_{\perp} ~
+~\gamma_{\perp} ~\otimes ~\gamma^*_{\perp}
\auto\label{m0int}
$$
where the $\otimes$ factor indicates that the two $\gamma$ - matrices 
operate on distinct fermion lines. The diagonal nature of this interaction 
implies that it can not produce either the interaction of Fig.~3.8(a) or
that of Fig.~3.8(b).
The exchange of an on-shell massive gluon with mass $M_C$ produces, 
however, an additional interaction
$$
{ \gamma \cdot \hat{k} ~\otimes ~ \gamma \cdot \hat{k} \over M_C^2}
\auto\label{mMint}
$$
where $\hat{k}$ is the momentum of the gluon.  
As is shown in Fig.~3.9,
\begin{center}
\leavevmode
\epsfxsize=4in
\epsffile{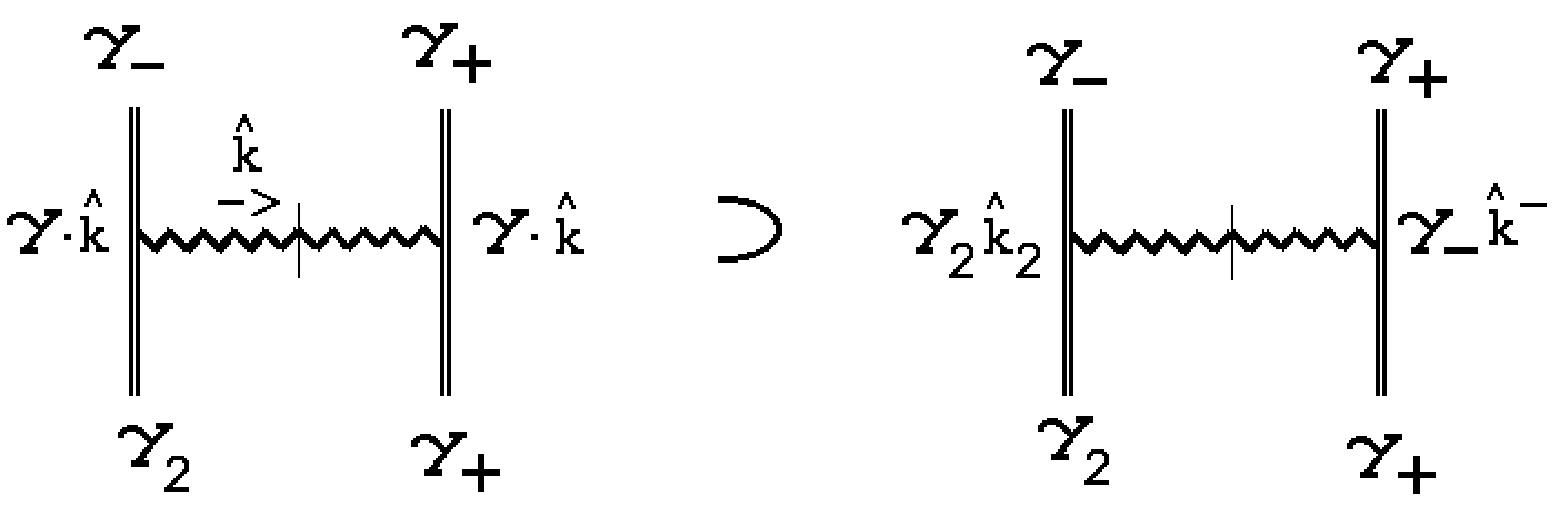}

Fig.~3.9  The Exchange of an On-Shell Massive Vector
\end{center}
the new interaction contains the needed coupling. (It also 
contains the transverse state coupling of Fig.~3.8(a).)
The $\gamma$ - matrix and momentum structure of the effective vertex
involving Fig.~3.9 is then as illustrated in Fig.~3.10.
\begin{center}
\leavevmode
\epsfxsize=5.5in
\epsffile{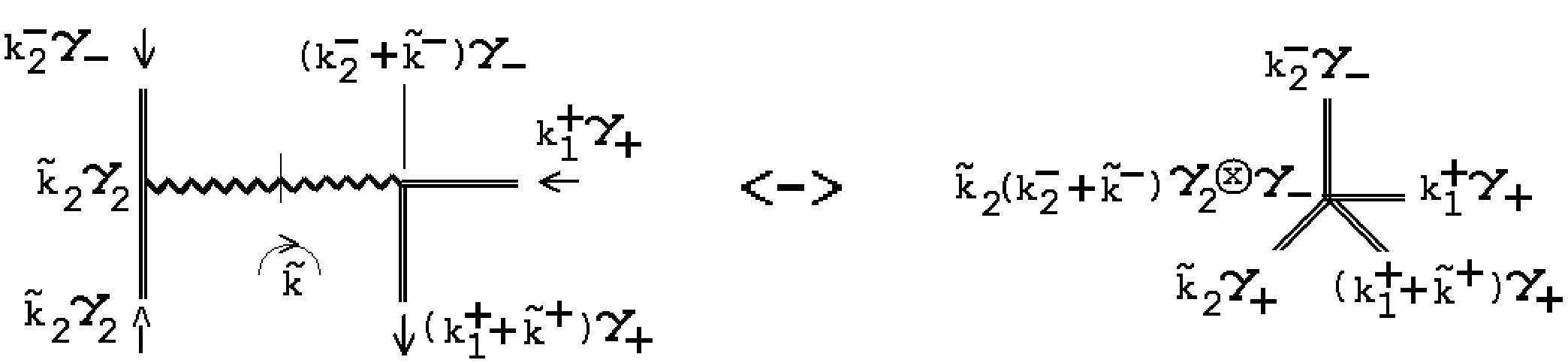}

Fig.~3.10  Generation of a Second Effective Vertex.
\end{center}
We have defined $\tilde{k}$ such that $k_1^+$  
flows directly into the quark line without flowing along the exchanged gluon
line. (This will give the final high-energy behavior most directly.) In this 
case the $\hat{k}$ appearing in Fig.~3.9 is identified with $\tilde{k}
+ k_2^-$ and so, as illustrated, the component of (\ref{mMint}) that we need
is
$$
{ \tilde{k}_2~(k_2^- + \tilde{k}^-)~~\gamma_2 \otimes  \gamma_-  \over M_C^2}
\auto\label{mMint1}
$$

In Fig.~3.11 we combine together the anomaly triangle diagram numerators  
and the $\gamma$ - matrix dependence of the above effective vertices
for the triangle diagram of Fig.~3.6. 
As illustrated, the resulting numerator factor is
$$
\eqalign{ \gamma_2~ [ k_2^- \gamma_-]~ \gamma_5
\gamma_3 ~ [k_1^+ \gamma_+]~ \gamma_-~  [k_1^+ \gamma_+]~ \gamma_-~
  ~& = ~-2~ \gamma_5 \gamma_2 \gamma_-\gamma_3 \gamma_+  \gamma_- 
  ~~ (k_1^+)^2 k_2^- \cr
& = ~- 4~\gamma_- \gamma^2_5 ~ ~(k_1^+)^2 k_2^- ~~+ ~\cdots \cr
 & = ~- 4~\gamma_-  ~ ~ (k_1^+)^2 k_2^- ~~+ ~\cdots }
\auto\label{gma12}
$$
which includes the anomaly numerator, together with an additional 
$\gamma_- $ that couples to the produced   
quark/antiquark pair. 
\begin{center}
\leavevmode
\epsfxsize=3.7in
\epsffile{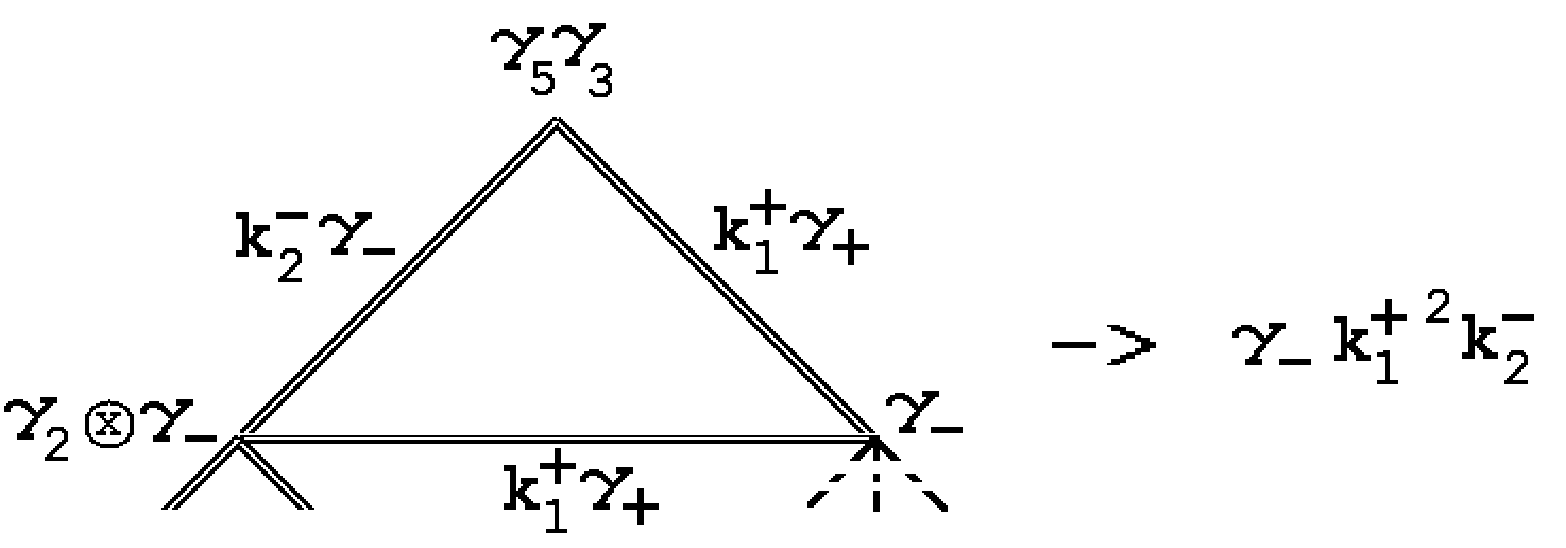}

Fig.~3.11  The Full Anomaly Numerator.
\end{center}
The role of the $\gamma_2$ - coupling in 
producing (\ref{gma12}) is clear. 

We will put the exchanged gluon on mass-shell via the $\tilde{k}^+$ -
integration. Including the numerator factor of $(\tilde{k}^- + k_2^-)$
that appears in (\ref{mMint1}) this integration has the form
$$
\int {d \tilde{k}^+~(\tilde{k}^- + k_2^-) ~\gamma_-
\over 2(\tilde{k}^- + k_2^-) \tilde{k}^+~ -~\tilde{k}_{\perp}^2~ -~M_C^2 }~
\times ~ \cdots ~~~\sim ~~ \gamma_- 
\auto\label{msh1}
$$
The momentum dependence of the quark/antiquark effective vertex is
then simply the remaining factor of $\tilde{k}_2 $ in (\ref{mMint1}).
The denominator of the reduced
diagram coincides with that of the triangle diagram,
and so the anomaly pole is generated straightforwardly. The full 
anomaly pole amplitude produced by Fig.~3.5(a) is therefore 
$$
~\gamma_-  ~ { \bigl[~(k_1^+)^2 k_2^- \tilde{k}_2~\bigr] \over M_C^2~q^2}
\auto\label{Fan1}
$$
The presence of a massive gluon is clearly crucial
for the generation of this amplitude.
(Since we are not going to sum diagrams nor include
color factors in our discussion we will also
(effectively) ignore all numerical factors.)

\subhead{3.4 The Four-Current Amplitude and the Contributing Diagrams}

A major purpose of the approach developed in
this paper is to avoid, as much as possible, the multi-regge
theory that has been a feature of our previous papers. Our intention is
to focus directly on properties of the anomaly and thus to  
arrive, as directly as is possible, 
at the dynamical interactions of pions (and nucleons).
Having the above pion couplings in hand, it might be anticipated
that we could obtain a pion scattering amplitude by considering 
a four axial current amplitude 
$$
M_{{\mu}_1{\mu}_2{\mu}_3{\mu}}~(p_1,p_2,p_3,p_4)~=~
<{A}_{{\mu}_1}^1(p_1){A}_{{\mu}_2}^2(p_2)
{A}_{{\mu}_3}^3(p_3){A}_{{\mu}_4}^4(p_4)>
\auto\label{4ca}
$$
in which the currents carry
flavor quantum numbers such that pion (or nucleon) 
scattering could appear. If there is confinement (of SU(2) color)
and chiral symmetry breaking, we
expect to find a contribution to the current amplitude of the
form (with a momentum conserving $\delta$-function removed) 
$$
M_{{\mu}_1{\mu}_2{\mu}_3{\mu}}~\centerunder{$\longrightarrow$}
{\raisebox{-5mm}{$p_1^2,p_2^2,p_3^2,p_4^2 ~ \to 0$}} 
~ {p_{1{\mu}_1}p_{2{\mu}_2}p_{3{\mu}_3}
p_{4{\mu}_4}
 \over
p_1^2~p_2^2~p_3^2~p_4^2 } ~~A(s,t)~~+~~~\cdots
\auto\label{mcp}
$$
where $s=(p_1 + p_3)^2$, $t=(p_1 + p_2)^2$ and, up to a normalization factor,
$A(s,t)$ is the pion scattering
amplitude. The omitted terms are less singular as $p_i^2 \to 0, i=1,..,4$.

We would not expect, of course, to be able to find the pion amplitude 
$A(s,t)$ at finite momentum. Instead, we might anticipate that  
combining the regge limit ($s \to \infty$ , $t$ fixed)
with the mass-shell limit ( $p_i^2 \to 0, i=1,..,4$ ) would enable us
to exploit the infinite momentum properties of the anomaly discussed in the 
previous Section. We would look for the appearance of 
pion poles via the anomaly pole interactions discussed above.
Isolating the anomaly pole dynamically (i.e. within a larger diagram)
is, however, highly non-trivial. To
proceed without multi-regge theory we will have to follow
a procedure which may appear contrived, if not artificial. It will,
nevertheless, have the significant advantage of taking 
us directly to the high-energy pion scattering amplitude. While
we will briefly explain how the procedure would be fully 
justified within a complete multi-regge analysis, we will be able
to stay away from the full calculation. 
We will indeed consider a four-current amplitude but the currents
will not be simple local operators.
We will also describe the formation of amplitudes  
in terms of diagrams that can be thought of, initially, as   
feynman diagrams. However, many of the 
integration regions in the diagrams will be  
cut-off, or even removed altogether. Before 
amplifying on our procedure, or discussing the justification,
we first describe the kinds of diagrams that will be involved. 

To have all the necessary anomaly effects present 
the diagrams must, unfortunately perhaps, be extremely
complicated. Even though almost all of this complexity will gradually
drop away as we proceed towards a physical pion scattering amplitude.
The simplest class of diagrams which combine all the anomaly interactions
are those shown in Fig.~3.12(a).
As indicated, the diagrams contain both massless and massive gluons 
together with massless quarks. 
From diagrams of the form shown in Fig.~3.12(a), 
we will obtain pion scattering via pomeron exchange 
as illustrated in Fig.~3.12(b). 
The $F_i$ amplitudes contain diagrams that will
generate the flavor anomaly and a pion pole as described above. 
The $U$ amplitudes contain diagrams that will
generate the U(1) anomaly, as described in \cite{arw011} and \cite{arw01}.
The $U$ amplitudes 
will provide the coupling of the pion to the ``pomeron''
that is exchanged in the regge limit.
\newline\parbox{3.4in}{\begin{center}
\epsfxsize=3.4in
\epsffile{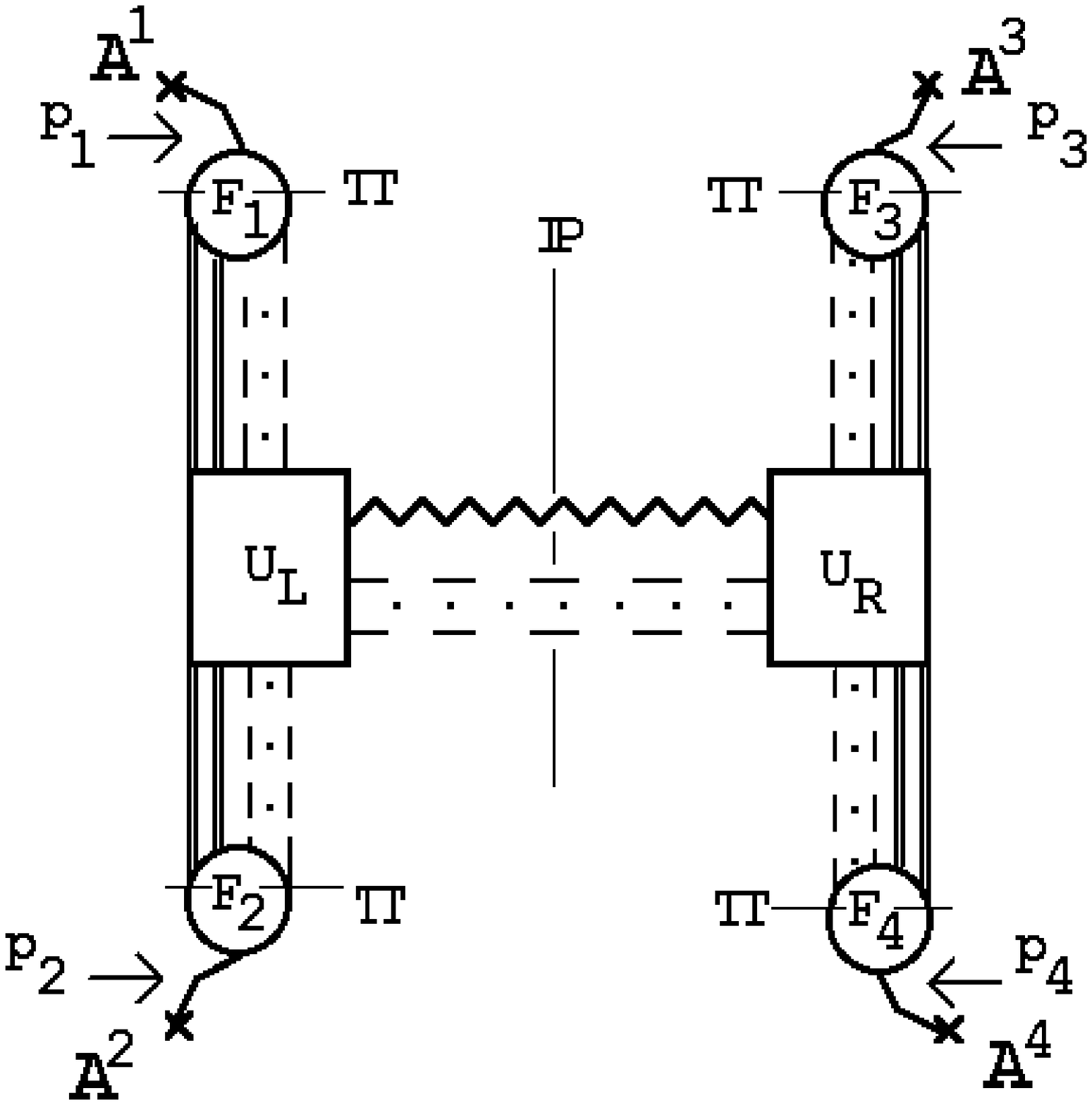}
\newline$~$
\newline $~$
\epsfxsize=3.2in
\epsffile{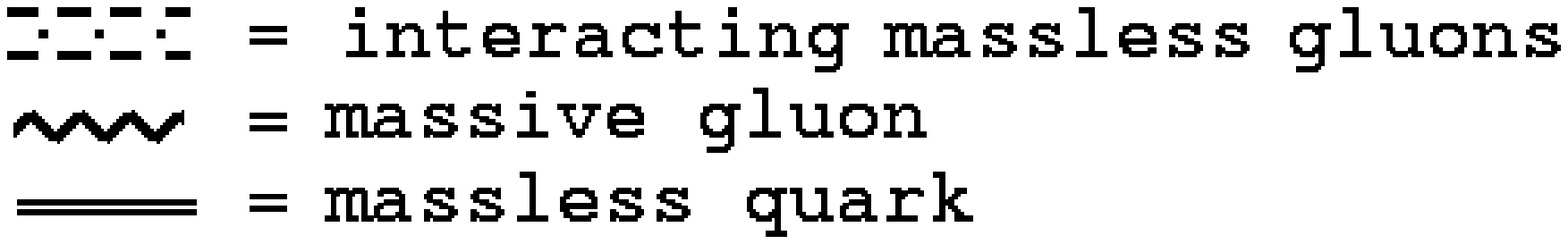}

(a)
\end{center}}
\parbox{2.5in}{
$~$
\newline $~$
\newline $~$\newline $~$
\newline$~$
\epsfxsize=2.5in
\epsffile{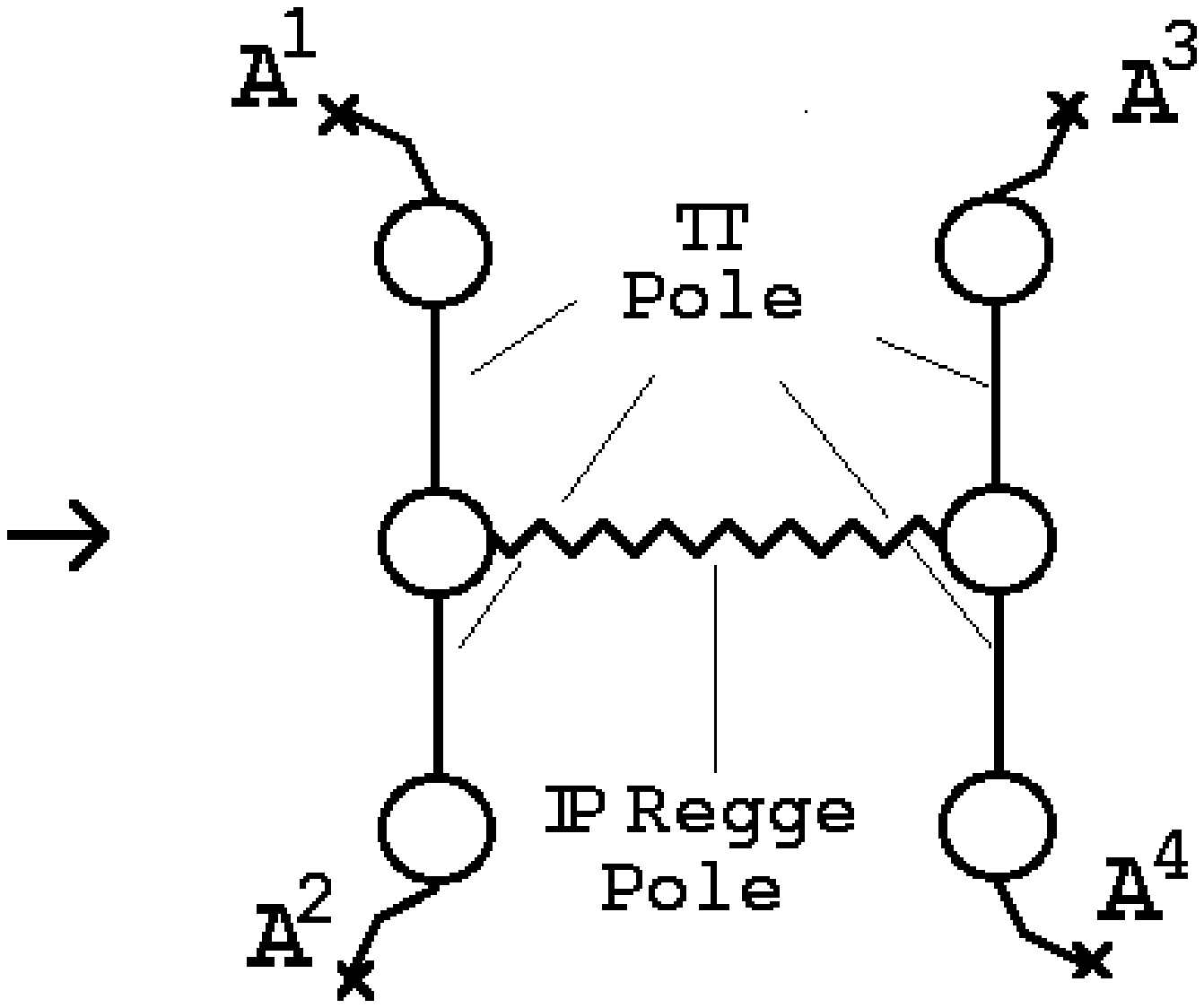}
\newline $~$
\newline $~$\newline $~$
\newline $~$\newline $~$
\newline $~$
\centerline{(b)}}
\begin{center}
Fig.~3.12 (a) The Simplest Diagrams (b) $\pi-\pi$ Scattering via 
Pomeron Exchange.
\end{center}

Having discussed the diagrams that generate the flavor anomaly
in the previous subsection it will be helpful, at this point, 
to give the structure of the 
diagrams contributing to $U_L$. Apart from the substitution
of a quark/antiquark pair for a gluon, these are essentially the 
diagrams discussed in \cite{arw011}. 
The simplest diagrams have the form shown in Fig.~3.13.
\begin{center}
\leavevmode
\epsfxsize=2.3in
\epsffile{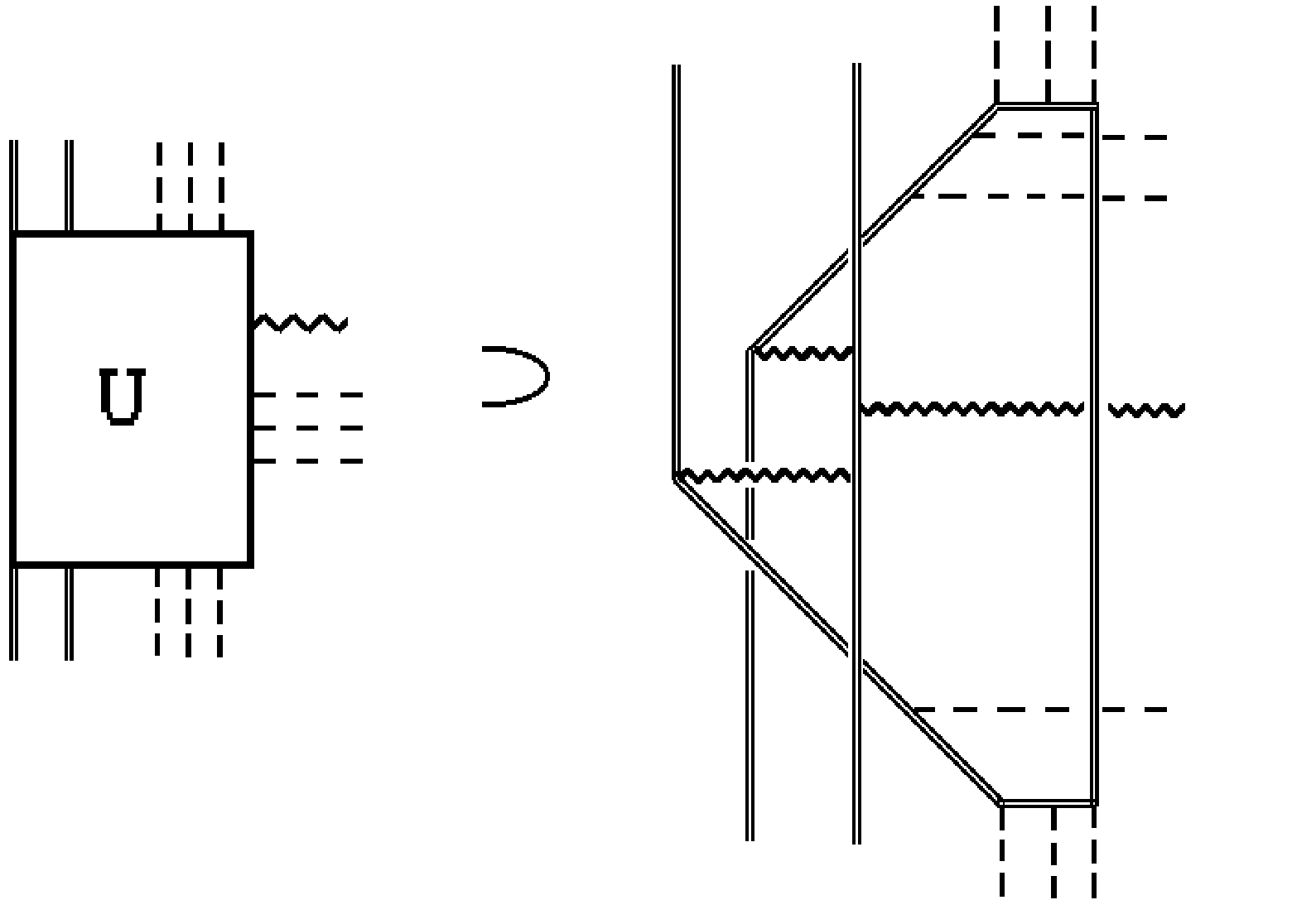}

Fig.~3.13 One of the Simplest Diagram Contributing to $U_L$.
\end{center}
As illustrated schematically in Fig.3.14, if the hatched lines 
are placed on-shell the diagram of Fig.~3.13 reduces
to a triangle diagram containing the anomaly. 
\begin{center} 
\parbox{1.4in}{
\begin{center}
\leavevmode
\epsfxsize=1.1in
\epsffile{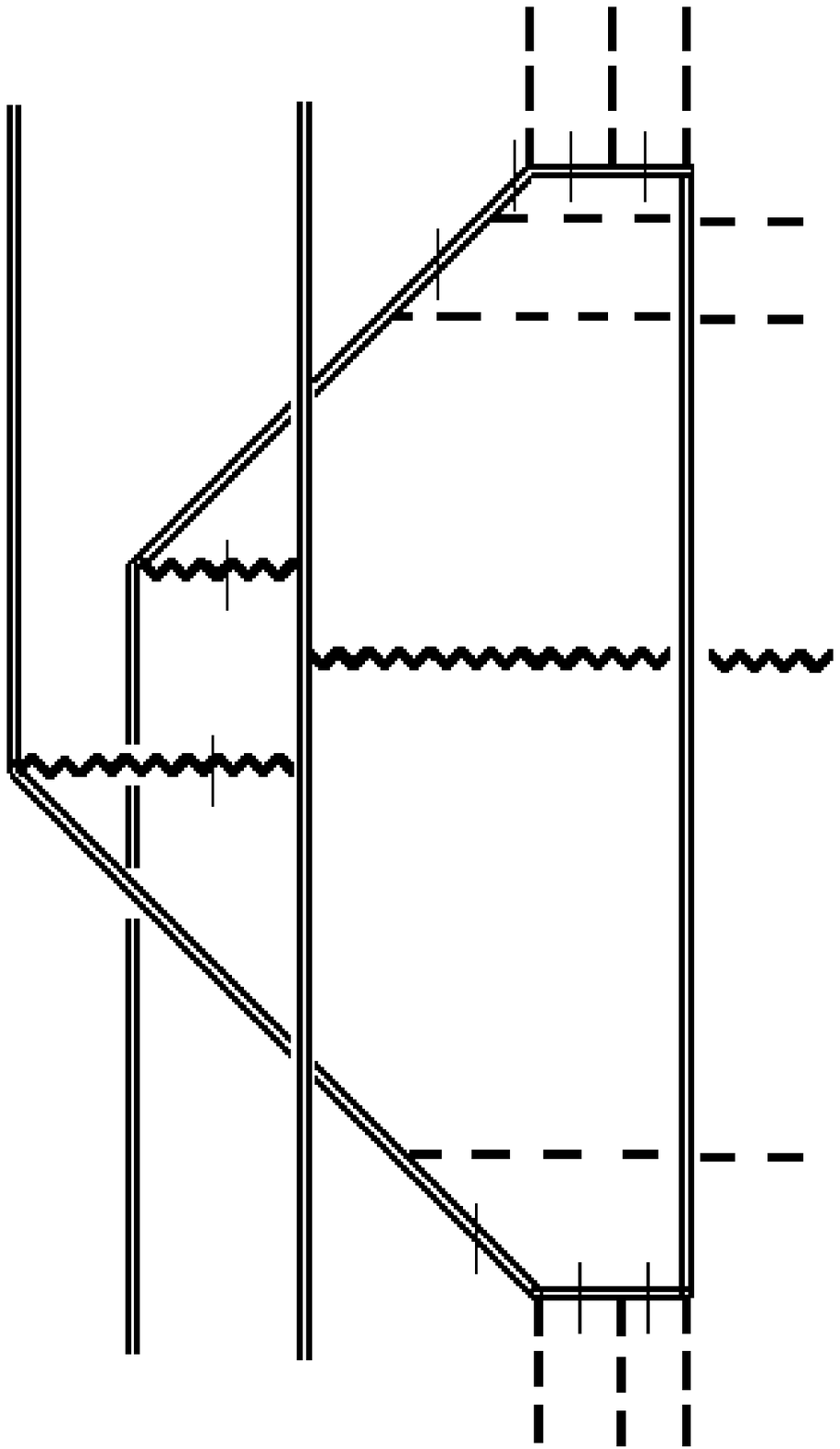}
\end{center}
}
\parbox{1.8in}{
\leavevmode
\epsfxsize=1.7in
\epsffile{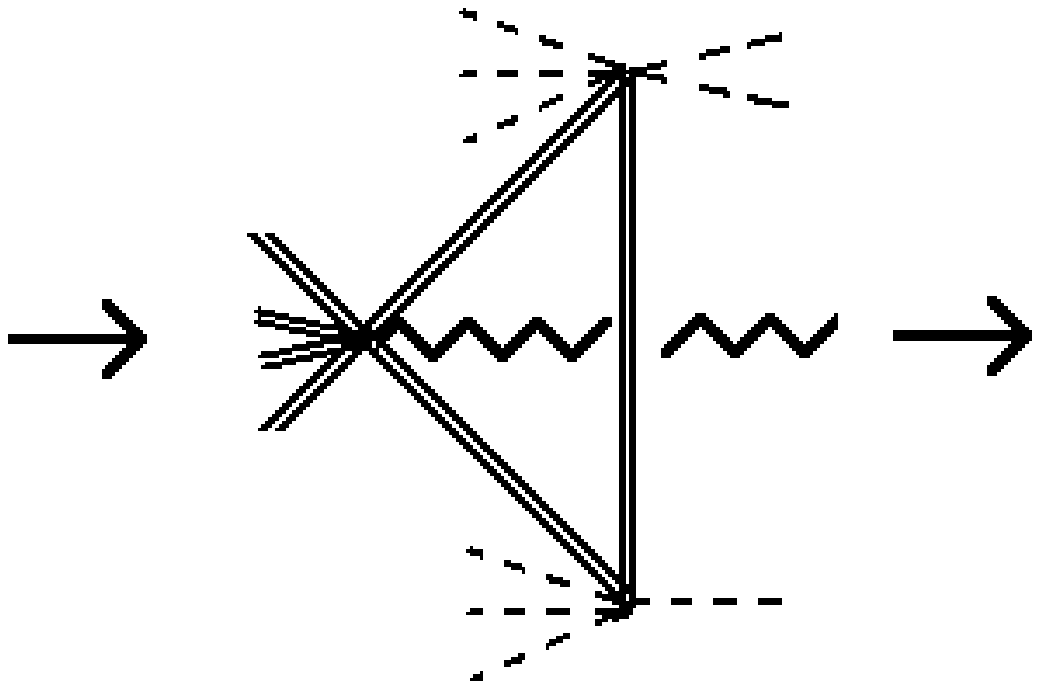}
}
\parbox{0.9in}{\begin{center}{\large U(1) Anomaly}\end{center}}
\newline
\parbox{4in}{
\leavevmode
\epsfxsize=3.8in
\epsffile{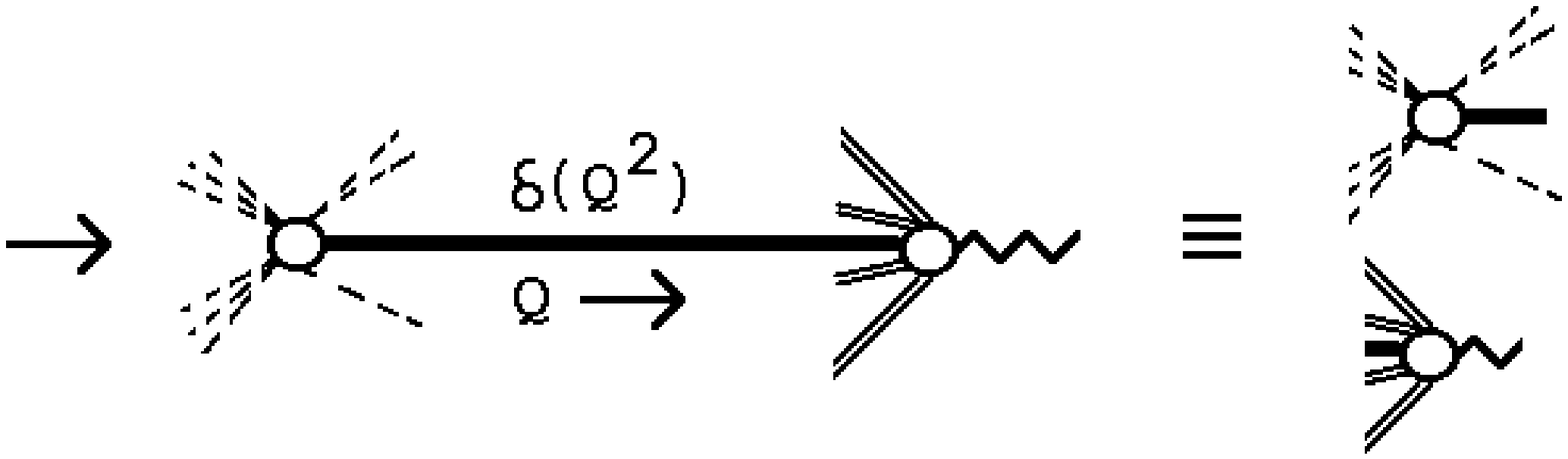}
}
\end{center}
\begin{center}
Fig.~3.14 Producing the U(1) Anomaly.
\end{center}
A crucial feature of this reduction is that the ``anomaly pole'' is integrated
over and so, as illustrated, is manifest
as a $\delta$-function that factorizes the transverse momentum dependence of 
the wee gluon interaction and the ``parton interaction''
of the quarks and massive gluons. To give more details of this 
reduction we will need the kinematics used to discuss 
the full diagrams of Fig.~3.12(a). 

Let us first assume that (schematically) the diagrams of Fig.~3.12(a) 
are generated by full feynman diagrams involving local 
axial currents. In the 
next subsection we will expose the subtleties which imply that 
this can not be the case. This will lead directly to an 
amended procedure, which we then follow. We would like
each $F_i$ amplitude to be an anomaly pole amplitude derived, in principle,
from underlying diagrams within which, a loop integration
is restricted to the region (\ref{intr01}). In this region   
a light-cone momentum circulates which is essentially
the corresponding external (pion) momentum. This momentum is ``large''
compared to the zero mass of the gluons. The central idea would 
be that, in the combined regge and mass-shell
limit, the dominant contribution to the full amplitude 
is obtained from this region of integration.
We would argue that the  
internal large light-cone momenta will combine with
the external regge limit to produce similar results to a multi-regge limit
in that we will be allowed to treat all the massless gluons as if they
were exchanged in a regge kinematic regime. As a result many propagators
will be placed on-shell, including those that reduce the $F$ - 
amplitude to an effective triangle diagram that 
contains the flavor anomaly as described in the previous sub-section. 
Similarly, within the $U$ - amplitude lines will be 
placed on-shell by both the external regge limit and the internal 
``regge limit'' of the  
massless gluons such that the triangle diagrams appear that
contain the (U(1)) anomaly. As with the flavor anomaly,
the internal integration can be restricted to a light-cone region
such that the anomaly interaction is separated out. 

Provided the massless gluon configurations reduce to transverse
momentum diagrams as we have just described we would expect, a priori, 
that the violation of gauge invariance associated with isolating 
the anomaly pole will produce the 
logarithmic scaling divergence discussed in subsection 3.2. 
We would
expect this divergence to occur separately for all odd-signature
massless gluon combinations, since interactions which iterate this 
divergence are absent in this case. Therefore, in
the ``dominant'' (divergent) contribution from diagrams of
the form of Fig.~3.12(a), all the massless gluons
should carry zero transverse momentum. This, in turn, would appear to
self-consistently justify keeping only the anomaly pole part of the 
$F$ and $U$ amplitudes. 

\subhead{3.5 Dynamical Isolation of the Anomaly Pole }

As we saw in Section 2, the light-like kinematic 
configurations in which the 
anomaly pole appears in the triangle diagram are extremely special.
Consequently, as we noted above, isolating it's occurrence within larger 
diagrams is very non-trivial. In particular, 
if we use the full uncut diagram of Fig. 3.6
as an axial current coupling, except that 
the internal loop integration is restricted
to the region (\ref{intr01}), then we have 
the following problem with the above schematic procedure. When the 
uncut diagram appears as part of a much bigger diagram, as it should do in 
the diagrams of Fig.~3.12(a),
the integration restriction is not actually sufficient to induce the regge 
kinematics we want. Even if it were, the light-like
momentum configurations produced by multigluon 
transverse momentum divergences,
although very close to those in which the anomaly pole
appears, would not be quite what is needed.

These problem are caused because when the diagram that gives
the pion coupling of Fig.~3.5(a) is a component of a larger diagram,
the light-cone
momentum denoted by $\tilde{k}^{'-}$ in Fig.~3.15(a)
should be integrated over. 
\begin{center}
\leavevmode
\epsfxsize=5in
\epsffile{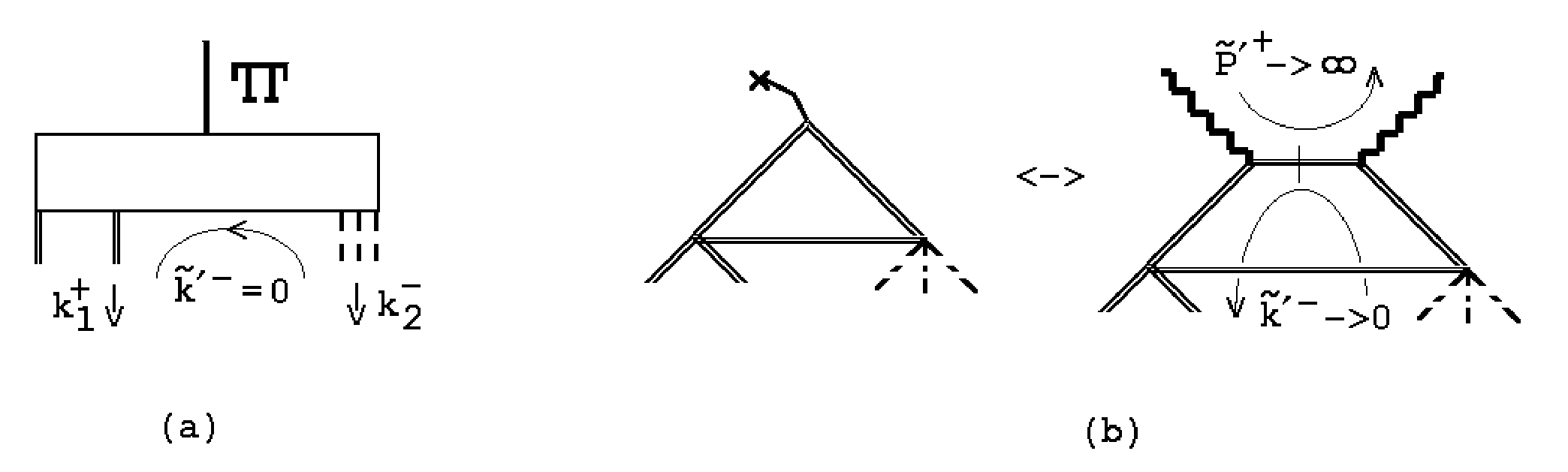}

Fig.~3.15 (a) $\tilde{k}^{'-} = 0$ (b) The $\tilde{P}^{'+} 
\to \infty$ Coupling.
\end{center}
The presence of this 
momentum has two effects. First it gives a mass $ \sim
\tilde{k}^{'-} k_1^+$ to the quark/antiquark pair that prevents
the appearance of the pion anomaly pole. Secondly, if it flows
through any of the massless gluon propagators, it will combine with the  
light-like momentum flowing in from the $U$ amplitude to remove
the transverse momentum divergence of the 
massless gluon state.
  
To remove these problems we make the momentum 
restriction that $\tilde{k}^{'-}  = 0$, i.e. $\tilde{k}^{'-}$
is not integrated over.
This will allow us to follow explicitly the schematic
procedure outlined in the previous subsection.
In effect, though, it is this restriction that generates 
the logarithmic transverse momentum
divergences which are the cornerstone of our dynamics. Within this
presentation, it may therefore appear artificial and perhaps even
unphysical at first sight.
However, this restriction would automatically
appear if the current was not a simple local operator but was instead a 
non-local current component that originates from a further external
infinite momentum limit as illustrated in Fig.~3.15(b). (As would
be exactly the case if we used multi-regge theory
to first obtain the pion as a spacelike reggeized state.)
In this case, the axial
current component is an effective point coupling
derived by placing an intermediate quark state
on-shell, via an integration over $\tilde{k}^{'-}$. 
(Using the $\tilde{k}^{'-}$ - integration for this
purpose leaves intact the full loop integration generating the anomaly.)

Probably, the feature that local axial currents are not only not needed 
but are not wanted in our formalism is a deep matter of principle. 
It seems to be essential that our pion be extracted as a wee-parton
component of additional infinite momentum external states. 
(Effectively exploiting the ``triviality of the infinite
momentum vacuum'' to the maximum.)
The additional external states should be vector particles with the 
appropriate polarizations to induce an axial current vertex.
It is interesting (and perhaps also a deep feature of our procedure) that 
the quantum numbers involved imply these particles could 
actually be $W's$ and the $ Z^0$. 

Although it would be a more complicated calculation,
there would be other advantages in making the further infinite momentum
limit part of our discussion. In particular it would eliminate the
need to appeal to the
phase-space restriction involved in generating the anomaly pole to justify
placing the hatched lines of the anomaly generating diagrams on-shell.
Indeed, it should now be clear that if we want to proceed systematically
we can not really avoid multi-regge theory and 
we are paying a heavy price by trying to do so.
If we simply studied the limit producing the
multi-regge amplitude of Fig.~3.16
\begin{center}
\leavevmode
\epsfxsize=4.5in
\epsffile{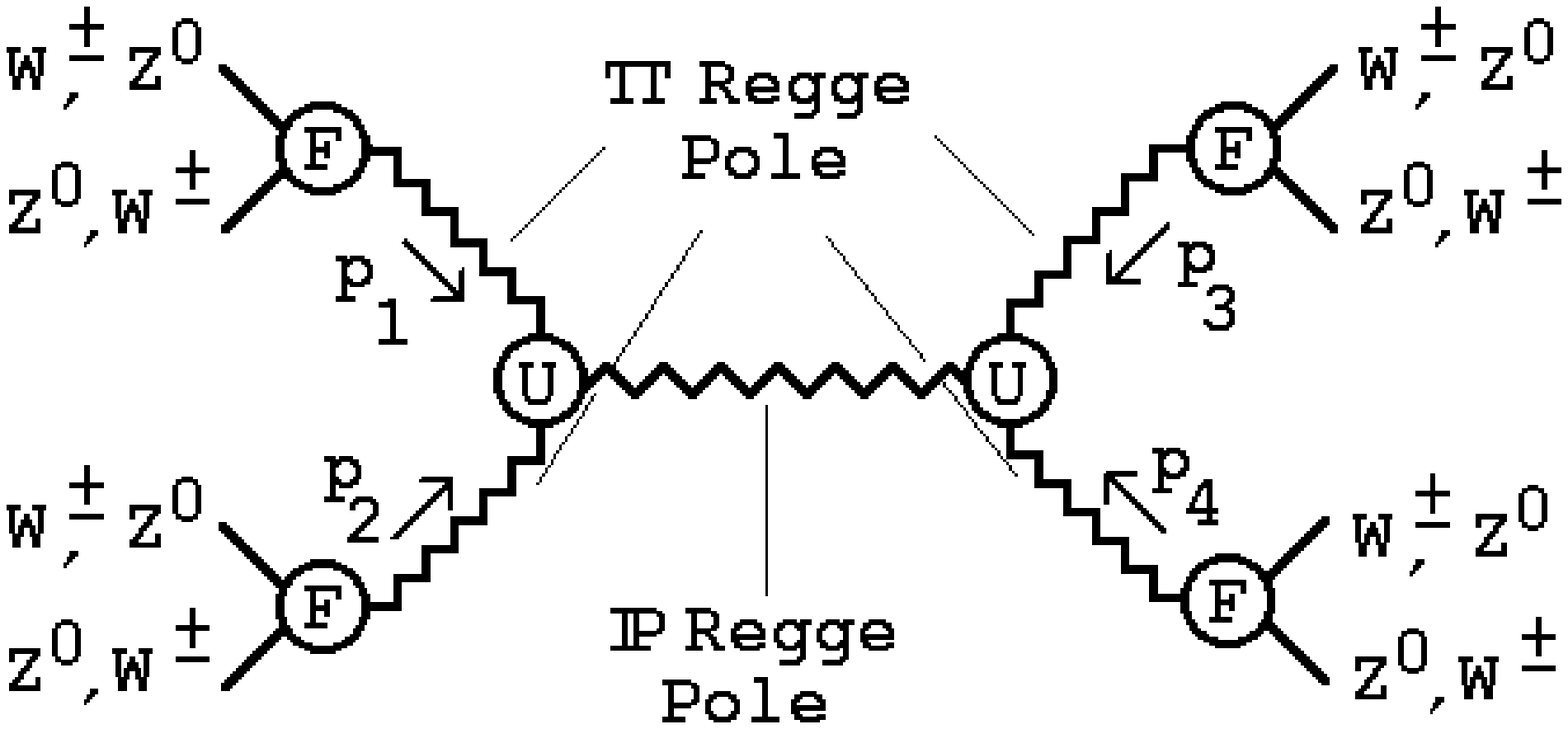}

Fig.~3.16 A Multi-Regge Amplitude.
\end{center}
no questionable procedures would be necessary. 
The appearance of the reduced
triangle diagram in the coupling of the external $W$ and $Z_0$ states 
would be (complicated but) straightforward in principle.
The anomaly (or pion) pole would directly appear in conjunction with
the transverse momentum divergences. 
The one subtlety that would remain 
would be the interplay between the ultra-violet
and infra-red contributions of both the chiral and
U(1) anomalies. However, we will not elaborate on this here.

It is important to note that, in this paper,
we will isolate the anomaly pole (in principle by a 
phase-space restriction)
in all the anomaly subdiagrams within our amplitudes.
In \cite{arw98} we proposed
starting with initial states that had effectively the same
wee gluon content as the pions we create via the anomaly pole. However, we 
then allowed them to scatter into arbitrary multi-reggeon states and 
argued that the anomaly interactions generate 
an overall logarithmic infra-red divergence that
selects the allowed physical states and amplitudes. 
In the present discussion, we will
require pion poles in both the initial and final states. Nevertheless, for
subtle reasons the overall divergence will remain logarithmic.

As a final point, before we proceed to the 
construction of actual amplitudes, we 
note that that we will impose a cut-off in all
transverse momenta. This has a dual purpose. Firstly, to obtain contributions
from ``relatively simple'' (gauge-dependent) feynman diagrams
to gauge-invariant transverse momentum diagrams (that have contributions,
of course, from
many feynman diagrams). Secondly we will want to exploit the infra-red scaling
properties of multigluon 
transverse momentum diagrams that lead to infra-red divergences 
as discussed in the above sub-section.
Our final results will be presented 
entirely in terms of transverse momentum diagrams or, at a later stage,
reggeon diagrams. 

\subhead{3.6 Light-Cone Momenta and Lorentz Frames}

Light-cone momenta are clearly a central feature of our discussion. 
In discussing the various components of diagrams of the form of
Fig.~3.12(a) we will need to allow for a variety of light-cone
momenta, both externally and as integration
variables. In particular, to introduce the triple-regge U(1) anomaly 
interaction the wee gluons in an outgoing pion must be associated with 
a light-cone whose space direction is orthogonal to that of the incoming
wee gluon light-cone. To describe this 
we will need to introduce some new light-cone
notation. In addition we will need to introduce a set of Lorentz frames in 
which the various external momenta take specific forms. 

We begin in what we will call the ``finite momentum frame'' ${\cal F}_L$
for the left-hand
part of Fig.~3.12(a). In this frame we write
$$
\eqalign{p_1 ~&=~{k}^{1^+}~+~{q}^{1^-} \cr
&=~{k}_{1^-}~+~{q}_{1^+} \cr
&=({k \over \sqrt{2}},{k \over \sqrt{2}},0,0)~
+~({q \over \sqrt{2}},- {q \over \sqrt{2}},0,0 )}
\auto\label{kin1l}
$$ 
The notation is straightforward in that ${k}^{1^+}$ is 
a vector with raised
index component along the light-cone defined by the
positive $\{1\}$ - axis (and all other othogonal components are 
zero). Simiilarly ${q}^{1^-}$ is a vector with raised
index component along the light-cone defined by the
negative $\{1\}$ - axis. 
The same vectors can be labeled via lowered index components as usual.
We similarly write 
$$
\eqalign{p_2 ~&=~- ~{k}^{2^+}~-~{q}^{2^-} \cr
~&=~- ~{k}_{2^-}~-~{q}_{2^+} \cr
 &= ~- ({k \over \sqrt{2}},0,{k \over \sqrt{2}},0)
~-~({q \over \sqrt{2}},0,-{q \over \sqrt{2}},0)}
\auto\label{kin2l}
$$ 
where now  ${k}^{2^+}$ is a vector with raised
index component along the light-cone defined by the
positive $\{2\}$ - axis while  ${q}^{2^-}$ is a vector with raised
index component along the light-cone defined by the
negative $\{2\}$ - axis.
Since
$$
p_1^2 ~= ~p_2^2~=~2kq 
\auto\label{kin21l}
$$
we see that 
$$
q \to 0 ~~\implies ~~p_1^2 ,~ p_2^2 ~\to ~0
\auto\label{kin021l}
$$

In the ``infinite momentum frame''  ${\cal F}_I$, 
in which we will consider the 
complete scattering process, the momenta $p_1$ and $p_2$ are obtained from 
their finite momentum frame forms by applying a boost $a_z(\zeta)$ along
the $z$-axis. If $C=cosh \zeta$ and  $S=sinh \zeta$ then
$$
p_1 ~=~ (C ~{k +q \over \sqrt{2}},{k -q\over \sqrt{2}},0,S~{k+q \over \sqrt{2}})
\auto\label{kin3l}
$$ 
and
$$
p_2 ~=~ - (C~ {k +q  \over \sqrt{2}},0, 
{k -q \over \sqrt{2}},S~{k+q \over \sqrt{2}})
\auto\label{kin4l}
$$

Similarly, in the ``finite momentum frame''  ${\cal F}_R$
the momenta entering the right-hand
part of Fig.~3.12(a) have the form
$$
\eqalign{p_3 ~&=~{k}^{2^+}~+~{q}^{2^-} \cr
 &=({k \over \sqrt{2}},0,{k \over \sqrt{2}},0)
~+~({q \over \sqrt{2}},0,{- q \over \sqrt{2}},0)}
\auto\label{kin5}
$$ 
and
$$
\eqalign{p_4 ~&=~-~{k}^{1^+}~-~{q}^{1^-} \cr
 &=~- ({k \over \sqrt{2}},{k \over \sqrt{2}},0,0)
~-~({q \over \sqrt{2}},-{q \over \sqrt{2}},0,0)}
\auto\label{kin6}
$$ 
and so we also have 
$$
p_3^2 ~= ~p_4^2~=~2kq
\auto\label{kin21r}
$$

For the right-hand momenta, however, the infinite momentum frame ${\cal F}_I$
is reached from 
the finite momentum frame ${\cal F}_R$ 
by applying a boost $a_z(-\zeta)$ along
the $z$-axis.  ${\cal F}_R$ is therefore reached from ${\cal F}_L$ by a boost 
$a_z(-2 \zeta)$. In  ${\cal F}_I$
$$
p_3 ~=~ (C ~{k +q \over \sqrt{2}},0,{k-q  \over \sqrt{2}},
-S~{k+q \over \sqrt{2}})
\auto\label{kin3r}
$$ 
and
$$
p_4 ~=~- (C~ {k +q \over \sqrt{2}},{k-q \over \sqrt{2}},0, 
-S~{k+q \over \sqrt{2}})
\auto\label{kin4r}
$$
Evaluating all momenta in  ${\cal F}_I$ we have 
$$
s~=~(p_1+p_3)^2 = ~(p_2+p_4)^2 
~\centerunder{$\longrightarrow$}{\raisebox{-5mm}{
$q \to 0$}}
~~(C^2 + S^2)k^2 ~
\centerunder{$\sim$}{\raisebox{-5mm}{
$C \to \infty$}} ~2C^2 k^2
\auto\label{kin5r}
$$
$$
t~=~(p_1+p_2)^2 ~\centerunder{$\longrightarrow$}{\raisebox{-5mm}{
$q \to 0$}}
~ ~- k^2 ~~~~~~~~~~~~~~~~~~~~~~~~~~~~~~~~~~~~~
\auto\label{kin6r}
$$
Therefore, we now have three external momentum scales, in addition to one
mass scale, in our discussion, i.e.
$$
q^2 ~~<< ~~ M_C^2~~<< ~~ k^2 ~~ << s
\auto\label{sca3}
$$
The mass-shell limit is now $q\to 0$ and the regge limit
$ s/t \to \infty$ is obtained as $C \to \infty$. In the following we will
combine these limits by taking 
$$ 
q ~\sim ~1 / C ~\to ~0 ~, ~~~ ~~ q~C ~>> M_C
\auto\label{sca4}
$$
  
\subhead{3.7 Constructing Amplitudes}

To construct amplitudes corresponding to the diagrams of 
Fig.~3.12(a) we proceed as outlined in the above subsections.
We first consider Fig.~3.5 as a one
loop feynman diagram within $F_1$ and ignore the hatches.
We consider the analagous diagram within $F_2$ 
and connect the two diagrams with
the $U_L$ diagram of  Fig.~3.13 to obtain the
full diagram shown in Fig.~3.17. 
\begin{center}
\leavevmode
\epsfxsize=4in
\epsffile{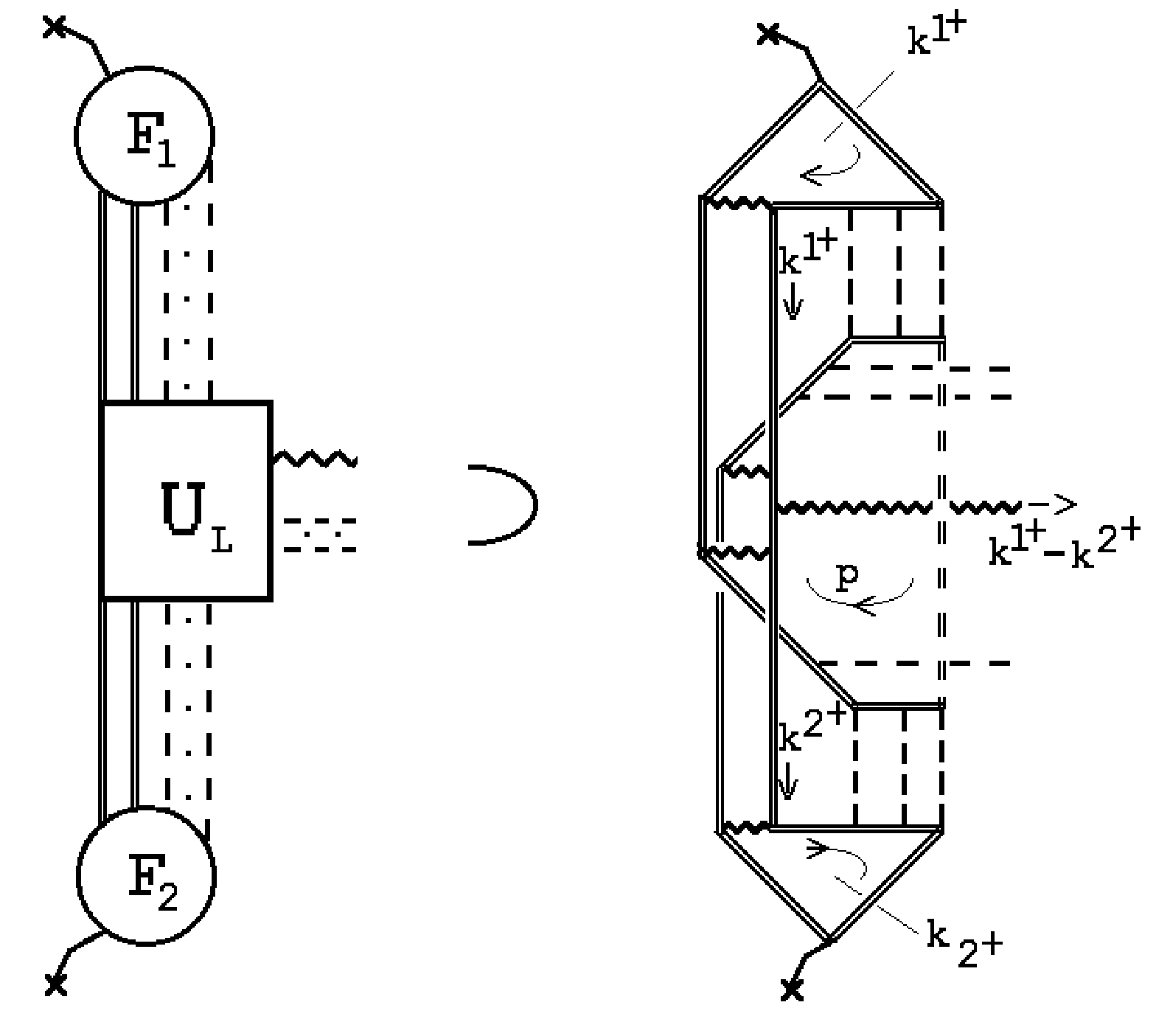}

Fig.~3.17 Connecting the $F_1$ and $F_2$ Amplitudes.
\end{center}
(The double-dashed line carries zero-momentum within the
anomaly configuration that will be discussed later.)
If we then treat this diagram as a subdiagram and join it 
with it's own reflection we obtain a complete diagram
of the form shown in Fig.~3.12(a). The 
left and right-hand subdiagrams will be 
joined only by the exchanged 
gluons (three massless and one massive), which will carry finite transverse
momentum (in all three Lorentz frames). The relevant parts of the 
left and right-hand subdiagrams
will have analagous forms in the ${\cal F}_L$ and ${\cal F}_R$ frames,
respectively, and will be in a relative regge limit in the
${\cal F}_I$ frame. The combination of the regge limit with the phase
space retrictions we impose will, as anticipated,
place a large number of lines on-shell such that
the central quark loop within $U_L$ reduces
to a triangle diagram as illustrated schematically in Fig.~3.14. The crucial
element will be, of course, that this diagram also contains the anomaly
pole. 
To understand this we must determine all the effective vertices that are
produced by the reduction to transverse momentum integrals.

To discuss the diagram of Fig.~3.17, we will begin in the  
${\cal F}_L$ frame and as we evaluate each part of the diagram we will
discuss the effect of transforming to the ${\cal F}_I$ frame. 
In the ${\cal F}_L$ frame  
$p_1$ and $p_2$ are given, respectively, by (\ref{kin1l}) and (\ref{kin2l}).
We direct the large light-cone momenta $k^{1^+} $ and 
$k^{2^+}$ through the diagram as shown and restrict 
the integration in both $F_i$ diagrams 
to the momentum region corresponding to (\ref{intr01}). Note that 
$$
\eqalign{k^{1^+}~-~k^{2^+} ~&= ~({k \over \sqrt{2}},{k \over \sqrt{2}},0,0)~
-~({k \over \sqrt{2}},0,{k \over \sqrt{2}},0)\cr
&=~(0,{k \over \sqrt{2}},-{k \over \sqrt{2}},0)}
\auto\label{splk}
$$
is a spacelike momentum lying in the $\{x,y\}$ - plane.
We introduce notation for all the loop
momenta of Fig.~3.17 in Fig.~3.18. 
We show only that part of the
diagram involving $F_1$ and part of $U_L$. The part containing $F_2$ 
can obviously be discussed analagously.
The hatched lines are those placed on-shell
by longitudinal momentum integrations and each 
of the hatches is labeled by the index for the momentum involved. 
We discuss each integration separately as follows. 
\begin{center}
\leavevmode
\epsfxsize=4in
\epsffile{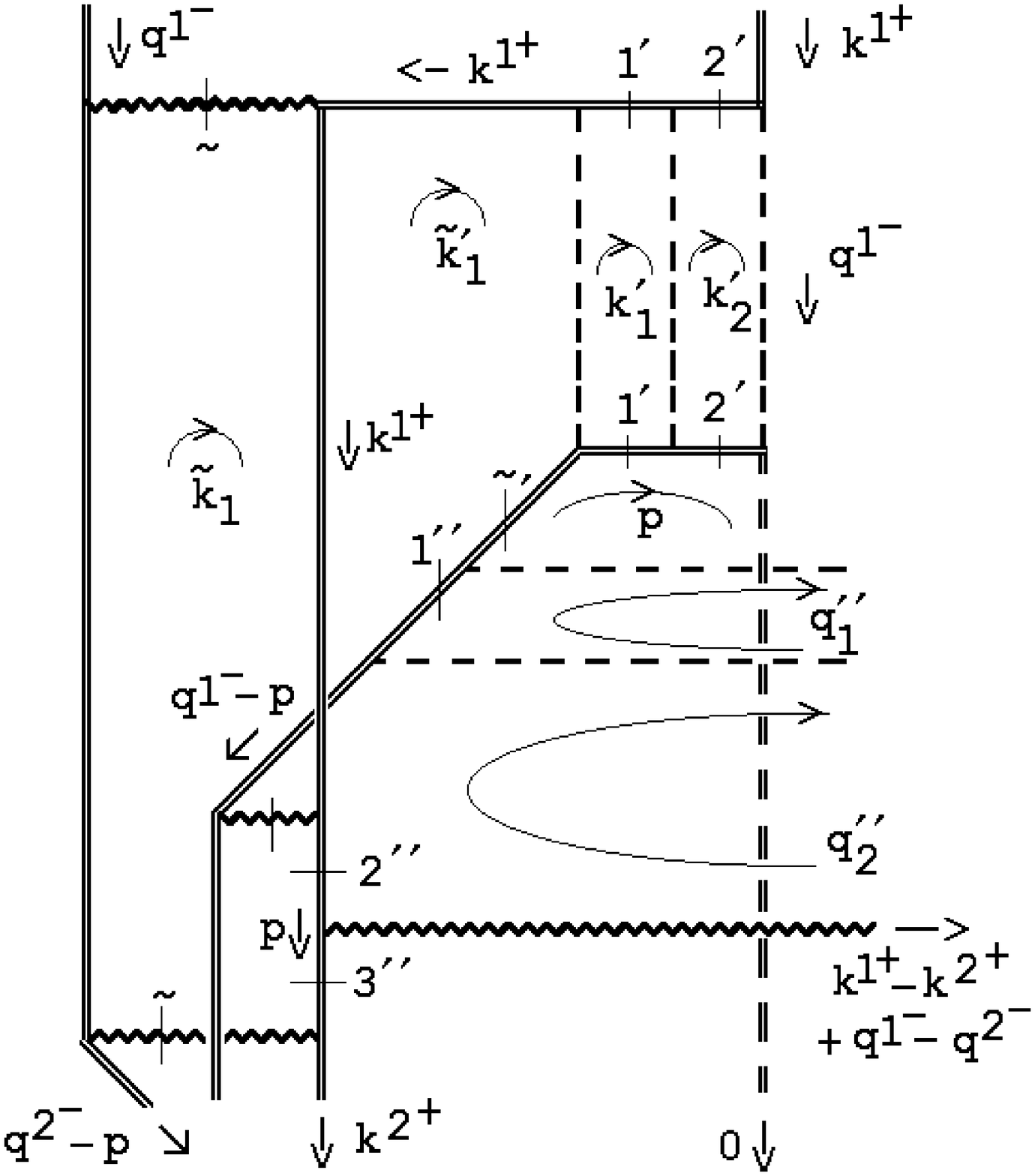}

Fig.~3.18 Notation for Fig.~3.17.
\end{center}

\subhead{3.8 The Massless Gluons}

Provided the loop momentum $p$ in the 
lower part of Fig.~3.18 is much less than $k_{1^+}$ (as will be the case
in the anomaly pole contribution we will extract) 
the $k_i'$ - integrations over 
the momenta of the vertical 
massless gluon exchanges can be reduced to transverse momentum diagrams 
by placing the hatched fermion lines on-shell. We can illustrate this 
(very well-known) procedure as follows. 
Using conventional light-cone co-ordinates, which in the notation of
subsection 3.6 correspond to 
light-cone vectors ${k'}^{1^+}$ and ${k'}^{1^-}$, we can write 
$$
\eqalign{&\int d^4k_1'd^4k_2'~~ \bigg\{{\gamma^{\mu}
~\gamma \cdot (k_1' - k^{1^+} + q^{1^-}) 
~\gamma^{\nu}~\gamma \cdot (k_2' - k^{1^+} + q^{1^-}) 
~\gamma^{\tau}
\over (k_1' - k^{1^+} + q^{1^-})^2 
~(k_2' - k^{1^+} + q^{1^-})^2}\biggr\} ~~\otimes\cr
& \biggl\{{\gamma_{\mu}~
\gamma \cdot (p - k_1'  - q^{1^-})~ \gamma_{\nu}
~\gamma \cdot (p - k_2' - q^{1^-})~ \gamma_{\tau} \over
 (p - k_1'  - q^{1^-})^2 ~
(p - k_2' - q^{1^-})^2}\biggr\}~ {1 \over
(k_1'- \tilde{k}'_1)^2 (k_2' - k_1')^2 (k_2' + q^{1^-})^2 }\cr
& \sim 
\biggl\{\gamma^{\mu}~{\int {dk_1'}^- \gamma_+ k^{1^+} \over
({k_1'}^- k^{1^+} + \cdots )^2 }~\gamma^{\nu}~
{\int {dk_2'}^- \gamma_+ k^{1^+} \over
({k_2'}^- k^{1^+} + \cdots )^2}~\gamma^{\tau}\biggr\}\cr
&~~~~~~~\otimes
~\biggl\{\gamma_{\mu}~{\int {dk_1'}^+ \gamma_+ p^+ \over
({k_1'}^+ p^{+} + \cdots )^2} ~\gamma_{\nu}~
{\int {dk_2'}^+ \gamma_+ p^{+} \over
({k_1'}^+ p^{+} + \cdots )^2}~\gamma_{\tau}\biggr\}\cr
&~~~~~~~~~~~~~~~\times ~\int d^2k_{1\perp}'d^2k_{2\perp}'~~ {1 
\over {(k_{1\perp}'- \tilde{k}_{1\perp}'})^2 
(k_{2\perp}' - k_{1\perp}')^2 (k_{2\perp}')^2 }\cr
&\sim ~~ \gamma^+ \otimes \gamma_+
~~ \int d^2k_{1\perp}'d^2k_{2\perp}'d^2k_{3\perp}' 
~~{\delta^2(\tilde{k}'_{1\perp} ~-~ \sum_i k_{i\perp}) 
\over {(k_{1\perp}'- \tilde{k}_{\perp}'})^2 
(k_{2\perp}' - k_{1\perp}')^2 (k_{2\perp}')^2 } }
\auto\label{3mg}
$$
which is a transverse state of the kind discussed in 
subsection 3.2. This illustrates how the integrals (\ref{rdve}) arise.
Also the $\gamma^+~ (~= \gamma_-)$ factor is the effective vertex appearing
in Fig.~3.7. 

(\ref{3mg}) is, as anticipated, infra-red divergent. As we discussed
at length in subsection 3.2, if the three gluon state carries color
all of the divergences will exponentiate in higher orders. If it
carries color zero the only divergence which will not exponentiate is the
the overall divergence that potentially occurs when $\tilde{k}'_{1\perp}$
is integrated over and the $k'_{i\perp}$ are scaled uniformly to zero. 
If this divergence is present and we isolate it's contribution, the 
massless multigluon propagators will contribute only at zero momentum 
and there will be no effect in transforming their contribution from 
the ${\cal F}_L$ frame to the ${\cal F}_I$ frame. (While the contribution
of the anomaly amplitudes to which the multigluon states couple
will depend on the small light-cone momentum $q^{1^-}$, the contribution
of the transverse propagators and interactions will be independent of
this momentum.)   
To discuss the exact nature of the divergence we must include 
the effective vertices provided by the anomaly amplitudes and the 
contribution of the quark/antiquark state.

For the reasons
discussed in subsection 3.5, we do not integrate over $\tilde{k'}_1^-$ and so  
at $\tilde{k'}_{1\perp} = 0$
the effective vertex provided by the $F_1$ amplitude will be the anomaly pole
amplitude of (\ref{Fan1}). In frame ${\cal F}_I$ we will use 
(\ref{bay5}) which gives the pion coupling
after the current momentum factor has been removed. In the present notation 
the full effective vertex provided by the $F_1$ amplitude is then
(without the current momentum factor) 
$$
\tilde{k}_2~~{\hbox{\Large $\epsilon$}_{\sigma\delta 3 - }~ 
(k^{1^+})^{\sigma} (q^{1^-})^{\delta} \over q^2}
~~\sim  ~~ \tilde{k}_2~ {\bigl[~k~q~C ~ \bigr] \over M_C^2~q^2} 
\auto\label{bay51}
$$
which, when $qC$ is kept finite, gives a finite pion pole residue.
Note that, since this vertex is independent of $\tilde{k'}_{1\perp}$, 
it will not affect the divergence at $\tilde{k'}_{1\perp} = 0$ . This
is a crucial consequence of the absence of a vector Ward identity for
the anomaly pole contribution.

In the regge limit, the $q''_{1}, q''_2$ and $q''_3$ integrations 
will, in analogy with (\ref{3mg}), be reduced to 
transverse momentum integrals in 
the $\{x,y\}$ - plane by placing on-shell the labelled hatched lines.
($q''_3$ is the momentum of the horizontal massless gluon line attached
to the bottom of the diagram.)
Since the  ${\cal F}_L$,  ${\cal F}_I$ and ${\cal F}_R$ 
frames differ only by boosts acting
in the $\{z,t\}$ - plane, the $q''_{j}$ transverse momentum
integrations will be the same in each of the frames we discuss and will
produce the same infra-red divergence. 
If we continue to work in the ${\cal F}_L$ frame the combination
of the $k_i'$ and $q_j''$ longitudinal integrations generates the effective 
vertices shown in Fig.~3.19. 
\begin{center}
\leavevmode
\epsfxsize=5in
\epsffile{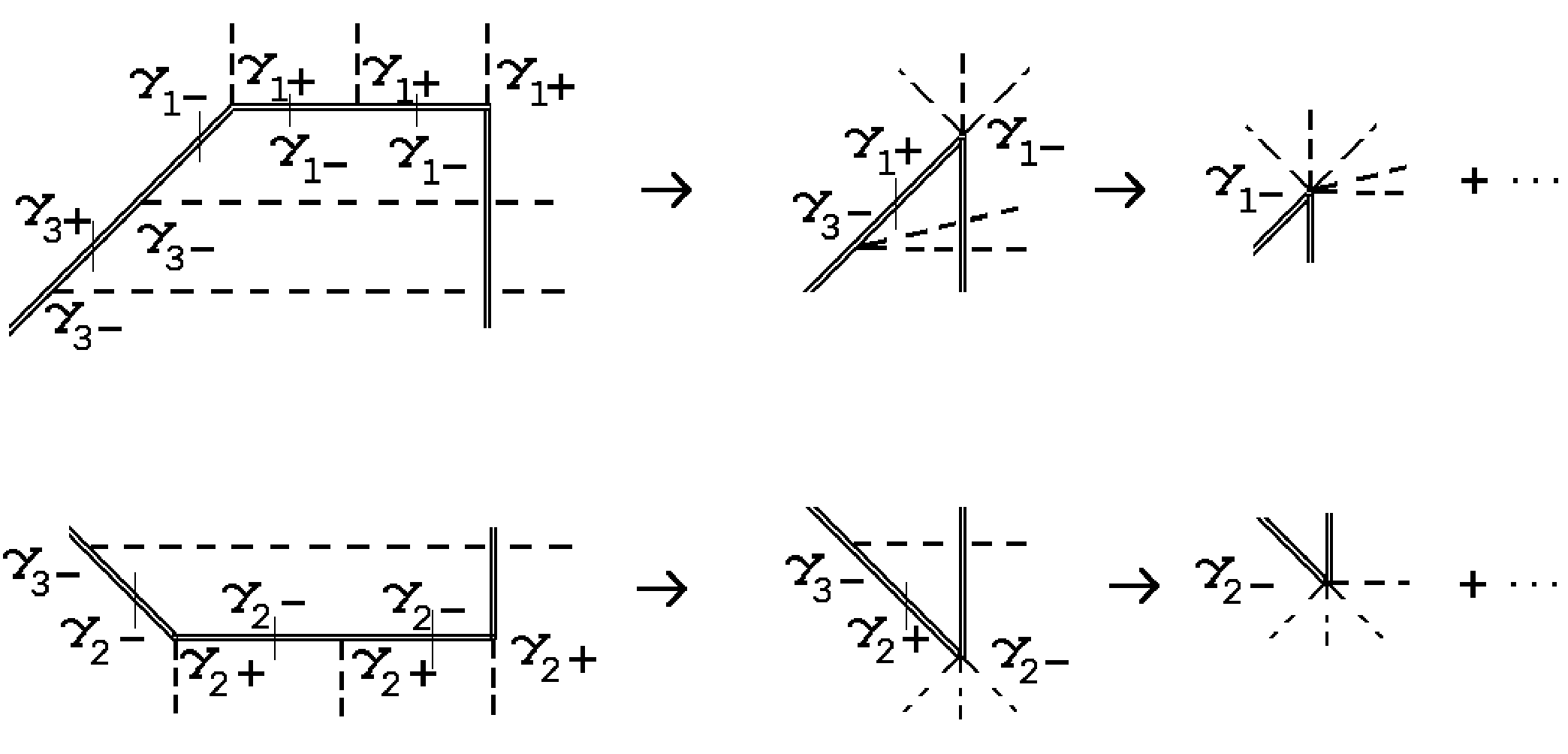}

Fig.~3.19 Massless Gluon Effective Vertices. 
\end{center}
We will combine these vertices to obtain the anomaly amplitude produced by
the $U_L$ loop shortly.

\subhead{3.9 Quark Transverse Momentum Integrals}

The reduction to transverse integrals of the quark loop integrations, 
over $\tilde{k}_1$ in Fig.~3.18
and over $\tilde{k}_2$ in the lower part of Fig.~3.17, is not straightforward.
This reduction should be responsible for 
placing all of the hatched massive vector propagators on-shell. However, the 
light-cone momenta flowing along these lines ($q^{1^-}$ and $q^{2^-}$)
is small (and zero on mass-shell) in ${\cal F}_L$, although it is finite 
in ${\cal F}_I$. Apparently, therefore, there is no regge limit kinematics
for us to exploit. Nevertheless, we need to place the relevant 
lines on-shell, both to
obtain a gauge-invariant result in which we understand the exponentiation
of infra-red divergences, and to utilize the anomaly couplings.

In the full multi-regge limit of Fig.~3.16 the 
$p_i$ momenta would be initially taken spacelike and (as we remarked earlier) 
quark transverse integrals would be obtained naturally. 
Assuming a reggeized pion
appears, the multi-regge amplitude will contain corresponding asymptotic 
dependence on invariant subenergies. This dependence  
should disappear as the scalar pions are placed
on mass-shell at $p_i^2 = 0, ~i=1,..,4$ 
and the on-shell amplitude should factorize out 
straightforwardly. Even though the full multi-regge amplitude  
is independent of the subenergies (at $p_i^2=0$) it is obtained by asymptotic
expansion around infinite subenergies. Correspondingly, any 
transverse integrals that are involved should be initially 
obtained at infinite subenergies. In effect, we are attempting to obtain 
these integrals directly at zero subenergies by appealing only to properties
of the anomaly poles generating the pions. 

In fact, even with the kinematic constraints
we have imposed, we will be able to place all the massive 
gluon lines on-shell (and so, consistently use the anomaly 
couplings). The result will be formally the same as carrying
out the large subenergy limit but only a limited range of transverse 
momenta will be involved. The discontinuity that is (effectively) taken 
will be that of an unphysical pseudothreshold (at zero subenergy), rather 
than a physical normal threshold. Presumably (although we will not
attemt to prove this) the unphysical chirality 
transition involved in the anomaly pole couplings can be
viewed as producing this contribution. (In the last Section we 
described the relationship between an unphysical singularity and
the anomaly pole and in 
\cite{arw011} we emphasized that triple-regge anomaly interactions
are due to unphysical multiple 
discontinuities containing pseudothresholds.)

As we have already noted, our discussion of the effective 
vertex of Fig.~3.10 applies directly to the placing on-shell of 
the upper (massive gluon)
line associated with the $\tilde{k}_1^{1^+}$ - integration in Fig~3.18. 
The $\tilde{k}_1^{1^-}$ - integration associated with the lower on-shell line
is very similar and massive gluon exchange
must again be involved. To see this we must establish which
$\gamma$ matrix couplings appear at the vertices. In fact,
these couplings are almost entirely determined by the requirement that 
the anomaly be present in the reduced $U_L$ diagram. 
The complete $\gamma$ - matrix 
structure of Fig.~3.17 that is not included in Fig.~3.19
is shown in Fig.~3.20.
The top and bottom trios of $\gamma$ - matrices in the initial
figure are those due to the effective vertices of $F_1$ and $F_2$
that are analagous to Fig.~3.10 together with the resulting propagator
components. We specifically choose the $\{3\}$ component for both the $F_1$
and $F_2$ external currents. This choice, together with the choice of the 
space directions for the large light-cone
momenta of $p_1$ and $p_2$, determines the relative structure of the
trios. The appearance of the $\gamma_{3^{\pm}}$ - matrices is a 
direct consequence of the regge limit. 
\begin{center}
\leavevmode
\epsfxsize=5.3in
\epsffile{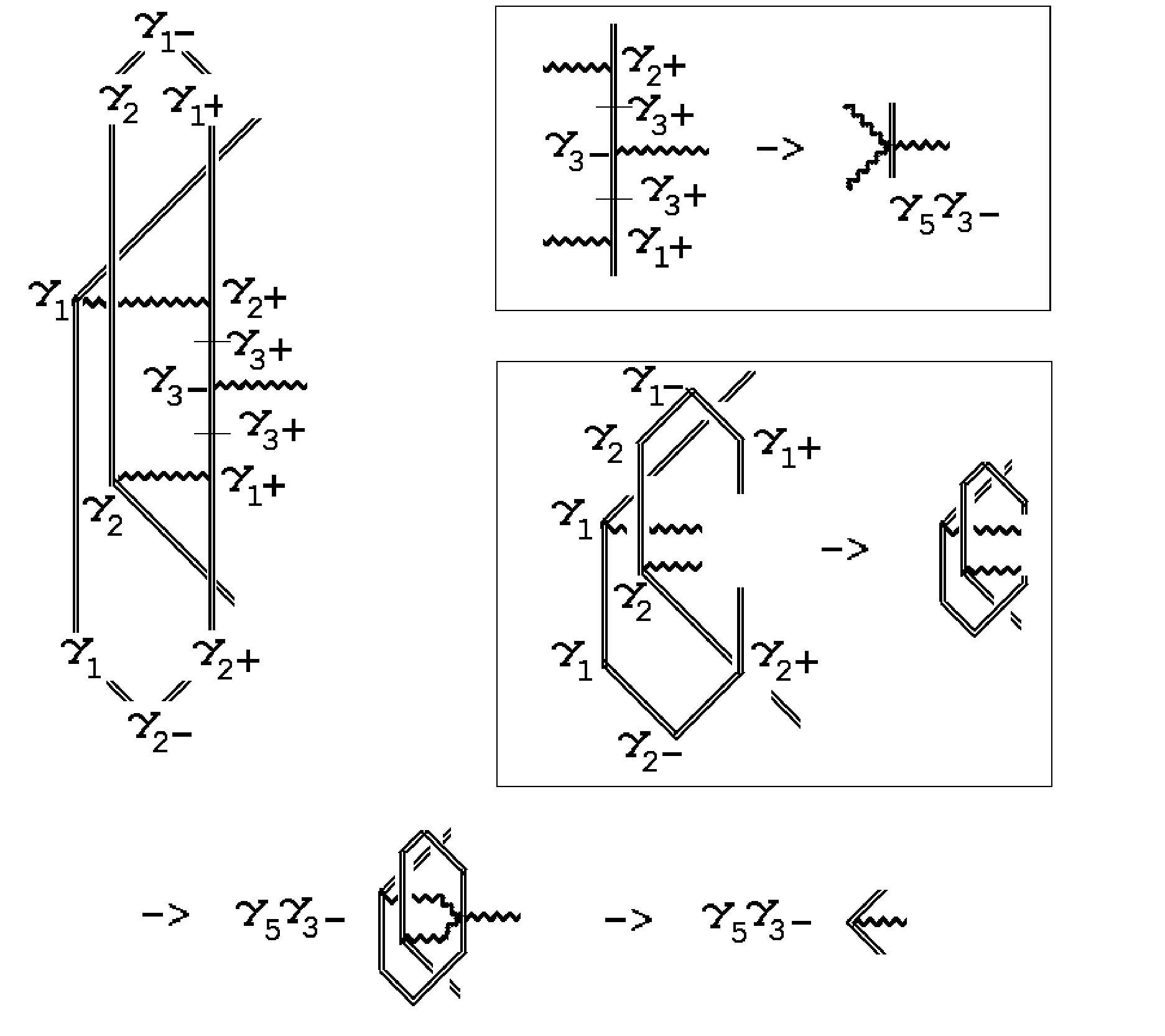}

Fig.~3.20 The $\gamma$ - Matrix Structure 
Generating the $\gamma_5$ - vertex. 
\end{center}
The upper inset in Fig.~3.20 shows how the identity (\ref{3ga})
generates a $\gamma_5$ - interaction. The participating $\gamma_{1^+}$
and  $\gamma_{2^+}$ matrices have to be produced by the longitudinal
momentum integrals (via massive gluon exchange), as we discuss below. 
The remaining $\gamma_{1}$
and  $\gamma_{2}$ matrices are needed to allow 
the reduction of the remaining product to the unit matrix (plus terms that
give zero when contracted with the massless gluon vertices), 
as the lower inset illustrates. 
It is not 
difficult to see that the requirement of a $\gamma_5$ - interaction,
together with a non-zero reduction of the remaining matrix product
determines the complete structure of Fig.~3.20.

In Fig.~3.21 we have isolated the $\gamma$-matrix structure and the relevant
momenta for the quark loops that couple $F_1$ and $F_2$ in Fig.~3.17.
Each of the $\gamma$-matrices in Fig.~3.21 is either a vertex
component of a massive gluon propagator  or is a numerator component of
a quark propagator. Although the $\gamma$-matrices contract, 
as we have already discussed, 
the corresponding momentum factors remain. We ignore the loop momentum
$p$ since it will be set to zero by the generation of the U(1) anomaly pole.
Also, since 
$\tilde{k}'_{1 \perp} = \tilde{k}'_{2 \perp} = 0$ after the transverse 
momentum divergence is extracted,
we first ignore both of these momenta.
\begin{center}
\leavevmode
\epsfxsize=4in
\epsffile{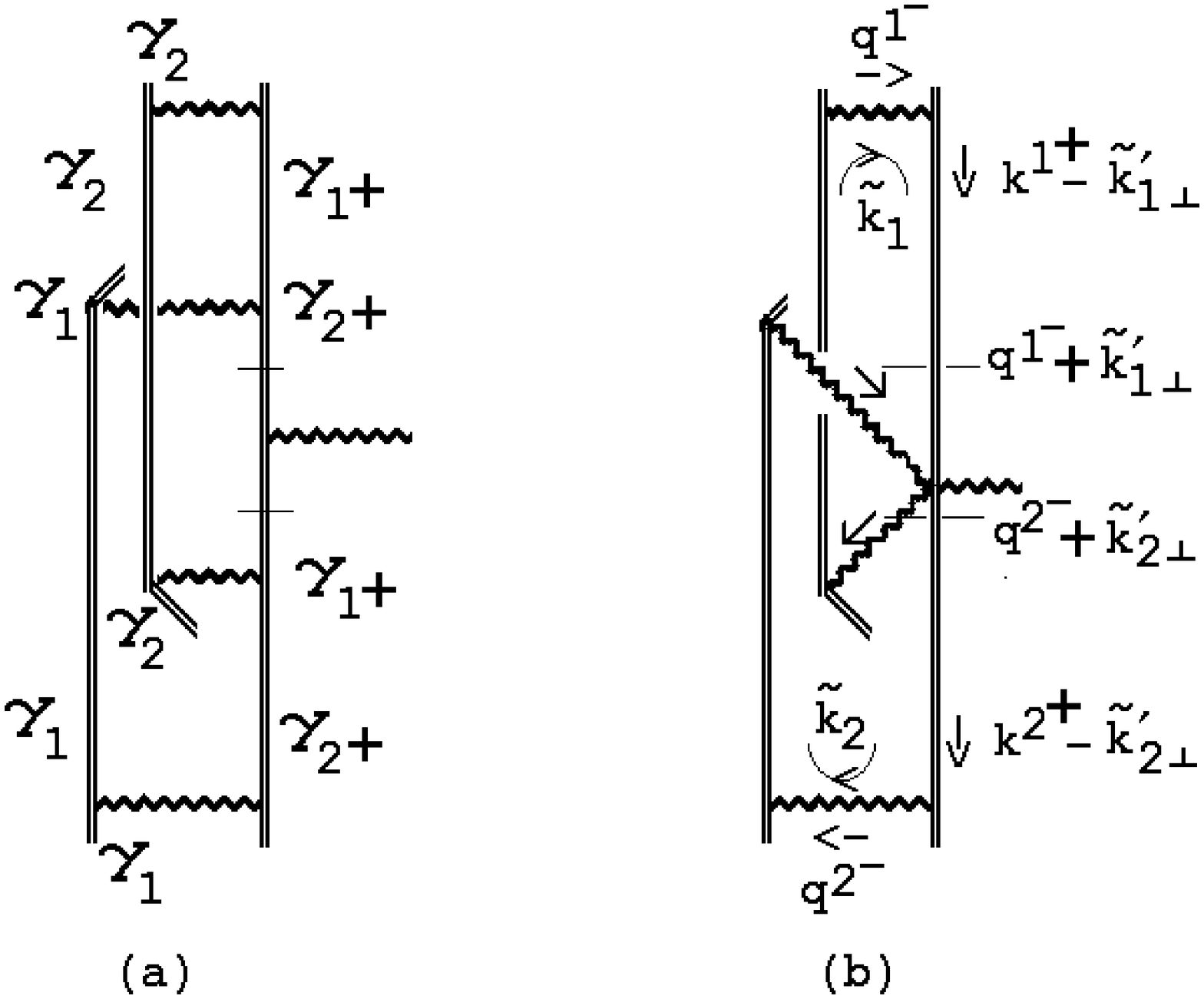}

Fig.~3.21 The Quark Loops (a) $\gamma$-matrices (b) Momenta 
\end{center}

The  $\tilde{k}_1^+$ integration is given by (\ref{msh1}) while
the  $\tilde{k}_1^-$ integration has the form
$$
\int {d \tilde{k}_1^-~(\tilde{k}_1^+\gamma_+ + q^{2^-}\cdot \gamma_+)
\over 2\tilde{k}_1^+ \tilde{k}_1^- +2 q^{2^-}\cdot\tilde{k}_1 
 -~\tilde{k}_{1\perp}^2~ -~M_C^2 }~
\times ~ \cdots ~~~\sim ~~ \gamma_+ ~ \times ~ \cdots
\auto\label{msh12}
$$
The two integrations give, respectively,
$$
 2\tilde{k}_1^+\tilde{k}_1^- + 2\tilde{k}_1\cdot q^{1^-}~
= ~\tilde{k}_{1\perp}^2~ +~M_C^2
\auto\label{msh11}
$$
and
$$
 2\tilde{k}_1^+\tilde{k}_1^- + 2\tilde{k}_1 \cdot q^{2^-}~
= ~\tilde{k}_{1\perp}^2~ +~M_C^2  
\auto\label{msh121}
$$
In the ${\cal F}_I$
frame, $q^{1^-}$ and $ q^{2^-}$ are boosted to become almost the same
light-like momentum $q^{3^+}$, i.e. 
$$ 
q^{1^-} ~\sim~ (Cq,-q,0,Sq)~~
\centerunder{$\sim$}{\raisebox{-4mm}{$q\to 0$}}
~~ 
q^{3^+}~, ~~~~ q^{2^-} ~\sim~ (Cq,0,-q,Sq)~~
\centerunder{$\sim$}{\raisebox{-4mm}{$q\to 0$}} ~~q^{3^+}
\auto\label{all}
$$
and so (\ref{msh11}) and (\ref{msh121})  
have a common solution as $q\to 0$ (which is why a pseudothreshold 
is involved) with 
$$
\tilde{k}_1^+ \sim \tilde{k}_1^- \sim\tilde{k}_{13}~~ \sim ~ ~
{M^2_C \over C~q}
 ~, ~~~~~\tilde{k}^2_{12} 
~~\centerunder{$<$}{\raisebox{-1mm}{$\sim$}}~~ M^2_C 
\auto\label{csol}
$$ 
Incorporating
all the remaining momentum factors given by the $\gamma$-matrices of 
Fig.~3.21(a) (together with a factor of $M_C^{-2}$ from the additional
exchanged gluon propagator)
the $\tilde{k}_{1\perp}$ transverse momentum integral has the form 
$$
\eqalign{ & {1 \over M_C^2}~\int d^2\tilde{k}_{1\perp} 
~\{F_1~ \hbox{numerator}\}
\times  \{\hbox{antiquark numerator}\}
\times \{\hbox{gluon numerator} \} \cr
&~~~~~~~~~~~~~~~\times \{\hbox{quark numerator}\}
~\big/~\{ \hbox{propagator denominator}\}^2 \cr
& =~ 
{1 \over M_C^2}~\int {d^2\tilde{k}_{1\perp} 
~\tilde{k}_{12}~\tilde{k}_{12}~ (\tilde{k}_{12} + q)~
 k^{1^+} \over \bigl(\tilde{k}_{1}^2 \bigr)^2 } }
\auto\label{qti11}
$$
$$
\sim ~ {q ~ k^{1^+} \over M_C^2}~
\biggl\{\int_{|\tilde{k}_{13}| \sim {M_C^2 \over Cq}}~d\tilde{k}_{13} 
\biggr\} ~\biggl\{\int_{{M_C^2 \over C~q}~
\centerunder{${\scriptscriptstyle < }$}{\raisebox{-1mm}{$
{\scriptscriptstyle \sim} $}} |\tilde{k}_{12}|
\centerunder{${\scriptscriptstyle <}$}{\raisebox{-1mm}{${\scriptscriptstyle
\sim}$}}~ M_C}~ {d\tilde{k}_{12}\over \tilde{k}_{12}^2 }\biggr\}
~~~~~~~~~~~~~~~~~~~~~~
$$
$$
\sim {q ~ k^{1^+} \over M_C^2} ~\biggl\{{M_C^2 \over Cq }\biggr\}
~\biggl\{{Cq \over M_C^2}\biggr\}
~~~\sim {q ~ k^{1^+} \over M_C^2}~~~~~~~~~~~~~~~~~~~~~~~~~~~~~~~~~~~ 
\auto\label{qti1}
$$
which gives a finite answer in the  ${\cal F}_I$ frame when $k^{1^+} 
\sim C \to \infty$, with $Cq$ kept finite. Note that the part of
the integrand in (\ref{qti11}) that does not vanish when $q=0$ is odd
with respect to  $\tilde{k}_{12}$ and hence integrates to zero. 
If this were not the case, the combination of light-cone momentum factors
from each fast quark numerator would give an amplitude 
increasing like $C^4 \sim s^2$.

Conversely, if we obtain only finite results of the form of
(\ref{qti1}) for each transverse momentum
integral, we will not obtain any increasing behavior as $s\to \infty$.
To obtain the maximally increasing amplitude, we must consider the 
$\tilde{k'}_{1\perp}$ dependence in more detail. If, for example, we  
direct $\tilde{k'}_{1}$
so that in the $\tilde{k}_{1\perp}$ integral we 
substitute $\tilde{k'}_{12}$  for $\tilde{k}_{12}$ in the antiquark 
numerator in (\ref{qti1}), this will give 
$$
{q ~k^{1^+}\over M_C^2}~~\to ~~ 
{\tilde{k'}_{12} ~k^{1^+}\over M_C^2 } ~~ \sim ~~k~C ~~ 
~\centerunder{$\to$}{\raisebox{-4mm}{$C \to \infty$}} ~~\infty
\auto\label{qti01}
$$
Alternatively we can keep the  $\tilde{k'}_{12}$
dependence of one of the denominators giving
$$
\int {d \tilde{k}_{12} ~\tilde{k}_{12}
\over ( \tilde{k}_{12}
- \tilde{k'}_{12})^2 }
 ~~~\centerunder{$\sim$}{\raisebox{-5mm}{$\tilde{k'}_{12} \to 0$}}
~~~\int {d\tilde{k}_{12}  ~\tilde{k'}_{12} 
\over \tilde{k}_{12}^2}
\auto\label{qti2}
$$
which again leads to (\ref{qti01}). If we keep the contribution of
the form of (\ref{qti1}) from the $\tilde{k}_{2\perp}$ integral
we will obtain a factor of $C$ from the left subdiagram (of Fig.~3.17).
Treating the transverse integrals from the right subdiagram 
in an analagous manner will give an amplitude increasing like $C^2 \sim s$.

Obviously we could also keep the factor of $q$ in the 
$\tilde{k}_{1\perp}$ integral and keep non-leading $\tilde{k'}_{21}$ 
behavior in the $\tilde{k}_{2\perp}$ integral. Either way
we gain one power of the energy while reducing the degree
of divergence of either the   $\tilde{k'}_{1\perp}$
or the $\tilde{k'}_{2\perp}$ integration. We will see that we can not
obtain a further power of the energy by
reducing the degree of divergence of all the $\tilde{k'}_{\perp}$ integrals
since there will then be no overall transverse divergence.
To see this we must consider 
the final part of the diagram that we have not yet discussed in detail.

\subhead{3.10 The U(1) Anomaly Amplitude and the Infra-Red Divergence} 

As illustrated in Fig.~3.22(a), combining Fig.~3.19 and Fig.~3.20 produces 
a triangle of $\gamma$-matrices which has the appropriate structure to  
give the anomaly.
\begin{center}
\leavevmode
\epsfxsize=5.5in
\epsffile{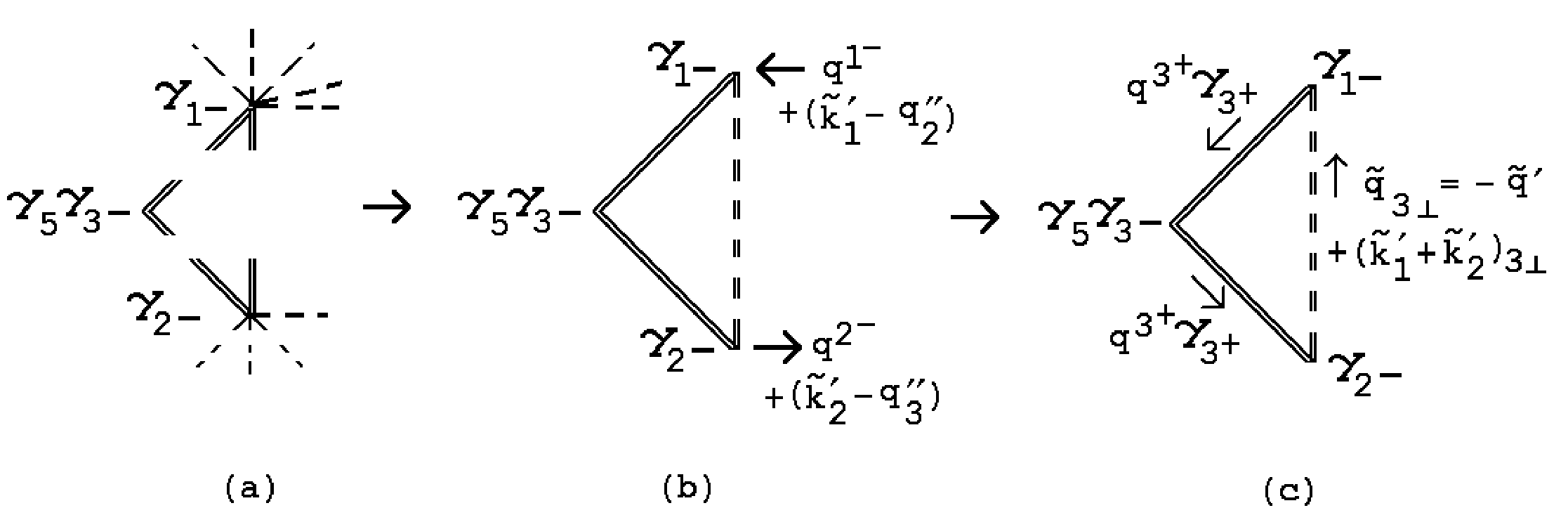}

Fig.~3.22 (a) The Triangle (b) Momenta in ${\cal F}_L $
(c) Momenta in ${\cal F}_I$ .
\end{center}
The large light-cone momenta $k^{1^+}$ and $k^{2+}$ flow in and out of
the $\gamma_5$ vertex and do not enter the triangle diagram. 
As shown in Fig.~3.22(b) the external 
momenta that flow through the diagram in ${\cal F}_L$ are $q^{1^-}$ and
$q^{2^-}$ together with the $\tilde{k}'_i$ and the $q''_j$
(all of which 
are zero in the infra-red divergence configuration for the massless gluons,
when the mass-shell limit is taken).  
In \cite{arw01} we discussed, at length, momentum 
configurations of this kind which produce the anomaly pole. For our 
present purposes it is simplest to go straight to the  ${\cal F}_I$ frame.
In this frame the timelike components of $p_1$ and $p_2$ 
that are $O(q)$ in the  ${\cal F}_L$ frame are boosted to give 
the finite light-like momentum $q^{3^+} \sim C q$, as in (\ref{all}), 
which then flows through the diagram
as in Fig.~3.22(c). In this last figure we have dropped
the small (transverse) momenta along lines where the finite light-like
momentum flows. If we define 
$\tilde{q}_{3\perp}$ to be the (small) momentum transverse to 
$q^{3^+}$  that is flowing through the (double) dashed vertical line then 
$$
\tilde{q}_{3\perp} = (\tilde{k'}_1 + \tilde{k'}_2)_{3\perp} - \tilde{q'}
\auto\label{q3p}
$$
where $\tilde{q'} = {q''}_2 + {q''}_3 $. Comparison with
the momentum configration (\ref{k+k-2})
shows that the anomaly ``$\delta$-function amplitude'' has the form
$$
(q^{3^+})^2 ~ \tilde{q}_{3}~ \delta(\tilde{q}_{3\perp}^2)
\auto\label{q3pa}
$$
which sets to zero momentum the double-dashed line.

The $\delta$-function in (\ref{q3pa}) couples the $q_i''$ and  
$\tilde{k'}_j$ infra-red divergences. As we saw above, these
divergences are also modified by the need to obtain non-zero 
quark transverse momentum integrals.
To consider the remaining  divergence we 
keep only the scaling infra-red divergence from each massless multigluon
state which, as discussed in subsection 3.2, is the only divergence that 
survives high-order exponentiation. 
The overall divergence that remains then has the form
$$
\eqalign{ & \int d^2 \tilde{q'}_{\perp} 
\{\hbox{multigluon scaling amplitude}\}\cr
& \int d^2 \tilde{k'}_{1\perp}d^2 \tilde{k'}_{2\perp}
\{\hbox{multigluon scaling amplitude}\}^2 \{\hbox{anomaly} \}
 \{\hbox{quark momentum} \}  \cr
& \int d^2 \tilde{k'}_{3\perp}d^2 \tilde{k'}_{4\perp} 
\{\hbox{multigluon scaling amplitude}\}^2
\{\hbox{anomaly} \} \{\hbox{quark momentum} \}
 }
$$
$$
\eqalign{ \sim ~ \int {d^2 \tilde{q'}_{\perp} \over \tilde{q'}^2_{\perp}}
 ~&\int {d^2 \tilde{k'}_{1\perp}d^2 \tilde{k'}_{2\perp}
\over \tilde{k'}^2_{1\perp} \tilde{k'}^2_{2\perp} }
~\delta\bigl(~( \tilde{q'}_{\perp} -
 \tilde{k'}_{1\perp}- \tilde{k'}_{2\perp})^2~ \bigr)
~( \tilde{q'}_{\perp} -
 \tilde{k'}_{1\perp}- \tilde{k'}_{2\perp})~  \tilde{k'}_{12} \cr
& \int {d^2 \tilde{k'}_{3\perp}d^2 \tilde{k'}_{4\perp} 
\over \tilde{k'}^2_{3\perp} \tilde{k'}^2_{4\perp} }
~\delta\bigl(~( \tilde{q'}_{\perp} -
 \tilde{k'}_{3\perp}- \tilde{k'}_{4\perp})^2~ \bigr)
~( \tilde{q'}_{\perp} -
 \tilde{k'}_{3\perp}- \tilde{k'}_{4\perp}) ~ \tilde{k'}_{32}\cr
\sim ~ \int {d^2 \tilde{q'}_{\perp} \over \tilde{q'}^2_{\perp}}&
}
\auto\label{ovdi}
$$
which is a simple logarithmic divergence as we anticipated.
We will not attempt to 
prove that this divergence can not be canceled by other diagrams that we
have not discussed. 

\subhead{3.11 The Physical Scattering Amplitude}

We keep as the physical scattering amplitude
the coefficient of the divergence (\ref{ovdi}) - the divergence being  
factorized off as a ``condensate'' that is to be part of the definition
of a physical pion state. (We have discussed how this is consistent
for reggeon states in \cite{arw01}.) The physical amplitude is then
given by
$$
\eqalign{& \prod_i\{F_i~ \hbox{anomaly pole amplitude}\}~
\{\hbox {quark $\tilde{k'}_{i\perp}$ integrals}\}\cr 
& \times ~~\prod_{j=L,R} \{U_j~ \hbox{anomaly amplitude} \}
~\{\hbox{massive gluon propagator}\} } 
\auto\label{phamp0}
$$
and so combining (\ref{bay51}), (\ref{qti1}), (\ref{qti2}) and (\ref{q3pa})
we obtain 
$$
\eqalign{&\biggl(~{k~ C~q \over M_C^2~q^2}~ \biggr)^4~
\biggl(~{(kC)\over M_C^2}~{(kCq) \over M_C^2 }~\biggr)^2~ 
~\biggl(~qC~\biggr)^4~ 
~~{1 \over t + M_C^2 }\cr
&~~~~~= ~\bigl(~{1 \over q^2}~\bigr)^4~ 
\bigl[~{C^2~q^2 \over M_C^2}~ \bigr]^4 ~~
\bigl[~{s~q^2 \over M_C^4}~ \bigr]~~
\bigl[~{t \over M_C^2}~ \bigr]^2~~
\bigl[~{s \over t+M_C^2}~ \bigr]
}
\auto\label{phamp}
$$
We have reorganized the result into the separate square brackets because
each represents a different physical effect, as we now briefly discuss.

The factor of $(1/q^2)^4$ in  (\ref{phamp}) is, of course, the contribution
of the four pion poles. 
All but the last two square brackets are finite constants 
when the limit $1/q \sim C \to \infty$ and so the pion 
scattering amplitude we obtain is (up to a normalization factor, of course)
$$
A(s,t)~=~ \bigl[~{t \over M_C^2}~ \bigr]^2~~
\bigl[~{s \over t+M_C^2}~ \bigr]
\auto\label{psca}
$$
It might be tempting to interpret the first factor 
as related to the Adler zeroes that should occur at zero four momentum
for each pion. However, our analysis has been carried through with 
the constraint that $k^2~ (= -t) ~>> M_C^2$ and so (\ref{psca}) can not 
be used at $t=0$. Because of the $\epsilon$-tensor that appears
in the current coupling, it is the transverse (with respect to the regge
limit) component of each of the $p_i$ that contributes to the factor
of $t$. It is natural, therefore, that
if the pions aquire a mass $m_{\pi}$ we will have, when all
pions are on-shell,
$$
\bigl[~{t \over M_C^2}~ \bigr]^2 ~\to~ 
\bigl[~O \bigl( ~{m_{\pi}^2 \over M_C^2}\bigr)~\bigr]^2 
\auto\label{mpi}
$$

The massive SU(2) singlet gluon reggeizes in higher-orders, with an 
infra-red finite trajectory $\alpha_g(t)$ that satisfies 
$\alpha_g(M_C^2)=1$. Also, since we consider the exchange of
four transverse momentum gluons, 
when we add all diagrams only the even signature
amplitude will survive. (Indeed, it is argued in \cite{arw01} that 
only even signature exchanges can couple via the anomaly.)
Therefore, as we add all diagrams and go to
higher-orders we anticipate that we will have 
$$
\bigl[~{s \over t+M_C^2}~ \bigr] ~\to~ \bigl[~{s^{\alpha_g(t)} ~+~
(-s)^{\alpha_g(t)}
 \over t+M_C^2}~ \bigr]
\auto\label{hor}
$$
and so there will be no pole at $t+M_C^2=0$. Nevertheless, reggeized gluon
exchange will provide the leading contribution to the pomeron.
It is interesting, of course, that only
the quark (or the antiquark) carries the light-like momentum of the pion
that produces the high-energy behavior. This is determined by the
generation of the anomaly pole via an internal light-cone 
momentum,  and we comment further
on this below.

The factor of 
$$
\bigl[~{s~q^2 \over M_C^4}~ \bigr]
\auto\label{osed}
$$
is off-shell energy dependence that could be naturally canceled
by off-shell propagators. Finally, we note that
the factor of
$$
\bigl[~{C^2~q^2 \over M_C^2}~ \bigr]^4
\auto\label{wpd}
$$
is a wee gluon contribution, with $Cq$ being the boosted longitudinal
momentum of wee gluons that in the finite momentum frame have 
vanishing momentum, orthogonal to the fast quark. In higher-orders
this contribution will 
include sums of $|M^0_N|^2 $ integrals as factors, 
where $M^0_N$ appears in (\ref{cld})
and contains diagrams of the form illustrated in Fig.~3.3. To say more
about this factor it is probably necessary to decouple the mass-shell
and regge limits by performing the full multi-regge calculation
discussed above.

\subhead{3.12 The Parton Amplitude and Color Confinement }

We have emphasized that the exponentiation of infra-red 
divergences already selects color zero transverse momentum states but that
this is not confinement because color zero massless
multigluon states still contribute
at zero $Q^2$. However, if the complete set of physical amplitudes is 
defined via the presence of the overall infra-red divergence we have
described then 
there will be both confinement and chiral symmetry breaking. This is because
the massless multigluon
states will contribute only via the condensate and the initial 
and final states must be Goldstone bosons for the divergence to be present.
Although we have not kept color factors we can make the following
comments about how color confinement is realized. 

In the original diagrams of the form of Fig.~3.12(a) 
the color factors have all of the 
complexity of the $\gamma$-matrix structure 
illustrated in Figs.~3.20 and 3.22. After removal of the color singlet 
divergent gluons, however,
the remaining amplitude necessarily describes SU(2) color zero scattering.
In this amplitude, we can interpret the flavor anomaly as unlocking the
quark content of a pion via a dynamical fluctuation of the Dirac sea 
(i.e. the zero momentum chirality transition). This fluctuation produces a 
``hard'' quark carrying all the 
light-cone momentum together with a wee gluon condensate
and an antiquark. The antiquark carries only a soft momentum and 
is also, essentially, a ``wee parton''. It 
has been produced out of
the Dirac sea via a chirality transition that is compensated for by 
the (effectively classical) background gluon field. 

The full amplitude can be represented, as in Fig.~3.23(a), 
by simple massive gluon exchange between the fast quarks.
This ``parton interaction'' 
produces all the transverse momentum that is exchanged.
The quark/gluon coupling is not, however, a normal perturbative
interaction. Although, as illustrated in Fig.~3.23(b),
it can be computed (``semi''-)perturbatively.
\begin{center}
\parbox{3.8in}{
\begin{center}
\leavevmode
\epsfxsize=3.5in
\epsffile{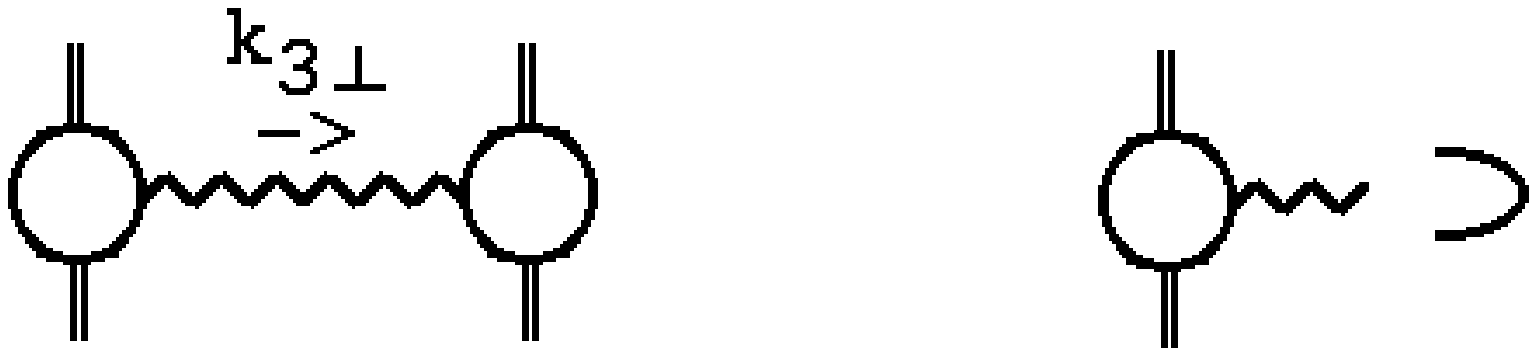}
\end{center}
}
\parbox{1.4in}{
\epsfxsize=1.3in
\epsffile{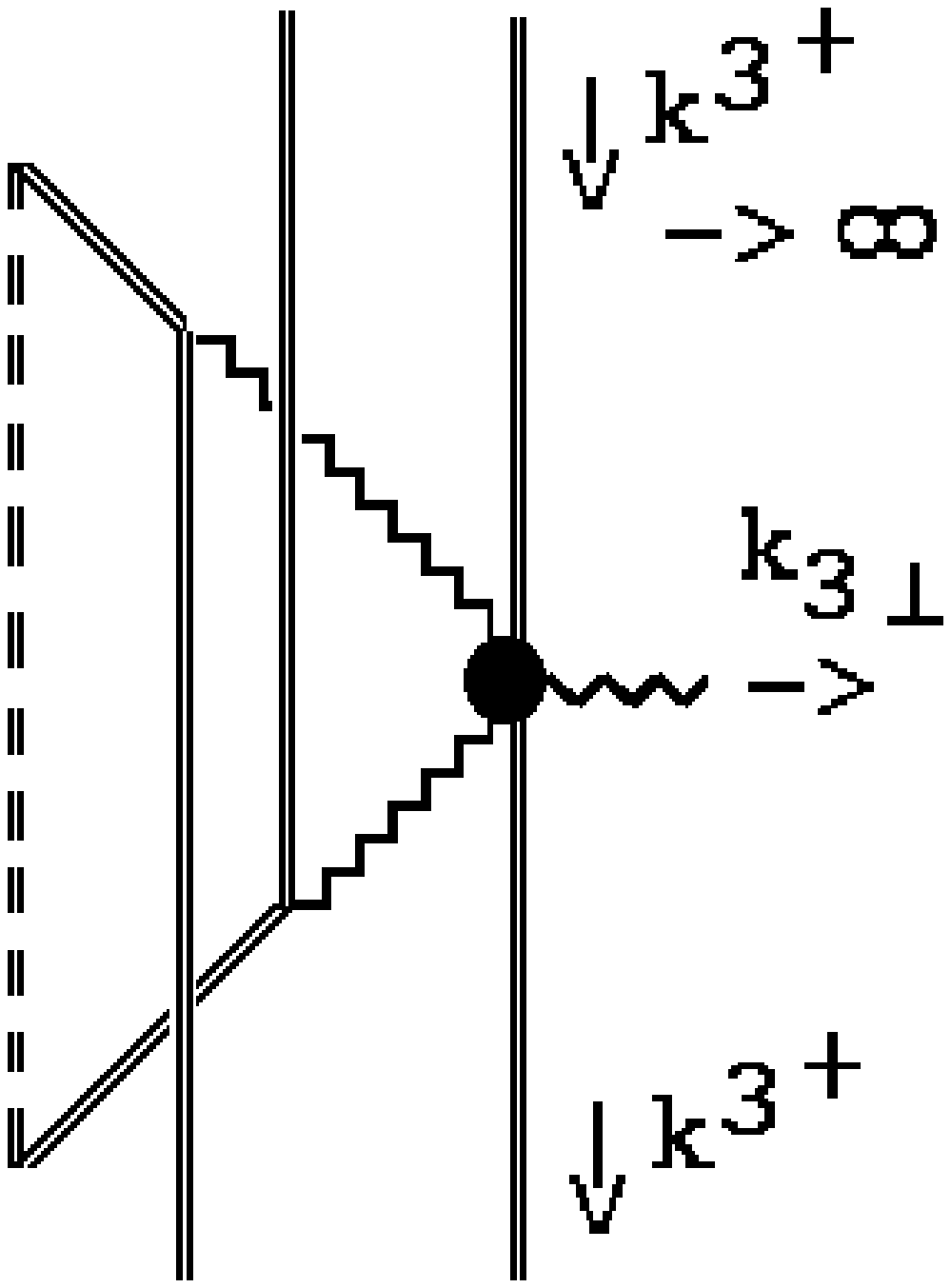}
}

(a) \hspace{2.5in} (b)$~~~~~~~~~~~~~~$

Fig.~3.23 (a) The Parton Amplitude (b) The Parton Interaction.
\end{center}
Accompanying
the hard quark interaction, there is a soft interaction in which
the slow antiquark, ultimately, is absorbed into the condensate. It is replaced
by another antiquark produced out of the condensate. The production
and absorption being mediated by a further
zero momentum quark chirality transition
(shown as the double-dashed line).
 During the interaction (which we have redrawn
compared to earlier figures to make its structure
more transparent) color and 
spin structure, but not momentum, is fed 
into the fast quark/gluon coupling (the massive
gluons can carry SU(2) color).  The spin structure input
transforms this coupling from a vector 
to an axial vector coupling. This being made possible by the chirality
transition of  the zero momentum quark. 
The input of color into the fast quark interaction helps
convert the odd-signature single gluon exchange to even signature.

An outgoing fast quark carries color, which is neutralized by a
(condensate produced) soft antiquark. The Dirac sea 
completes the confinement by locking the pairs back 
into a massless Goldstone boson pion 
via a final zero momentum chirality transition of the soft antiquark
that is accompanied
by the disappearance of the background ``classical gluon field''.  
Apparently then, in the infinite momentum frame, a physical pion
contains a hard elementary quark plus a color compensating 
``unphysical antiquark''  that is described by an antiquark field, but with the
Dirac sea shifted. Conversely the quark/antiquark constituents of a pion
can not be liberated without an accompanying gluon field that is responsible
for moving the Dirac sea back to it's perturbative location. 
That the dynamical participation of the Dirac sea frees and confines 
infinite momentum frame quarks (and also modifies interactions)
in this manner is natural if
in a finite momentum pion the 
quarks are confined by a non-perturbative adjustment of the Dirac sea, 
as proposed by Gribov\cite{vg1}. Since no strong force between quarks
is involved, Dokshitzer\cite{dok} has called this ``soft and gentle
confinement''. He has argued for some time that 
significant experimental evidence for this form of confinement is provided by 
the momentum properties of multihadron production. Since there appears to be
very little momentum reordering in the transition from quarks to pions,
confinement must take place in a soft and gentle manner. A readjustment of the 
Dirac sea of the soft quarks/antiquarks that combine with the hard quarks
to form hadrons should have just this property. 

\subhead{3.13 The Supercritical Pomeron}

In higher-orders more massive gluons will be exchanged and more wee
\newline quark/antiquark pairs will input additional structure into the 
interaction. An example of a coupling that will produce two pomeron exchange
is shown in Fig.~3.24(a). 
We also expect to find vertices, of the form shown in Fig.~3.24(b),
which include a pair of massive gluons produced by a wee gluon interaction
only.
\begin{center}
\leavevmode
\epsfxsize=3.3in
\epsffile{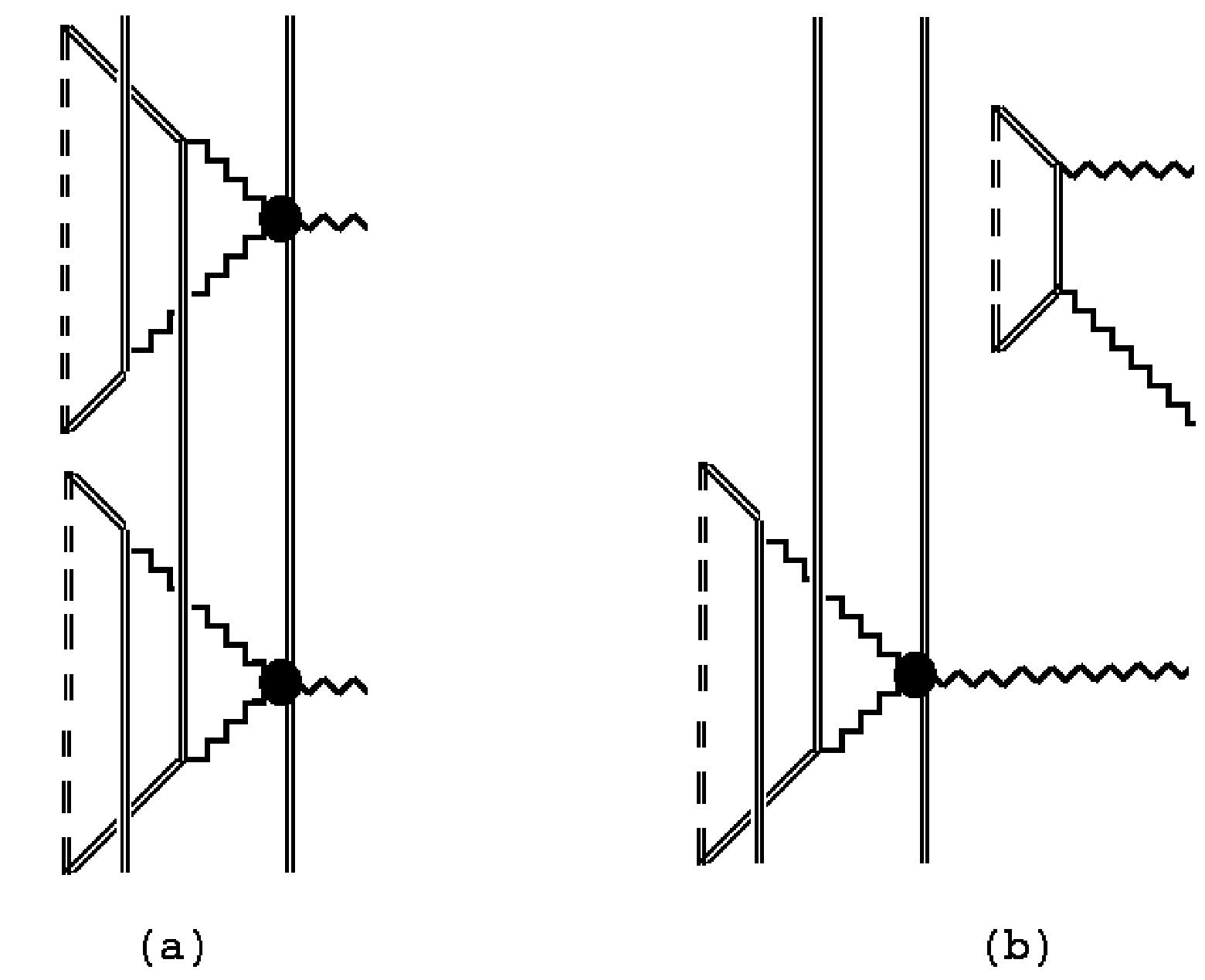}

Fig.~3.24 (a) A Two Pomeron Vertex (b) A Supercritical Vertex?
\end{center}
 To have the axial vector structure for the anomaly, 
the produced gluon represented by the diagonal element can not have the
polarization to be exchanged in the scattering process. The wee gluon 
interaction can, however, take place sufficiently far across the rapidity
axis that it leads to particle pole interactions within pomeron 
vertices, just
as is expected in the supercritical pomeron phase\cite{arw91}.
Since the pomeron is also exchange degenerate with the reggeized gluon,
all features of Supercritical RFT appear to be present.

\newpage 

\mainhead{4. DISCUSSION}
  
The analysis of this paper demonstrates clearly how
(at least the zero momentum part of) the spectral
flow of the Dirac sea, which does not enter in standard
perturbation theory, enters the (multi-)regge region interactions 
that describe the scattering of bound states.
The manifestation of this spectral flow is 
the chirality transition that a 
zero momentum propagator undergoes in producing the anomaly pole.
While we had formulated 
the basic physics of this phenomenon in our previous papers, we had been
unable to find a simple starting point to begin to calculate amplitudes
in a sufficiently well-defined way. In the new approach presented in this 
paper the wee gluon content of Goldstone bosons, produced by the flavor 
anomaly, provides this starting point. The
rotation of the wee gluons during the scattering process introduces,
essentially, the triple-regge kinematics 
needed for the U(1) anomaly interactions to appear. 
As we noted in \cite{arw01}, 
because of chirality conservation the anomaly interactions
cancel, even when the kinematics allow their presence, 
if the scattering states are elementary quarks or gluons. In contrast, since
the initial wee gluon coupling of
the pion pole involves a chirality transition, there is no reason
for the chirality transitions to cancel in the subsequent scattering.

As has become apparent,
our ``new approach'' is 
not actually logically separate from the multi-regge formalism used in 
our previous papers. Rather it is, essentially,
a short-cut that reproduces multi-regge results without doing the full 
calculation. The basic idea we have used 
is that the internal light cone momenta 
of the flavor anomaly couplings introduce all the large light-cone momenta
needed, in addition to the elastic scattering regge limit,
and so this avoids the introduction of complicated multi-regge limits.
This has enabled us to keep the kinematics ``relatively'' simple. However,
we have had to supplement our analysis with additional constraints that
appear artificial but really just introduce features that would be provided
directly by an underlying multi-regge limit.
The ``axial currents'' to which our Goldstone bosons have coupled    
are not local currents but rather 
effective local current components that would be 
produced by a non-local infinite momentum 
interaction. Such current components appear naturally within
a multi-regge amplitude. 

We have described the formation of amplitudes  
in terms of transverse momentum 
diagrams that can be thought of, initially, as originating from
particular feynman diagrams. However, many of the 
integration regions in the feynman diagrams are   
cut-off, or even removed altogether. Again multi-regge theory 
provides the underlying justification. 
To have all the necessary anomaly effects present 
the initial diagrams must be extremely
complicated. Remarkably, though, after infra-red divergences are extracted
and the anomaly contributions isolated, almost all of the 
complexity disappears and the physical pion scattering amplitude
has the very simple structure we have described. 
Although we have not discussed combining diagrams to obtain explicit
color and signature factors it is clear that, in first approximation,
the pomeron is a regge pole with 
the same trajectory as a massive, reggeized, gluon 
just as we anticipated in our multi-regge work. 

It is amusing (and there may also be deeper implications)
that a complete calculation of the multi-regge 
S-Matrix would not be necessary to obtain our results. It would be
sufficient to calculate the eight-point amplitude for 
$W^{\pm}$ and $Z^0$ vector mesons in which the scattering of
reggeized pions occurs. The reggeized pion scattering amplitude could be
factorized out and the on-shell amplitude we discuss would be obtained 
by continuing this amplitude from a spacelike to a light-like pion
mass. In the language of the present paper this implies
that the wee gluon structure of a pion is best understood if it is obtained
as a wee parton component of an infinite momentum, elementary, vector meson! 

We believe that in the multi-regge framework the existence 
of a reggeon condensate in color superconducting QCD
would clearly be a derived result. In addition,
reggeized Goldstone bosons
would be the only composite states obtained. The arguments of
\cite{arw98} imply that if the initial reggeon states are
color zero Goldstone bosons the overall logarithmic divergence
will produce final states only of this kind. That is, there should be
a completeness relation. In this case, 
the condensate (or rather the
infra-red divergences and anomalies that produce it) can be said
to be responsible for confinement and chiral symmetry breaking.
Conversely, the quark content of a pion or nucleon is ``liberated''
only if it is accompanied by an (effectively classical) gluon 
background ``condensate'' 
that is associated with a shift of the (zero momentum
part of) the Dirac sea. The implication being that, at infinite momentum, 
quarks are locked inside a hadron by a relatively simple spectral flow of the
Dirac sea. This form of confinement would have a natural connection
with the finite momentum Dirac sea confinement proposed by Gribov\cite{vg1}.

As we have described in more detail in 
other places\cite{arw98,arw93}, we expect that
SU(3) color is obtained by critical pomeron behavior\cite{cri} 
that randomizes the 
SU(2) direction of the condensate within SU(3), while
also decoupling the massive reggeized gluon, as it becomes massless. Thus
providing complete SU(3) confinement.
The shifting of the Dirac sea that produces confinement 
then becomes a completely dynamical part of the pomeron, and hadrons, that
has no simple ``classical'' component.
With the better understanding and explicit calculational ability that
the results of this paper demonstrate, we should
be able to directly identify the higher-order superconducting
pion amplitudes with those of supercritical RFT and so establish 
the connection 
between the critical pomeron\cite{cri} and QCD. A further implication
will be that the physical states of QCD (or 
rather those that scatter via the physical pomeron) are
either chiral symmetry breaking Goldstone bosons (pions) or contain, as a 
component, a two quark state that is a Goldstone boson in the color
superconducting theory (nucleons). Conversely, 
the very nature of the pomeron will be determined by chiral symmetry breaking. 

A basic implication of our general program has always been that
the regge limit of QCD, including those properties that are a consequence of
confinement and chiral symmetry breaking,
would be reachable by essentially perturbative
calculations - with the dynamical participation of the Dirac sea being 
the only extra ingredient. The results of this 
paper emphasize this implication. According to our results, 
the only  non-perturbative element in color superconducting
high-energy amplitudes is the wee-gluon condensate which can be directly
understood as a consequence of the all-orders summation of transverse
momentum infra-red divergences that couple via anomalies
(together with the introduction of ultra-violet cut-offs). We should note that 
the condensate is associated with wee gluon configurations that have the
same quantum numbers as the winding number current. Although, as with
the currents we use to obtain the flavor anomaly, they are really
infinite momentum local current components that result from
non-local interactions. Nevertheless, 
there could be a parallel with the Schwinger model where the existence
of a condensate can be obtained either by summing diagrams\cite{fr} or via 
non-perturbative topological contributions\cite{nm}. 
However, the topology would
have to be in the infinite momentum frame. (Perhaps a winding number for
Wilson loop operators in the transverse plane - this might give a direct
analogy with the Schwinger model.)

We have emphasized that, in order to 
construct high-energy superconducting QCD as we described,
it is necessary to introduce cut-offs both in the transverse momenta
and in the internal momenta of diagrams that generate anomalies.
In effect, these 
cut-offs regulate the relative infra-red/ultraviolet 
spectral flow of the Dirac sea that is due to the chiral 
and U(1) anomalies.
That all cut-offs can be 
consistently removed, and the necessary critical behavior
retained, is a highly non-trivial requirement which, as we have discussed
elsewhere\cite{arw93,arw00}, 
is likely to significantly restrict the quark content of QCD.
However, it is possible (if not likely) that the very existence of a hadron
S-Matrix within QCD requires that asymptotic freedom, and the 
consequent perturbation theory, have the
maximal applicability. Since parton model cross-sections rise asymptotically,
this is likely to imply that all physical 
cross-sections must
rise asymptotically. The critical pomeron is well-known 
to be the only description of such cross-sections that satisfies all 
($s$- and $t$-channel) unitarity properties. Consequently, the occurrence of
the critical pomeron in QCD may actually be a necessary requirement for
the existence of a hadron S-Matrix.

The non-perturbative formulation of a gauge theory is generally presumed
to be via some form of euclidean functional integral. In this 
framework color confinement (as it is usually formulated and studied)
is completely disjoint from perturbation theory. In fact, the general 
expectation is that there will be a ``non-perturbative'' pomeron that is  
crucially dependent on confinement and, as such, is
far removed from perturbation theory. However, 
the regge region involves a mixture of large light-cone and 
small transverse momenta and so appears only in Minkowski space.
As a consequence, if the 
euclidean path integral is the starting point, detailed properties of the
pomeron can only be determined by a complete non-perturbative solution of the
theory from which Minkowski space hadron scattering amplitudes can be
extracted and the regge limit taken. 
Something that seems unlikely to be possible for a very long time to come.
Indeed, given that a complete non-perturbative solution of QCD has been 
found, the pomeron would probably be one of the last things to be studied.
Note that, since light-cone momentum regions become all-important as 
the continuation to Minkowski space is made, the very existence of
this continuation is likely to be contingent on the existence 
of (unitarity?) boundedness properties in the regge region. 

We would like to emphasize that there is no guarantee that
a Minkowski region unitary S-Matrix can be derived from a 
non-perturbative euclidean 
path integral - particularly given the complexity\cite{vg} of the, 
large field, unphysical 
degrees of freedom that are present. (Indeed a commonly agreed
procedure to definitively eliminate these degrees of freedom has 
not yet been found.) The demonstrated perturbative unitarity\cite{gth} is 
only a formal property since infra-red divergences prevent
the existence of a finite S-Matrix. There is certainly no understanding of 
how the unitarity properties of the perturbative theory might translate
into unitarity with respect to a non-perturbative
physical spectrum that manifests confinement
and chiral symmetry breaking. Indeed it is our strong belief that the regge 
region must play a special role  
in unraveling this relationship within QCD. 
Since small transverse momenta are involved, 
the physical properties of confinement and 
chiral symmetry breaking must be evident in the t-channel unitarity condition. 
Conversely, if asymptotic freedom has 
maximal applicability, the involvement of large momenta should imply 
that the pomeron is not too far from perturbation theory. 
Therefore, the (multi-)
regge region should provide a unique possibility
to understand the relationship between perturbation theory and the physical
states appearing in the unitarity condition.

If the regge limit of QCD can be constructed by the essentially 
perturbative methods we describe then
the unitarity properties of massive quark and gluon reggeon 
diagrams translates into similar properties for the pomeron and hadron
reggeon diagrams. The unitarity of the critical pomeron will be clearly
related to the original perturbative unitarity of quarks and gluons.
Indeed it could also be that, since the construction
stays so close to perturbation
theory, the problem of eliminating large field unphysical degrees
of freedom will have been avoided.

\end{document}